\documentclass[11pt,a4paper]{article}

\usepackage{amsmath}
\usepackage{graphicx}
\usepackage{amsfonts}
\usepackage{subcaption}
\usepackage{float}
\usepackage{wrapfig}
\usepackage{url}
\usepackage{tabularx}
\usepackage{caption}
\usepackage[singlelinecheck=false]{caption}
\usepackage{multicol}
\usepackage[superscript,biblabel]{cite}
\date{}
\renewenvironment{abstract}
 {\par\noindent\textbf{\abstractname}\ \ignorespaces}
 {\par\medskip}
 \usepackage[utf8x]{inputenc}

\begin{document}

\title{\textit{Ab initio} study on the atomic and electronic structures of twisted InSe bilayer}

\author{Siow Mean Loh and Nicholas D. M. Hine$^{*}$ \\ Department of Physics, University of Warwick, \\ Coventry CV4 7AL, United Kingdom.} 
\maketitle
\noindent E-mail: \underline{n.d.m.hine@warwick.ac.uk} \\
\\
\begin{abstract} 
The electronic properties of the twisted InSe bilayer are studied by large-scale density functional theory. Spectral Function Unfolding reveals that the electronic structure of the twisted system can be described in terms of a combination of features of the bandstructures of the aligned InSe bilayer with different stacking configurations, enabling predictions of the band gap and the effective mass for holes. The effective mass for holes in the twisted InSe bilayer is shown to be influenced primarily by the interlayer distance. The intralayer and interlayer exciton binding energies are thus calculated based on a model recently developed by Ruiz-Tijerina et al. We apply similar analysis to the trilayer heterostructure InSe/hBN/InSe: its electronic structure is shown to be well-described by the superposition of band structures of two InSe monolayers with a small coupling through the hBN layer.
\end{abstract}


\begin{multicols}{2}
\section{Introduction}
The ideas behind the burgeoning field of twistronics have been applied in recent years to a range of different low-dimensional materials. In twisted bilayer graphene (TBG), the intervalley coherence between electrons at the two valleys leads to a Mott insulator phase and to superconductivity \cite{Origin_Mott_insulating_superconductivity_twisted_graphene-bilayer}. Tarnopolsky et al. proposed that flat bands appear at magic angles with a periodicity of $\Delta \alpha \simeq$3/2 ($\alpha \sim 1/\theta$) in TBG. They suggested that lowest Landau level in a torus is similar to the flat bands seen in TBG \cite{Origin_magic-angles_twisted_graphene-bilayer}. The Dirac electrons in twisted graphene bilayer were shown to be localised in regions corresponding to AA stacking, because of the moir\'e potential. Their velocities approach toward zero as the twist angle decreases to be smaller than 3$^{\circ}$, leading to the appearance of flat bands \cite{Localisation_Dirac-electrons}. Lopes dos Santos et al. showed that these localised states can be found in regions other than AA stacking if the twist angle is smaller than the magic angle (1.08$^{\mathrm{o}}$) based on the continuum model \cite{Localised-states_other-regions}. The same localisation in the AA stacking for the flat band can also be seen in twisted MoS$_{2}$ bilayer from a tight-binding model \cite{twisted_MoS2-bilayer}. 

Furthermore, Zhan et al. have shown that ultraflat bands occur at any small twist angles in twisted transition metal dichalcogenides (TMDCs) and the localisation of the charge density becomes obvious as the twist angle decreases based on a tight-binding model. They suggested that the strongly correlated states can be well studied as the twist angle below 7$^{\mathrm{o}}$ \cite{twisted_TMDCs}. The long-period moir\'e pattern induced by a small twist angle in twisted TMDCs leads to the significant changes of optical absorption spectrum \cite{twisted_TMDCs-1}. 

The anisotropic transport properties of the twisted black phosphorene bilayer can be tuned by varying the twist angle \cite{twisted_black-phosphorene-bilayer}. Furthermore, a spontaneous electric polarisation was seen for the same twisted system with weak interlayer coupling \cite{twisted_black-phosphorene-bilayer}. The interlayer coupling strength is determined by the twist angle. The band gap changes differently in strong and weak interlayer coupling systems: band gap decreases with increasing electric field for strong interlayer coupling systems, whereas there is a linear Stark effect for weak interlayer coupling systems \cite{twisted_black-phosphorene-bilayer}. In addition. ferroelectric-like domains were seen in the twisted hBN bilayer, both in theory and experiments, because there is a large surface potential originating from the out-of-plane B-N dipoles in the two different layers \cite{Ferroelectric_domains_twisted_hBN-bilayer}. 

Previous studies have shown that the carrier mobility is high in few-layer InSe and the variation of the band gap is large (more than 1 eV) for different thickness of InSe. Furthermore, an indirect-direct band gap transition can be seen by tuning the thickness of InSe due to the strong interlayer coupling \cite{Carrier_mobility_few-layer_InSe, Carrier_mobility_InSe, InSe_strong_interlayer_coupling}. The indirect-direct band gap transition is also seen when tuning the twist angle between two InSe bilayers from 3$^{\mathrm{o}}$ to 1.48$^{\mathrm{o}}$ with an increase of band gap at the same time \cite{Twisted_InSe_bilayer}. The charge density for the state at the valence band maximum (VBM) becomes more localised in the region with the smallest interlayer distance when the twist angle decreases to 1.48$^{\mathrm{o}}$. However, the charge density for the conduction band minimum (CBM) spread over the whole atomic structure \cite{Twisted_InSe_bilayer}. Structural relaxation is important for this system because the charge density localised in the region with the largest interlayer distance without structural relaxation. The formation energies for twist angles from 1.48$^{\mathrm{o}}$ to 3$^{\mathrm{o}}$ are similar, suggesting that the twisted systems with twist angle from 1.48$^{\mathrm{o}}$ to 3$^{\mathrm{o}}$ are all possibly grown in the experiment \cite{Twisted_InSe_bilayer}. 

In this work, we have studied the band structure and the projected density of states of the VBM, firstly for the InSe bilayer with different stacking configurations, and then for twisted InSe bilayer. The displacement (and residual force distribution) of each atom after geometry optimisation is also discussed. It is shown that the band structure from the primitive cell calculation can be used to describe the effective band structure of twisted InSe bilayer from the supercell calculation. The interlayer distance resulting from the different stacking configurations in twisted InSe bilayer is a major factor in terms of its influence on the band structure. Finally, the hBN-spaced twisted InSe bilayer is also introduced and discussed in this work. Its band structure is very similar to the superposition of two InSe monolayers, and also matches well with the band structure of InSe monolayer from the primitive cell calculation. This indicates that a single spacer layer of hBN is sufficient to recover essentially monolayer-like behaviour.

\section{Computational Methods}
The atomic and band structures of InSe bilayers with different stacking configurations were calculated within the primitive cells by Quantum Espresso simulation package \cite{Quantum_Espresso, Quantum_Espresso-1}. Scalar-relativistic ultrasoft pseudopotentials from the GBRV library \cite{GBRV_library} were used. Van der Waals (vdW) interactions were treated by adopting optB88-vdW functional \cite{optB88-vdW} as the exchange-correlation functional. The optimised norm-conserving Vanderbilt pseudopotentials (fully-relativistic) \cite{Vanderbilt_pseudopotential} for GGA-PBE from PseudoDojo \cite{PseudoDojo} was adopted for the calculations which considered spin-orbit coupling. The vdW correction was not considered in the calculation including spin-orbit coupling.

A large vacuum spacing of 20 \AA \, was used to avoid spurious interaction between periodic images of the system. Calculations were performed with a 1632.7 eV cut-off energy and a 10$\times$10$\times$1 Monkhorst-Pack \cite{Monkhorst-Pack} kpoint sampling grid. During the ionic minimisation process, electronic convergence was such that forces were converged to  5$\times$10$^{-5}$ eV/\AA \, and the total energy tolerance for all atoms was 3$\times$10$^{-7}$ eV.

Calculations on the twisted bilayer system must be performed at very large length scales to enable a low lattice mismatch between the coincident supercells representing each layer. The linear-scaling density functional theory (LS-DFT) package, ONETEP \cite{ONETEP, ONETEP_program}, was used to perform the calculations of the twisted InSe bilayer and InSe/hBN/InSe heterostructure in this work. The projector augmented waves (PAWs) \cite{PAW-1, PAW-2, PAW_in_ONETEP} with the GGA-PBE potentials from the GBRV library \cite{GBRV_library} were employed. The same vdW correction functional (optB88-vdW \cite{optB88-vdW}) was also adopted in this simulation package. The atomic structures with a vacuum spacing of 20 \AA \, were calculated with a 1200 eV cut-off energy for the psinc grid and 12 a$_{0}$ NGWF radius. The geometry optimisation force tolerance is 0.12 eV/\AA \, and enthalpy tolerance is 3$\times$10$^{-5}$ eV/atom. 

Python-based tools building on the atomic simulation environment (ASE) \cite{ASE} were used to generate the atomic structures and display the distribution of force. The visualisation for electronic and structural analysis (VESTA) \cite{VESTA} was also used to display the atomic structures and the isosurfaces of charge densities. 

Appropriately-sized supercells are found by searching for a pair of coincident supercells of each of the two layers $A$ and $B$ for which both in-plane lattice vectors of each layer match with minimal strain. One layer ($B$ here) is subjected to a rotation of angle $\theta$ and, to make the match exact, a near-unitary strain tensor is applied. This produces the following equations, similar to the method described in Stradi et al: \cite{Matching_twisted-bilayer}:
\begin{equation}
\begin{bmatrix}
\textbf{L}^{A}_{1} \\
\textbf{L}^{A}_{2}
\end{bmatrix}
= 
\begin{bmatrix}
\textbf{L}^{B}_{1} \\
\textbf{L}^{B}_{2}
\end{bmatrix},
\end{equation}
where
\begin{equation}
\begin{bmatrix}
\textbf{L}^{A}_{1} \\
\textbf{L}^{A}_{2}
\end{bmatrix}
= 
\begin{bmatrix}
l & m \\
n & k
\end{bmatrix}
\begin{bmatrix}
\textbf{a}^{A}_{1} \\
\textbf{a}^{A}_{2}
\end{bmatrix},
\end{equation}
and
\begin{equation}
\begin{bmatrix}
\textbf{L}^{B}_{1} \\
\textbf{L}^{B}_{2}
\end{bmatrix}
= 
\begin{bmatrix}
\sigma_1 & \lambda_1 \\
\lambda_2 & \sigma_2
\end{bmatrix}
\begin{bmatrix}
p & q \\
r & s
\end{bmatrix}
\begin{bmatrix}
\cos{\theta} & -\sin{\theta} \\
\sin{\theta} & \cos{\theta}
\end{bmatrix}
\begin{bmatrix}
\textbf{a}^{B}_{1} \\
\textbf{a}^{B}_{2}
\end{bmatrix},
\end{equation}
for integer $l,m,n,k$, $p,q,r,s$, $\sigma_i \simeq 1$ and $\lambda_i \simeq 0$. The finding of a supercell match proceeds by first choosing a range of $\theta$ in which to search, and a range of supercell size, then for each set of values of $l,m,n,k$ for which the determinant of the supercell matrix is in that range, the values of $p,q,r,s$ that most-closely match for layer $B$ are determined, then the values of $\sigma_i$ and $\lambda_i$ that make the match exact are found. If these are below a chosen threshold, the match is accepted, or if not, the angle is further refined until no further improvement can be obtained.

Generally, there is a trade-off between the supercell size and the magnitude of strain that must be accepted in order to make a coincident cell. The supercell size is expected to increase as the twist angle decreases ($\propto$ 1/$\theta$) \cite{Twisted_InSe_bilayer}. For the supercells used in this work, the values of strain $\sigma_i$ on the InSe layers are under 0.8\% and those of shear $\lambda_i$ are zero to numerical precision.

\section{Results and Discussions}
\subsection{InSe bilayers with different stacking configurations}
\begin{figure} [H]
\captionsetup[subfigure]{justification=centering}
\begin{subfigure}{0.093\textwidth}
  \centering
  \includegraphics[width=0.85\textwidth]{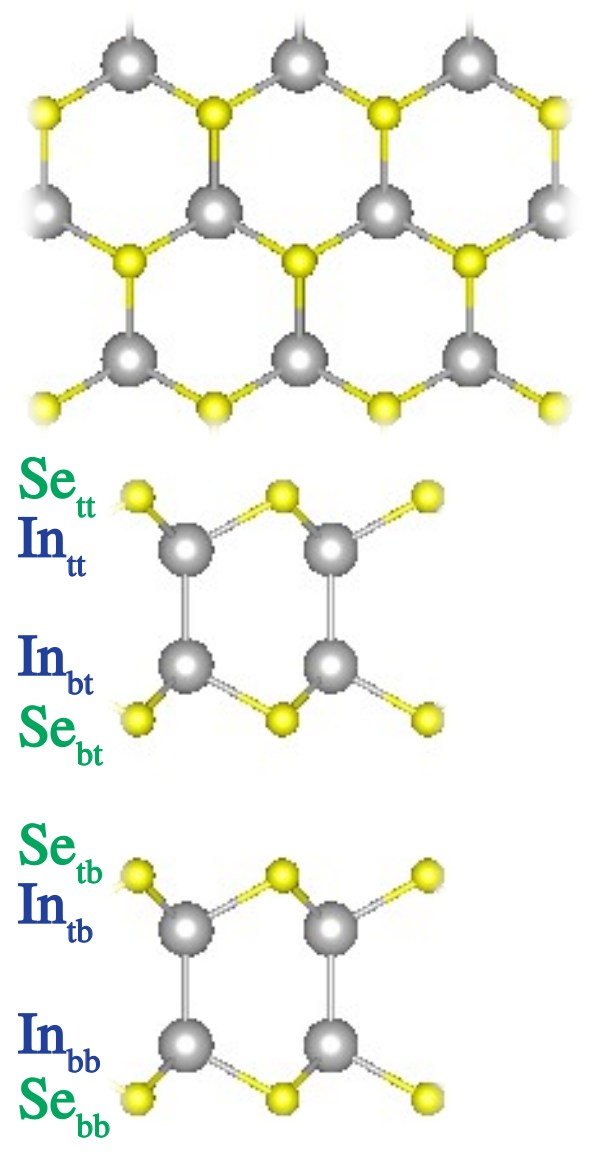}
  \caption{A-1}
\end{subfigure}
\begin{subfigure}{0.093\textwidth}
  \centering
  \includegraphics[width=0.9\textwidth]{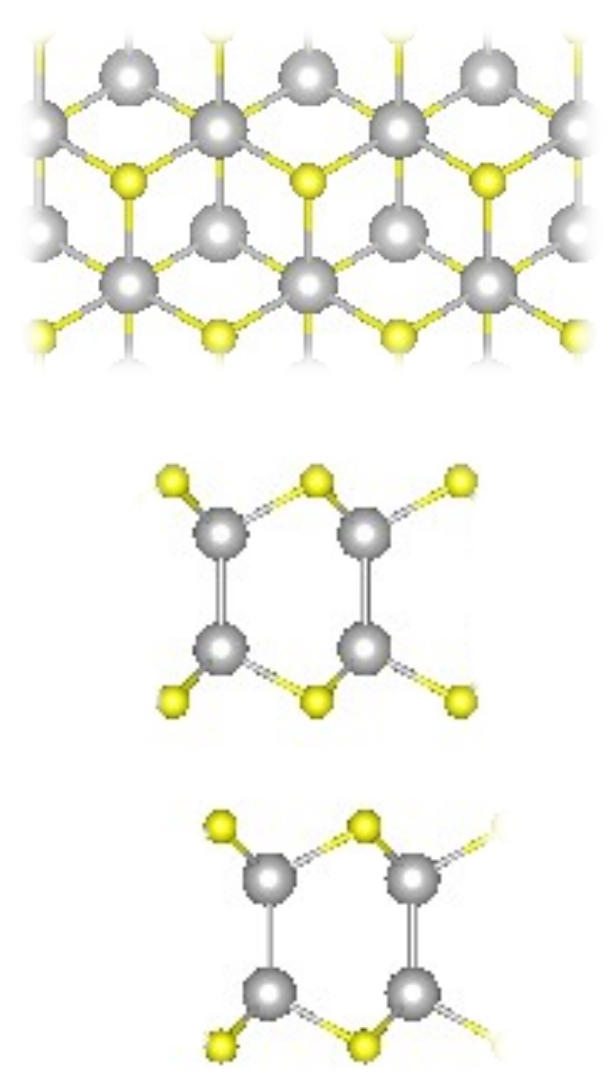}
  \caption{A-2}
\end{subfigure} 
\begin{subfigure}{0.093\textwidth}
  \centering
  \includegraphics[width=0.85\textwidth]{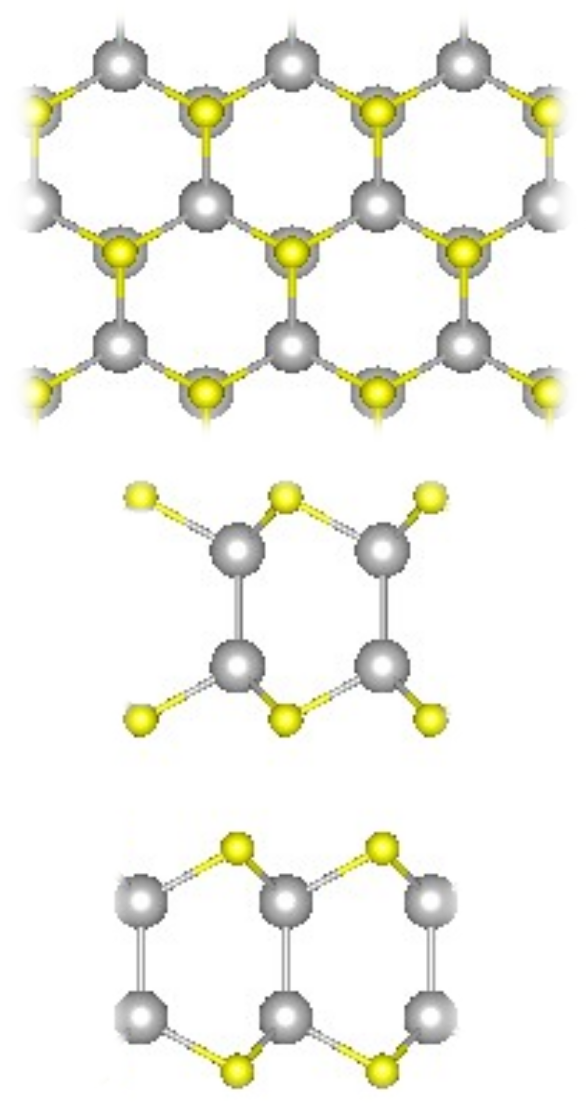}
  \caption{B-1}
\end{subfigure}
\begin{subfigure}{0.093\textwidth}
  \centering
  \includegraphics[width=0.85\textwidth]{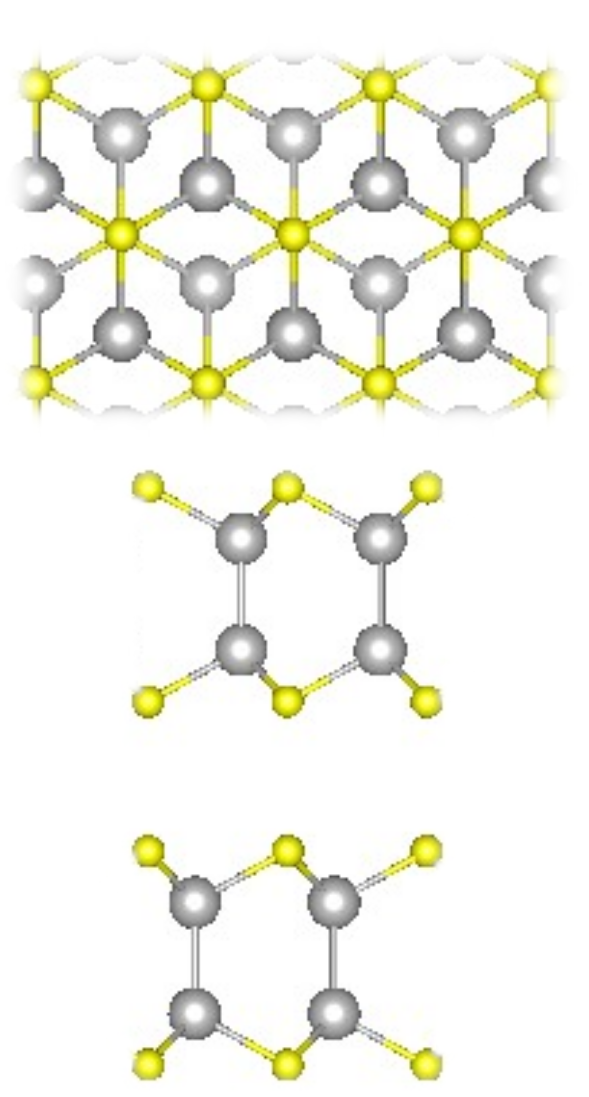}
  \caption{B-2}
\end{subfigure}
\begin{subfigure}{0.093\textwidth}
  \centering
  \includegraphics[width=0.85\textwidth]{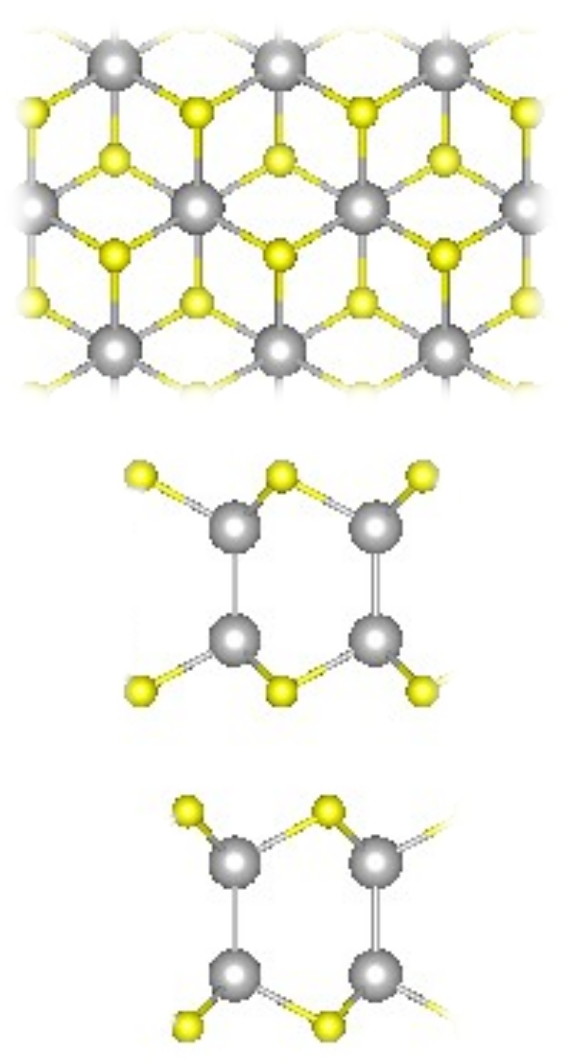}
  \caption{B-3}
\end{subfigure}
 \caption{ Schematic diagrams of InSe bilayers with five different stacking configurations, classified as two categories: A-type and B-type. In and Se atoms are shown in grey and yellow. In$_{ij}$ and Se$_{ij}$ ($i,j$= t,b) indicated in (a) refer to the vertical positions ($i$) of In or Se atoms in the top or bottom layers ($j$).}
\label{fig:InSe_AA-AB}
\end{figure}

When two aligned InSe monolayers are stacked on top of each other to form a bilayer, there are five possible stacking configurations, as addressed in previous studies \cite{Lattice_contant-Interlayer_distance_InSe_bilayer, Twisted_InSe_bilayer}. Fig. \ref{fig:InSe_AA-AB} shows the top and side views of these five stacking configurations. The A-1 and A-2 belong to the category we refer to as A-type, whereas the B-1, B-2 and B-3 are B-type. A stacking configuration in either category can be transformed into any of the others in the same category by a translation.

\begin{figure} [H]
\captionsetup[subfigure]{justification=centering}
\begin{subfigure}{0.24\textwidth}
  \centering
  \includegraphics[width=0.85\textwidth]{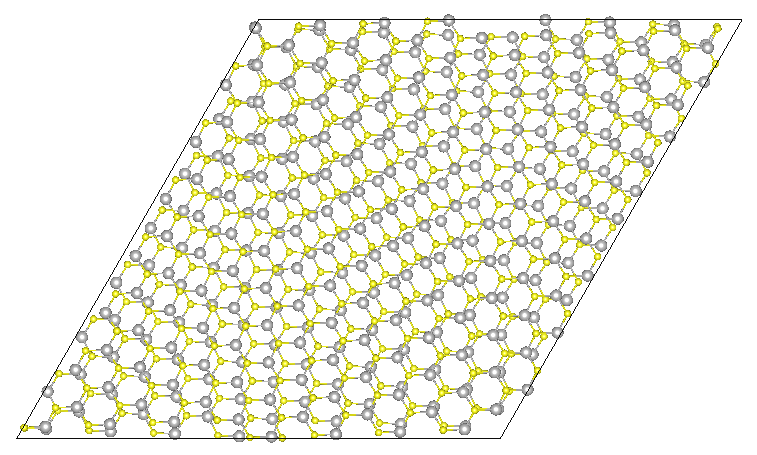}
  \caption{A-type twisted InSe bilayer}
\end{subfigure}
\begin{subfigure}{0.24\textwidth}
  \centering
  \includegraphics[width=0.85\textwidth]{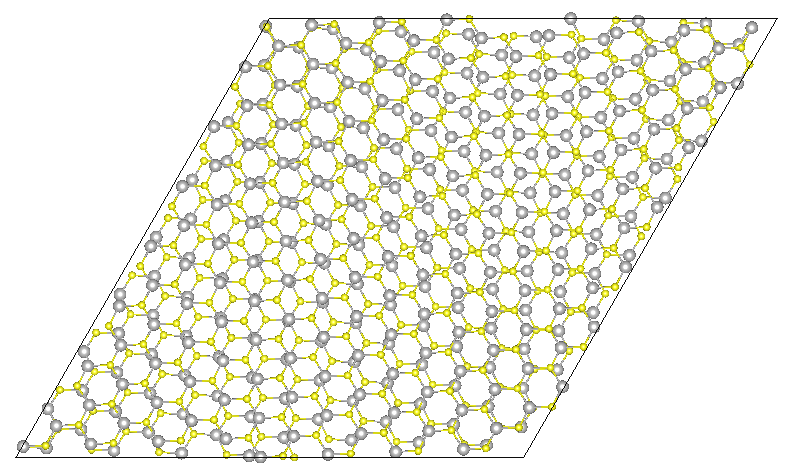}
  \caption{B-type twisted InSe bilayer}
\end{subfigure} 
\begin{subfigure}{0.24\textwidth}
  \centering
  \includegraphics[width=0.85\textwidth]{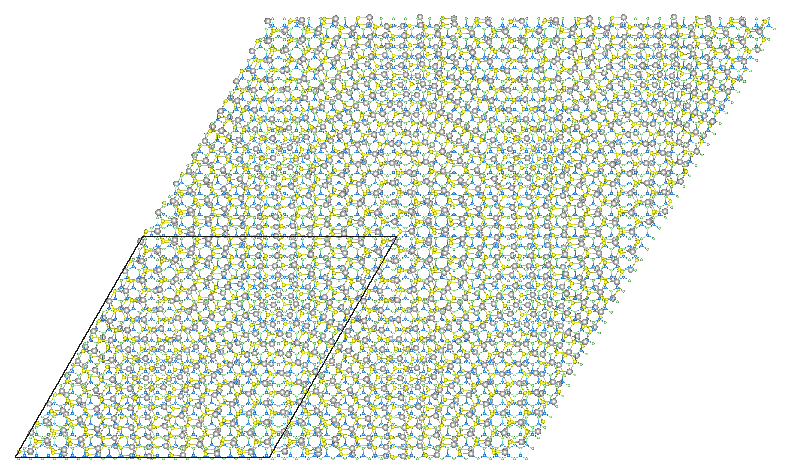}
  \caption{HBN-encapsulated A-type twisted InSe bilayer}
\end{subfigure}
\begin{subfigure}{0.24\textwidth}
  \centering
  \includegraphics[width=0.85\textwidth]{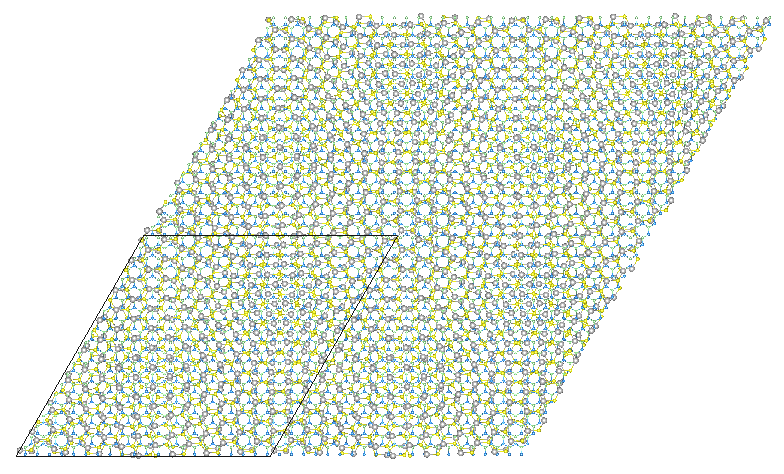}
  \caption{HBN-encapsulated B-type twisted InSe bilayer}
\end{subfigure}
 \caption{  (a,b) Schematic diagrams of the two types of twisted InSe bilayer (c,d) the hBN-encapsulated twisted InSe bilayer. (a) and (c) represent the atomic structures related to the A-type stacking configurations, whereas (b) and (d) represent the atomic structures related to the B-type stacking configurations. The unit cell is repeated to clearly demonstrate the moir\'e pattern. The twist angle between two InSe layers is 4.4$^{\mathrm{o}}$, while the twist angle between InSe and hBN layers is 32.2$^{\mathrm{o}}$. In, Se, B and N atoms are shown in grey, yellow, blue and green respectively.}
\label{fig:twisted_InSe-bilayer_hBN-trilayer}
\end{figure}

Fig. \ref{fig:twisted_InSe-bilayer_hBN-trilayer}(a) and (b) show the A-type and B-type twisted InSe bilayers, respectively.
Motifs corresponding to the different stacking configurations seen in fig. \ref{fig:InSe_AA-AB} can easily be identified in certain regions of the twisted system. For example, A-1 is seen at the corners of fig. \ref{fig:twisted_InSe-bilayer_hBN-trilayer}(a), whereas the A-2 is seen between the corners and the central region along the $\textbf{a}+\textbf{b}$ direction. Similar atomic structures for twisted InSe bilayers can be seen in a previous study \cite{Twisted_InSe_bilayer}. Fig. \ref{fig:twisted_InSe-bilayer_hBN-trilayer}(c) and (d) also shows the hBN-encapsulated twisted InSe bilayer which will be further discussed below.

A selection of atomic and electronic parameters are displayed for context in table \ref{InSe-bilayer_parameters}, specifically the lattice constant, interlayer distance, binding energy, band gap and effective mass for holes, for the InSe monolayer and for the five InSe bilayers in different stacking configuration. Previous studies \cite{ InSe-monolayer_lattice-constant_bandgap-2, B3_effective-mass} on InSe monolayer gave similar calculated results to those obtained here: $a$= 4.05, $E_{\mathrm{g}}$= 1.45 and $m_{\mathrm{h}}$= 2.1. The lattice constants and interlayer distances of the InSe bilayer with different stacking configurations are also consistent with previous studies \cite{Lattice_contant-Interlayer_distance_InSe_bilayer, Twisted_InSe_bilayer}. Note that the optimized lattice constants of InSe bilayers with different stacking configurations are very slightly different, but the difference is below 0.32\%. The lattice constant of InSe monolayer (4.059 \AA) was therefore adopted for all the calculations of twisted InSe bilayer.

The band gap and binding energy do not deviate significantly from a previous study \cite{Lattice_contant-Interlayer_distance_InSe_bilayer} with the using similar computational methodology. In table \ref{InSe-bilayer_parameters}, the binding energy is seen to be proportional to the interlayer distance, making the B-3 the most stable stacking configuration. It suggests that the repulsion (positive energy) between the atoms is smallest in B-3. This is consistent with the stacking configuration of  B-3, because only In atoms in the top and bottom layers are aligned, while the repulsion is strongest between pairs of aligned Se atoms. Yang et al. also proposed that stronger vdW bonding is associated with smaller interlayer distance \cite{Lattice_contant-Interlayer_distance_InSe_bilayer}.

In addition, the indirect and direct (calculated from the topmost valence band at $\Gamma$) band gaps decrease as the interlayer distance decreases, consistent with previous studies \cite{Lattice_contant-Interlayer_distance_InSe_bilayer, interlayer-distance_bandstructure}. The effective masses for holes in the most stable stacking configuration in this work (B-3) are similar to those of a previous study \cite{B3_effective-mass}. Along the sequence B-3 $\rightarrow$ A-2 $\rightarrow$ B-1 $\rightarrow$ B-2 $\rightarrow$ A-1 there are clear trends: as the interlayer distance increases, so do the binding energy, band gap and the effective mass for holes. In fact, decreases in hole mass are roughly proportional to decreases in interlayer distance, relative to A-1. The band gap and effective mass for holes are both smaller in all of the InSe bilayers than in the InSe monolayer, as a result of interlayer hybridisation.
\end{multicols}

\begin{table}[h!]
\centering
\begin{tabularx}{\textwidth} { 
  | >{\centering\arraybackslash}X 
  | >{\centering\arraybackslash}X 
  | >{\centering\arraybackslash}X 
  | >{\centering\arraybackslash}X 
  | >{\centering\arraybackslash}X
  | >{\centering\arraybackslash}X
  | >{\centering\arraybackslash}X | }
 \hline
   & monolayer & A-1 & A-2 & B-1 & B-2 & B-3 \\
 \hline
$a$ (\AA) & 4.059 & 4.058 & 4.067 & 4.059 & 4.058 & 4.071 \\
 \hline
$d$ (\AA) & & 9.180 & 8.449 & 8.514 & 9.178 & 8.389 \\
 \hline
$E_{\mathrm{b}}$ (meV) & & -179.52 & -269.74 & -264.23 & -179.84 & -274.22 \\
 \hline
$E_{\mathrm{g}}$ (eV) & 1.42 & 1.02 & 0.87 & 0.92 & 1.02 & 0.82 \\
 \hline
$E_{\mathrm{g}}^\Gamma$ (eV) & 1.50 & 1.08 & 0.94 & 0.99 & 1.08 & 0.90 \\
 \hline
$m_{\mathrm{h}}$ (m$_{0}$) & 2.22 & 1.99 & 1.03 & 1.10 & 1.93 & 1.05 \\
 \hline
$m_{\mathrm{h}}$ (m$_{0}$) (at $\Gamma$) & -0.83 & -0.96 & -0.70 & -0.74 & -0.94 & -0.66 \\
 \hline
\end{tabularx}
 \caption{Atomic and electronic parameters of the InSe monolayer and InSe bilayers with five different stacking configurations, showing lattice constant $a$, interlayer distance $d$ (In$_{\mathrm{bt}}$-In$_{\mathrm{bb}}$), binding energy $E_\mathrm{b}$, band gap $E_\mathrm{g}$, band gap $E_\mathrm{b}^{\Gamma}$ calculated at $\Gamma$, effective mass for holes $m_{\mathrm{h}}$ and effective mass for holes $m_{\mathrm{h}}^{\Gamma}$ at $\Gamma$ for the VBM. } 
\label{InSe-bilayer_parameters}
\end{table}

\begin{multicols}{2}
\subsection{Relationship between the electronic structure and the stacking configuration of InSe bilayer}
\begin{figure} [H]
    \centering
    \includegraphics[width=0.5\textwidth]{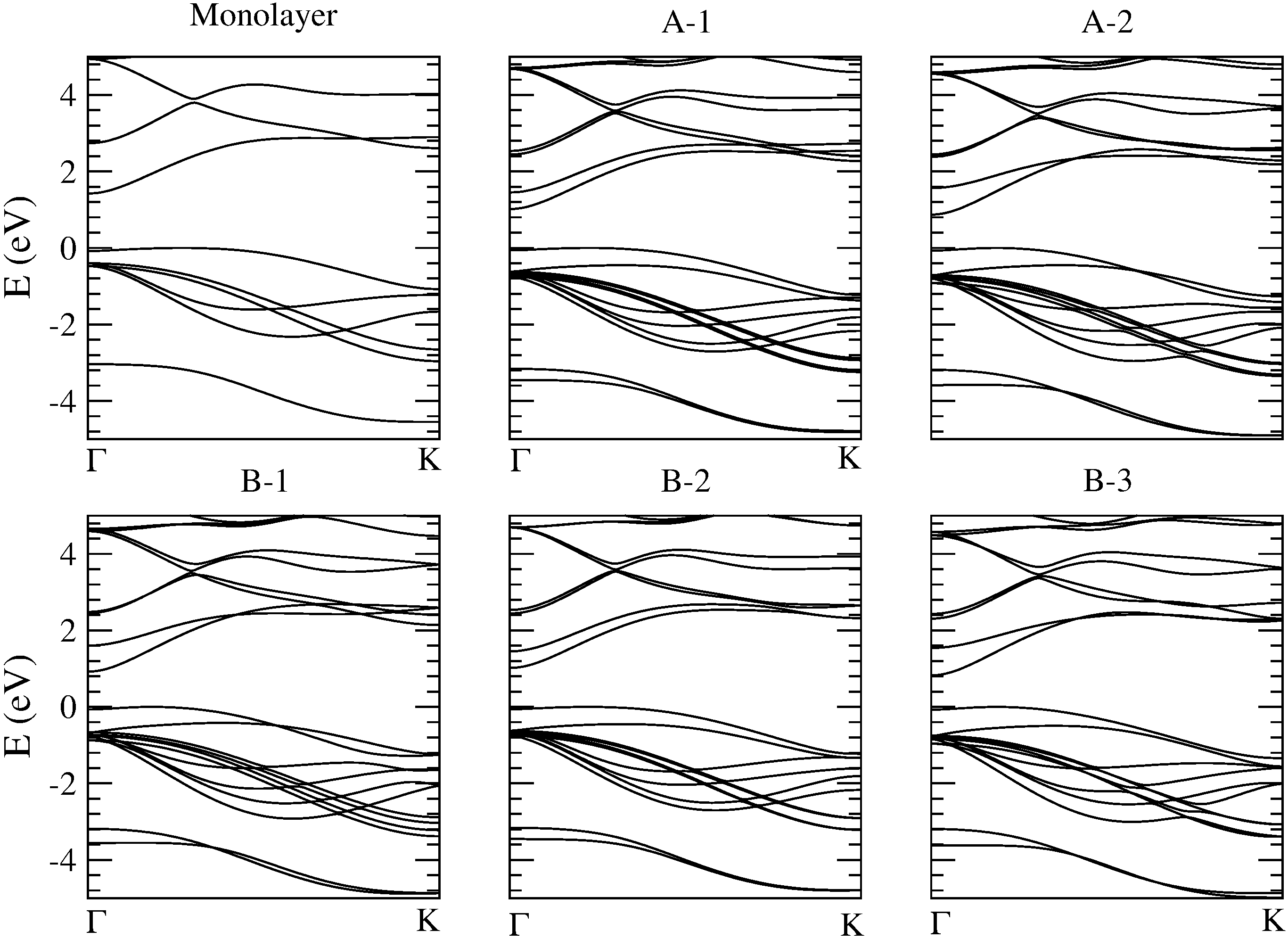}
 \caption{ Band structures of the InSe monolayer and the A-type and B-type InSe bilayers. The zero of energy is set to the highest energy of the valence band.} 
\label{fig:InSe-bilayer_bandstructure}
\end{figure}

Fig. \ref{fig:InSe-bilayer_bandstructure} displays the band structures of InSe monolayer, the A-type and B-type InSe bilayers. We mainly focus on the behaviour of the bands near the VBM in this work. Crucially, band crossings leading to a degeneracy are observed between the topmost two valence bands for stackings A-2 and B-1, whereas these are not seen for A-1 and B-2, while there is a near-crossing for B-3. This is consistent with their stacking configurations: band crossing with a degeneracy happens when there is alignment between In and Se atoms and is not observed for alignment between two Se atoms in the top and bottom layers. It can be seen that it is the topmost VB of the InSe monolayer that hybridises to form these two bands in the InSe bilayer. According to the projected band structure of the InSe monolayer in fig. \ref{fig:InSe-monolayer_pdos} and table \ref{PDOS_InSe-monolayer-VBM}, the topmost VB is dominated by the $p_{\mathrm{z}}$ orbitals of Se before 0.74 $\Gamma$-$\textbf{K}$, and by the $p_{\mathrm{z}}$ orbitals of In after. The band repulsion is strongest in A-1 and B-2 and weaker in A-2 and B-1, presumably because the interlayer distance is the smallest for cases with two aligned atoms, leading to there being no band crossing in A-1 and B-2 and band crossing in A-2 and B-1. The ordering of the strength of the band coupling which can be inferred as resulting from pairwise alignment of atoms is therefore Se-Se $>$ Se-In $>$ In-In. That has the consequence that the splitting of the bands increases as one approaches $\Gamma$, due to the increasing orbital contribution of $p_{\mathrm{z}}$ of Se atoms. Similar band structure effects can also seen in a previous study \cite{Lattice_contant-Interlayer_distance_InSe_bilayer} but
without discussing the orbital contribution along the kpoint path in detail. It is worth noting that there is no $p_{\mathrm{z}}$ orbital contribution from Se atoms at $\textbf{K}$ for the InSe monolayer.

The band energies of topmost band (labelled VBM) and the band below it (labelled VBM-1) in the InSe bilayer can be modelled, near a crossing, with a simple two-band model and a hybridisation parameter $\Delta(\mathbf{k})$. The eigenvalues of following determinant give the resulting energies after hybridisation \cite{twisted_hBN-bilayer, interlayer-coupling_graphene-hBN}:
\begin{equation}\label{bands-split}
\begin{vmatrix}
\epsilon_{\mathrm{VB}}(\mathbf{k})-\epsilon & \Delta(\textbf{k})/2 \\
\Delta(\textbf{k})/2 & \epsilon_{\mathrm{VB}}(\mathbf{k})-\epsilon
\end{vmatrix}
= 0 \, .
\end{equation}
where two $\epsilon_{\mathrm{VB}}$ are the energies of the VBMs at $\textbf{k}$ from top and bottom InSe monolayers. $\Delta(\textbf{k})$ describes the interband coupling between two InSe monolayers. The main contributor to $\Delta(\textbf{k})$ is that the orbital contributions vary along a kpoint path.

\begin{figure} [H]
    \centering
    \includegraphics[width=0.5\textwidth]{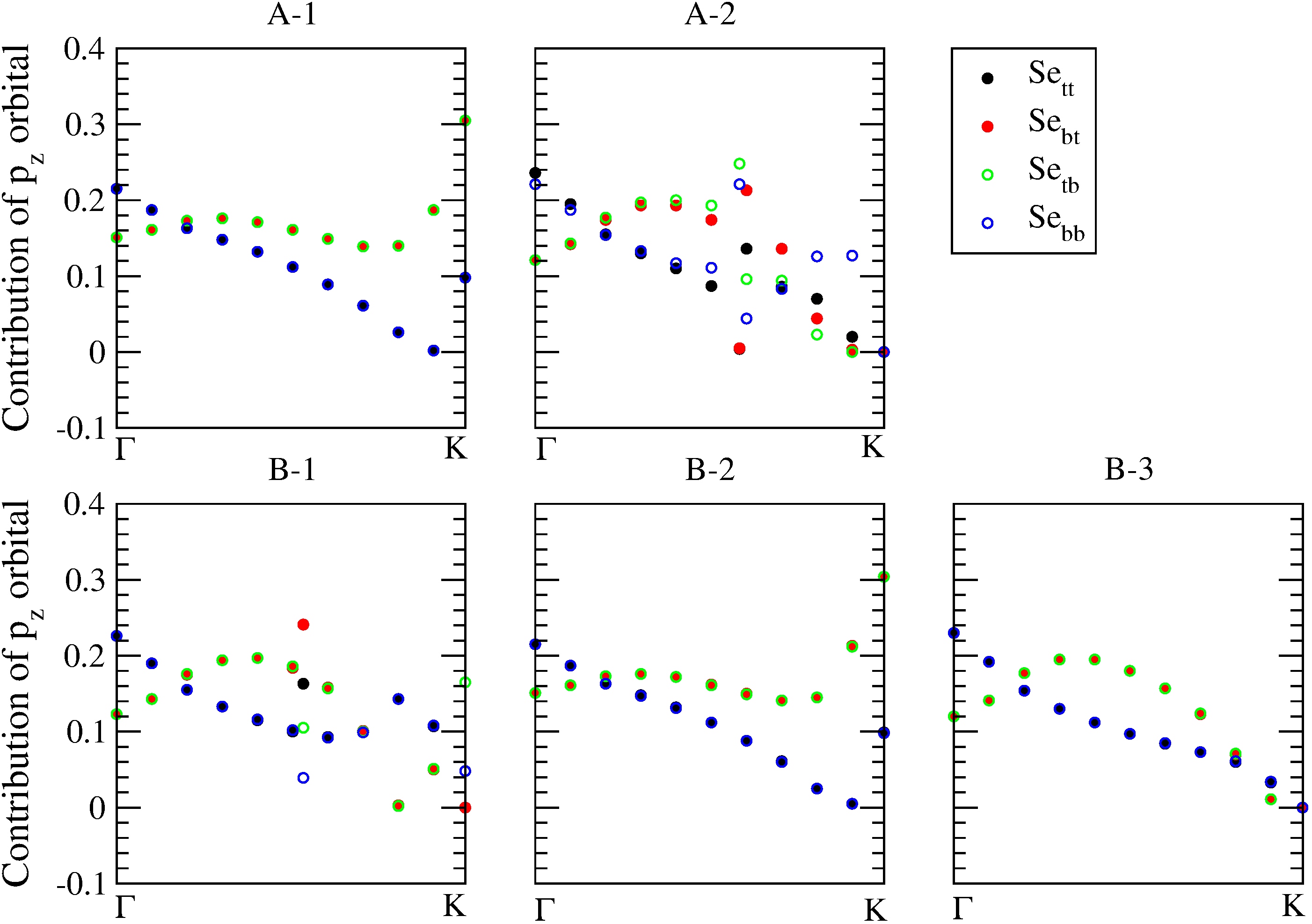}    
 \caption{ The contribution of $p_{\mathrm{z}}$ orbital for the VBM along $\Gamma$ to $\textbf{K}$ from all Se atoms in the unit cells of A-type and B-type InSe bilayers. The label of each Se atom is indicated in fig. \ref{fig:InSe_AA-AB}.} 
\label{fig:InSe-bilayer_pdos-VBM}
\end{figure}

The projected density of states (PDOS) for the A-type and B-type InSe bilayers (fig. \ref{fig:InSe-bilayer_pdos}) closely resembles that of the  InSe monolayer (fig. \ref{fig:InSe-monolayer_pdos}). The details of the $p_{\mathrm{z}}$ orbital contributions of Se atoms are shown in fig. \ref{fig:InSe-bilayer_pdos-VBM}. In fig. \ref{fig:InSe-bilayer_pdos-VBM}, the classification of $p_{\mathrm{z}}$ orbital contributions of Se atoms is consistent with the previous discussion: A-1 is similar to B-2 and A-2 is similar to B-1, whereas B-3 is distinct. The ordering of Se atom (from black to blue curves) is according to the vertical position (from top to bottom) of Se atom in InSe bilayer. The red and black curves are not seen in the A-1, B-2 and B-3 because they have same magnitudes as the green and blue curves, respectively. This means that the $p_{\mathrm{z}}$ orbitals of the two outermost Se atoms (Se$_{\mathrm{tt}}$ (black) and Se$_{\mathrm{bb}}$ (blue)) are the same. This is also true for the two innermost Se atoms (Se$_{\mathrm{bt}}$ (red) and Se$_{\mathrm{tb}}$ (green)). In general, the $p_{\mathrm{z}}$ orbitals of the outermost Se atoms are only larger than the innermost Se atoms near $\Gamma$, excepting the discontinuous behaviour around the intermediate kpoint path and $\textbf{K}$ for the A-2 and B-1. 

The discontinuous behaviour that occurs just beyond the midpoint of the kpoint path is the result of to the interband coupling between the VBM and the VBM-1: where a peak occurs for the VBM, there is a trough for the VBM-1 (not shown). The discontinous behaviour in A-2 and B-1 is the same but in reverse, for example, there are peaks for Se$_{\mathrm{tb}}$ (green) and Se$_{\mathrm{bb}}$ (blue) around the kpoint of band crossing in A-2, whereas these two curves display troughs for B-1. The discontinuous behaviour is also reversed for the VBM and VBM-1 in the A-2 or B-1 stackings. Furthermore, the divergence in the PDOS weights on approach to  $\textbf{K}$ for A-1 and B-2 also originates from the interband coupling for the two bands below the VBM. From the analysis of PDOS, the $p_{\mathrm{z}}$ orbital contributions around $\textbf{K}$ for these two bands do not approach zero (with values of $\simeq$ 0.4), whereas the $p_{\mathrm{z}}$ orbital contributions for the VBM decrease to zero as approaching toward $\textbf{K}$ in InSe monolayer. The $p_{\mathrm{z}}$ orbital contributions of these two bands (below the VBM) fall to zero and the VBM becomes non-zero for the A-1 and B-2. The interlayer coupling between the VBM and VBM-1  related to two Se atoms (Se$_{\mathrm{tb}}$ and Se$_{\mathrm{bb}}$) can also be seen in the B-1 stacking. 

The $p_{\mathrm{z}}$ orbital contributions of the outermost Se atoms (black and blue curves) have the same behaviour as InSe monolayer (if ignoring the sudden increase near $\textbf{K}$) because they decrease from $\Gamma$ to $\textbf{K}$. The $p_{\mathrm{z}}$ orbital contributions of the innermost Se atoms (red and green curves) first increase to $\simeq$ 0.3-0.4 $\Gamma$-$\textbf{K}$, then decrease along the kpoint path (if ignoring the sudden increase in the A-1, B-1 and B-2 stackings). The $p_{\mathrm{z}}$ orbital contributions of Se atoms in the B-3 are similar to other stacking configurations but without the weird behaviours mentioned above. There is a more rapid changes for the $p_{\mathrm{z}}$ orbital contributions if the interlayer distance is smaller due to the increase of interlayer coupling strength. It is worth noting that the CBM of InSe monolayer and InSe bilayer are predominated by the delocalised $s$ orbitals of In atoms (see table \ref{PDOS_InSe-monolayer-CBM}). 

\section{Residual forces in twisted InSe bilayer}
Due to the need for large-scale calculations to study the twisted InSe bilayer at smaller twist angles, we prepare structures with initial corrugations in the z positions, based on the interlayer distances from the structures shown in table \ref{InSe-bilayer_parameters}). A fast Fourier transform (FFT) is used to Fourier interpolate from a small grid corresponding to the known stackings in the corner and midpoints, up to a grid fine enough to provide heights for each layer at each atom position. The magnitude of the corrugation required to pre-optimise the InSe bilayer to generate the twisted InSe bilayer is relatively small (fig. \ref{fig:interlayer_distance-corrugation}), so the effect of corrugation on the effective band structure is marginal (fig. \ref{fig:compare_with-without_corrugation}). Two neighbouring bands, such as the VBM and VBM-1, are only separated slightly when considering the corrugation in the atomic structure. Corrugation is thus not expected to result directly in the occurrence of flat bands as in twisted hBN bilayer \cite{twisted_hBN-bilayer}. 

\begin{figure} [H]
\captionsetup[subfigure]{justification=centering}
\begin{subfigure}{0.24\textwidth}
  \centering
  \includegraphics[width=\textwidth]{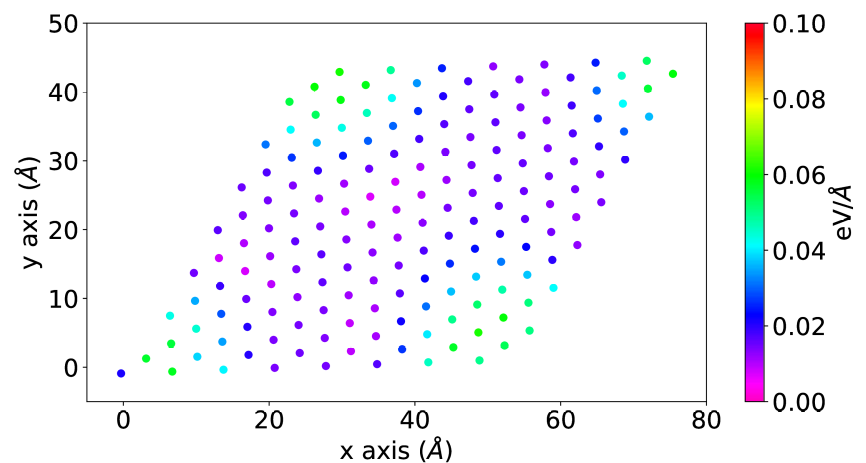}
  \caption{In$_{\mathrm{bt}}$}
\end{subfigure}
\begin{subfigure}{0.24\textwidth}
  \centering
  \includegraphics[width=\textwidth]{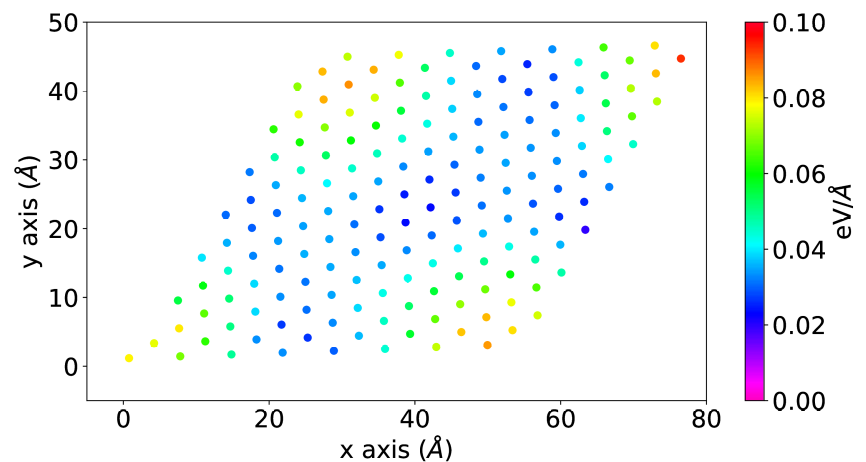}
  \caption{Se$_{\mathrm{bt}}$}
\end{subfigure}
\begin{subfigure}{0.24\textwidth}
  \centering
  \includegraphics[width=\textwidth]{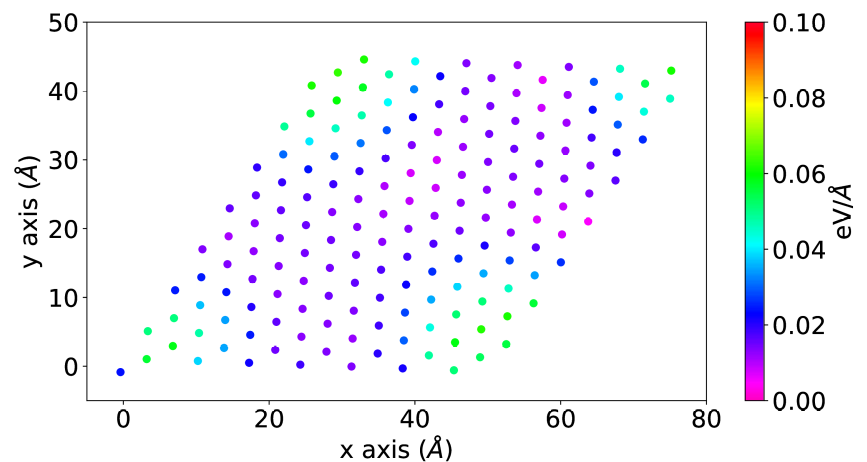}
  \caption{In$_{\mathrm{tb}}$}
\end{subfigure}
\begin{subfigure}{0.24\textwidth}
  \centering
  \includegraphics[width=\textwidth]{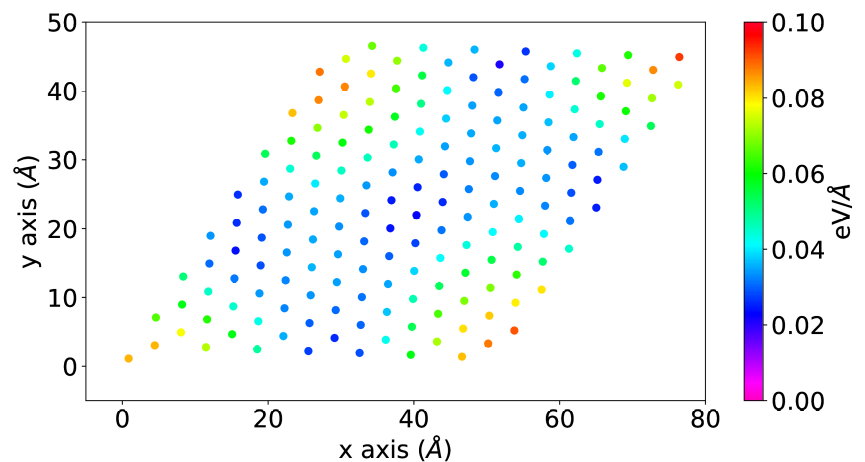}
  \caption{Se$_{\mathrm{tb}}$}
\end{subfigure}
 \caption{ Residual total force in the innermost In and Se atoms for the A-type twisted InSe bilayer with a twist angle of 4.4$^{\mathrm{o}}$ after geometry optimisation.}
\label{fig:Force_distribution_after-relaxation}
\end{figure}

An initial atomic structure is shown in fig. \ref{fig:Corrugation_distribution} for the A-type twisted InSe bilayer with a twist angle of 4.4$^{\mathrm{o}}$. The corresponding residual total force associated with the innermost atoms (In$_{\mathrm{bt}}$, Se$_{\mathrm{bt}}$, In$_{\mathrm{tb}}$ and Se$_{\mathrm{tb}}$) after the calculation of geometry optimisation is displayed in fig. \ref{fig:Force_distribution_after-relaxation}. The distribution of residual total force in fig. \ref{fig:Force_distribution_after-relaxation}  look similar to fig. \ref{fig:Corrugation_distribution}. The outermost atoms (In$_{\mathrm{tt}}$, Se$_{\mathrm{tt}}$, In$_{\mathrm{bb}}$ and Se$_{\mathrm{bb}}$) do not exhibit similar distributions in fig. \ref{fig:Corrugation_distribution} and their forces are smaller than the innermost atoms (not shown). This phenomenon is suggested to be intrinsic for any twisted bilayer system. The direction of the residual force provide the information of how the atoms move in full structural relaxation until no residual force is retained.

The residual total forces are mainly from the forces along the z direction (compare fig. \ref{fig:Force_distribution_z} and fig. \ref{fig:Force_distribution_xy}), implying the larger rearrangement of vertical positions of innermost atoms compared to their in-plane positions (for at least the twist angle considered here). The similarity of the residual force distributions along the x, y and z directions are seen  between the neigbouring innermost In and Se atoms, whereas the forces are larger in Se atoms because the distance between two innermost Se atoms is smaller than other combinations of atoms. In addition, the two InSe layers have opposite residual forces along all directions (see fig. \ref{fig:Force_distribution_z} for the residual forces in z direction). 

\subsection{Electronic structure of twisted InSe bilayer}
The electronic parameters such as the band gap and the effective mass for holes calculated from the band structure of twisted InSe monolayer (has a twist angle relative to the referenced untwisted InSe monolayer) along the $\Gamma$-$\textbf{K}$ of the untwisted InSe monolayer do not change significantly (fig. \ref{fig:bandstructure_InSe-monolayer_rotation}). The main factor that affects the band structure is the interlayer distance, or we can say the stacking configuration, because the interlayer distance is the particular result of the interaction between atoms in a stacking configuration. Shang et al. \cite{interlayer-distance_bandstructure} also showed that the band structure changes with the interlayer distance. The resulting interlayer distance of the twisted InSe bilayer is determined by the proportions of different stacking configurations in the different regions of twisted InSe bilayer. According to table \ref{InSe-bilayer_parameters}, the interlayer distances are within [8.449, 9.180] and [8.389, 9.178] for the A-type and the B-type twisted InSe bilayers, respectively.

The inclusion of spin-orbit coupling (SOC) in the calculation of the effective band structure leads to the splittings of bands. The splittings of the VBM and CBM are only pronounced near $\textbf{K}$ and around 0.5 $\Gamma$-$\textbf{K}$ to $\textbf{K}$, respectively, for InSe monolayer, twisted InSe bilayers and the hBN-encapsulated twisted InSe bilayers (see fig. \ref{fig:bandstructure_AA-stacking_soc_no-soc} and fig. \ref{fig:bandstructure_AB-stacking_soc_no-soc}). Thus, its effect on the band gap and effective mass for holes can be ignored. It is interesting to see that the avoided crossing of bands around 4 eV is eliminated by the SOC for all systems. The computation effort can be reduced by not considering the SOC in the calculations. In addition, the projection of the effective band structure on the top and bottom layers have larger difference at larger twist angle, whereas the difference is very small at small twist angle (see fig. \ref{fig:bandstructure_InSe-monolayer_rotation}, fig. \ref{fig:bandstructure_AA-stacking_top_bottom} and fig. \ref{fig:bandstructure_AB-stacking_top_bottom}).

The encapsulation of the hBN layer aims to change the electronic properties of twisted InSe bilayer. The inclusion of twist angle between InSe and hBN layers does not change the band structure \cite{InSe-hBN_stacking}. The insertion of the hBN layer only decreases the separation between neigbouring bands such as the VBM and VBM-1 in the twisted InSe bilayers, by increasing the distance between two InSe layers (reducing the interaction between two InSe layers). Similar result can be seen in the hBN-encapsulated black phosphorus \cite{BP-hBN-BP}. It is worth noting that the anti-crossing bands appear near $\textbf{K}$ within the range of [-2.0, -1.0] eV. The effective band structure of the hBN-encapsulated twisted InSe bilayer is suggested to approach the band structure of InSe monolayer. This is because the projection of the hBN-encapsulated twisted InSe bilayer onto the hBN layer and its PDOS both show that the VBM of the hBN layer is at least 1.5 eV below the Fermi level, thus the VBM of twisted InSe bilayer is not affected by the hBN (fig. \ref{fig:InSe-hBN-InSe_projected_on_hBN} and fig. \ref{fig:InSe-hBN-InSe_pdos}). 

\begin{figure} [H]
\captionsetup[subfigure]{justification=centering}
\begin{subfigure}{0.5\textwidth}
  \centering
  \includegraphics[width=\textwidth]{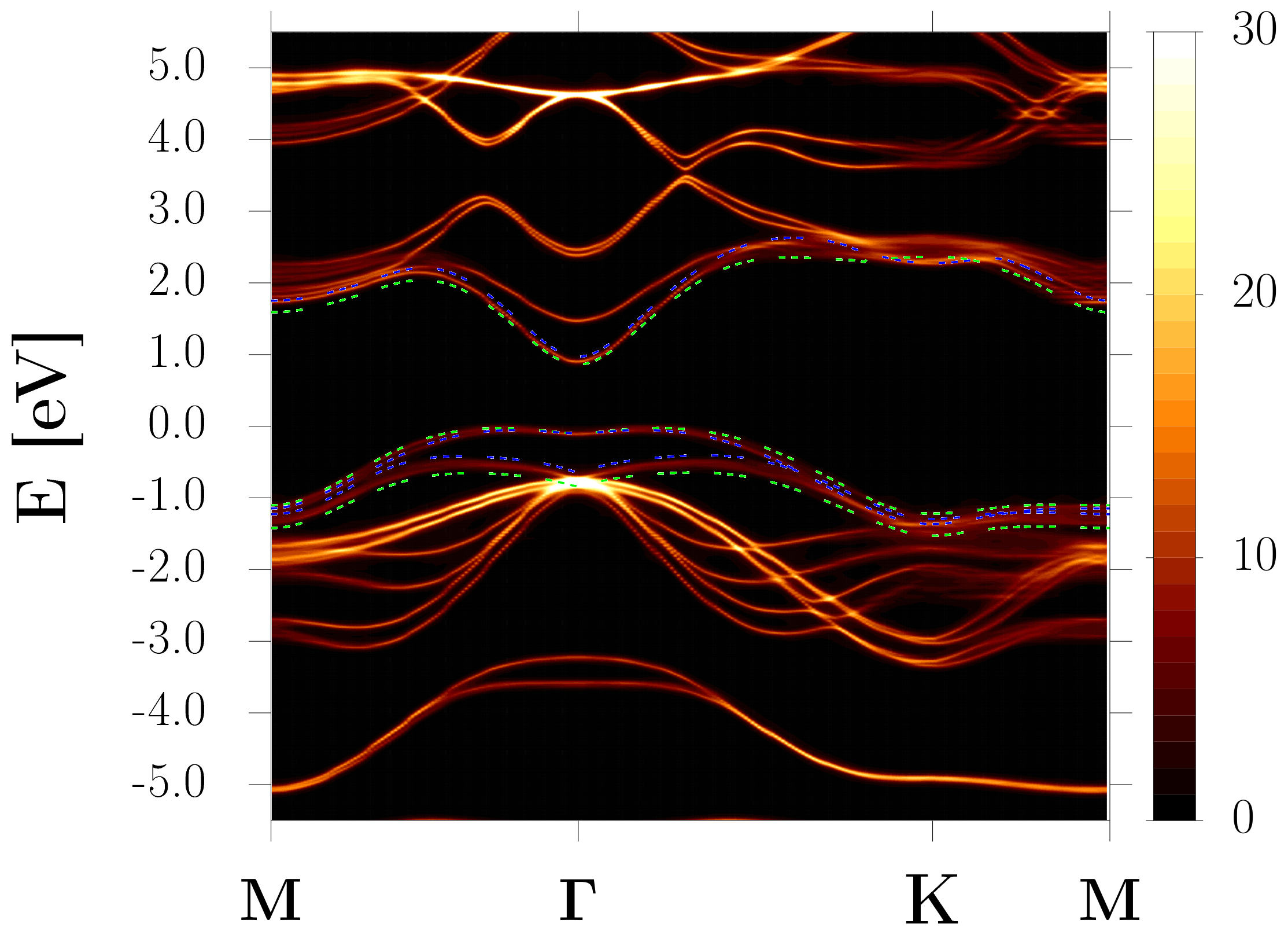}
  \caption{A-type twisted InSe bilayer ($\theta$= 4.4$^{\mathrm{o}}$)}
\end{subfigure}
\begin{subfigure}{0.24\textwidth}
  \centering
  \includegraphics[width=\textwidth]{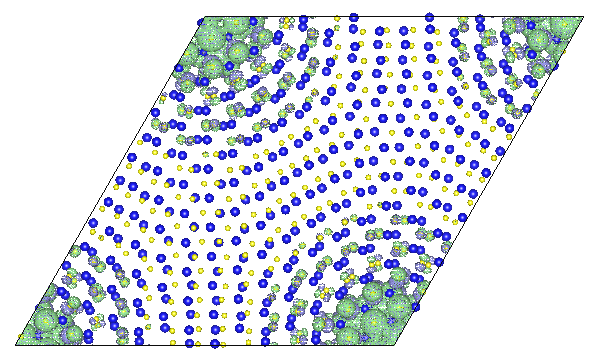}
  \caption{Corresponding to the A-1 stacking}
\end{subfigure}
\begin{subfigure}{0.24\textwidth}
  \centering
  \includegraphics[width=\textwidth]{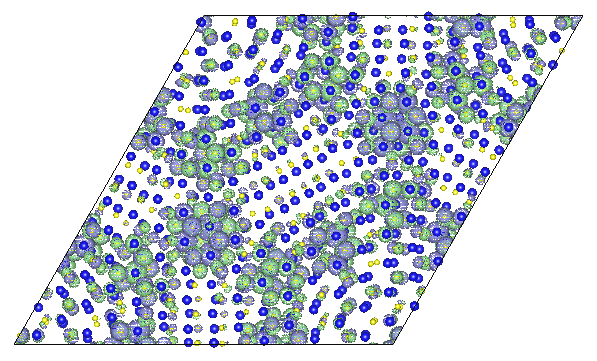}
  \caption{Corresponding to the A-2 stacking}
\end{subfigure}
 \caption{ (a) Effective band structure of the A-type twisted InSe bilayer  ($\theta=4.4^{\mathrm{o}}$) without inclusion of spin-orbit coupling.  Green and blue curves refer to the bands of the A-1 and A-2 stackings from primitive cell calculations, respectively. The interlayer distances for the primitive cell calculations are the same as the supercell calculation. The charge densities for the VBM localised around the regions corresponding to the (b) A-1 (c) A-2 stacking without considering SOC. }
\label{fig:bandstructure_twisted_InSe-bilayer-AA-stacking_ONETEP_QE}
\end{figure}

We are interested to know the effect of the different stacking configurations seen on the atomic structure of the twisted system on its band strcuture (fig. \ref{fig:twisted_InSe-bilayer_hBN-trilayer}(a) and (b)). In fig. \ref{fig:bandstructure_twisted_InSe-bilayer-AA-stacking_ONETEP_QE}, the VBM of twisted InSe bilayer can be well-described by the same band of aligned InSe bilayer from the primitive cell calculation with the same interlayer distance. In fig. \ref{fig:bandstructure_twisted_InSe-bilayer-AA-stacking_ONETEP_QE}(a), the green and blue curves represent the bands in the A-1 and A-2 stackings, from the primitive cell calculations. The isosurfaces of charge densities in fig. \ref{fig:bandstructure_twisted_InSe-bilayer-AA-stacking_ONETEP_QE}(b) and (c) show the charge density localised around the regions corresponding to the A-1 and A-2, respectively. The finite-width bands in the effective band structure was calculated by using the spectral function unfolding method. They are contributed by the A-1, A-2 and their combinations when analysing through the isosurfaces of the charge densities. The argument also applies to the twisted InSe bilayer with considering SOC in the calculation (fig. \ref{fig:bandstructure_twisted_InSe-bilayer-AA-stacking_ONETEP_QE_4p4deg-soc}). The result is partly different from twisted hBN bilayer \cite{twisted_hBN-bilayer} (its stacking configurations is similar to InSe bilayer) because the VBM is contributed by the A-1 stacking. However, the VBM is contributed by the A-2 stacking in twisted InSe bilayer at smaller twist angles in the previous study \cite{Twisted_InSe_bilayer}. It is suggested that the charge densities for the VBM in our calculations may further localised in the A-2 if the twist angle is smaller than 3$^{\mathrm{o}}$. The flat bands is suggested to appear at small twist angles (magic angles are unnecessary) for twisted bilayer system composed of polar 2D semiconductor \cite{twisted_hBN-bilayer}. Zhao et.al. proposed that the flat bands can be obtained by reducing the sublattice symmetry of twisted system, which means increasing the difference of the atoms in the unit cell \cite{twisted_hBN-bilayer}. 

The length from $\Gamma$ to the lowest energy kpoint of the VBM equals to the length of kpoint path of the top InSe layer inside the first Brillouin zone of the bottom InSe layer: these are 0.886 $\Gamma$-$\textbf{K}$ and 0.867 $\Gamma$-$\textbf{K}$ for a twist angle of 17.9$^{\mathrm{o}}$ and 27.8$^{\mathrm{o}}$, respectively (see fig. \ref{fig:Bandstructure_ONETEP_QE}(c)-(f)). This phenomenon can also be seen in fig. \ref{fig:bandstructure_InSe-monolayer_rotation} for the band structures of various twist angles considered in InSe monolayer. Furthermore, we have noticed that there is approximately no difference between the effective band structures of the A-type and B-type twisted systems at the same twist angle in fig. \ref{fig:Bandstructure_ONETEP_QE} because of the similarity of the band structure between A-type and B-type InSe bilayers in fig. \ref{fig:InSe-bilayer_bandstructure}. This is similar to the results in previous literature \cite{Twisted_InSe_bilayer-1}.

It is worth noting that the effect due to the difference of the lattice constant for different stacking configurations seen in some regions of twisted InSe bilayer can be ignored  (fig. \ref{fig:compare_lattice-constant}). The flatness of the VBM around the highest energy kpoint increases (so the effective mass for holes increases) if the interlayer distance increases (fig. \ref{fig:compare_interlayer-distance}), consistent with table \ref{InSe-bilayer_parameters}. However, the interlayer distances of the A-type and B-type twisted InSe bilayers with various twist angles considered in this work are 8.85$\pm$0.04 \AA \, (fig. \ref{fig:interlayer-distance_twist-angle}), similar to the small variation of interlayer distance for twist angles between 13.2$^{\mathrm{o}}$ and 32.2$^{\mathrm{o}}$ in previous studies \cite{Twisted_InSe_bilayer-1}. The small variation of interlayer distance at different twist angles implies that the proportion of each stacking configuration in the same type of twisted InSe bilayer only change slightly. 

The curvatures of the VBMs which determine the effective mass for holes with different twist angles considered in this work seem kept at a constant (fig. \ref{fig:VBM_twist-angle}). Different band foldings at different twist angles lead to the variation of flatness of the VBM, as seen in the previous literature \cite{Twisted_InSe_bilayer}. The defect states are proposed to be formed in polar 2D semiconductor at sufficiently small twist angles and this leads to the appearance of flat bands due to the weak interactions with other host and defect states \cite{twisted_hBN-bilayer}. Furthermore, the studies of twisted MoS$_{2}$ bilayer showed that the inhomogeneous interlayer hybridisation and the local strain can also lead to the flat bands \cite{twisted_MoSe-bilayer}. 
We may possibly see the flat bands in twisted InSe bilayer if further decreasing the twist angles to smaller than 3$^{o}$, as seen in the previous study \cite{Twisted_InSe_bilayer}. The variation of band gap is small, it is 0.95$\pm$0.03 eV for all twist angles in both the A-type and B-type twisted systems considered in this work. This small variation of band gap does not deviate from the results of previous studies \cite{Twisted_InSe_bilayer-1}.

\section{Calculated exciton binding energy in twisted InSe bilayer and twisted InSe/hBN/InSe heterostructure}

\begin{figure} [H]
\captionsetup[subfigure]{justification=centering}
\begin{subfigure}{0.24\textwidth}
  \centering
  \includegraphics[width=\textwidth]{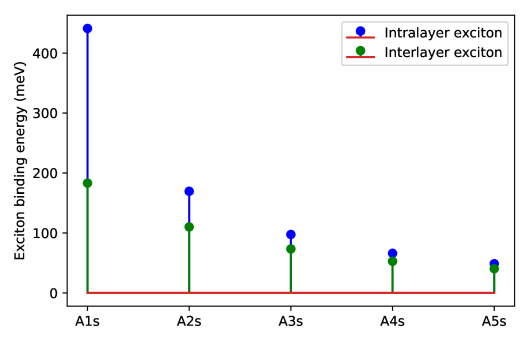}
  \caption{In the vacuum environment}
\end{subfigure}
\begin{subfigure}{0.24\textwidth}
  \centering
  \includegraphics[width=\textwidth]{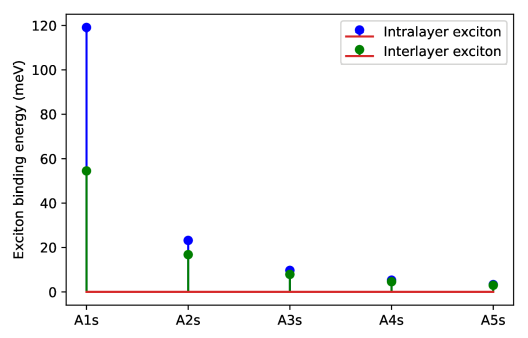}
  \caption{In the hBN environment}
\end{subfigure}
 \caption{ Calculated exciton binding energies of twisted InSe bilayer in the environment of (a) vacuum (b) hBN. The results are also shown in table \ref{Exciton_binding-energy_vacuum} and table \ref{Exciton_binding-energy_hBN}.} 
\label{fig:Exciton_binding-energy-1}
\end{figure}

According to the work of Viner et al. \cite{Viner2021} (detailed calculations are shown in  supporting information), we can estimate the intralayer and interlayer exciton binding energies of twisted InSe bilayer based on the effective mass for holes obtained here. In the previous literature \cite{PBE-GW_consistent}, the band dispersion of GGA-PBE and GW were in good agreement  except for the GW have larger band gap than the GGA-PBE. The calculated exciton binding energies in the environments of vacuum and hBN are shown in table \ref{Exciton_binding-energy_vacuum}, table \ref{Exciton_binding-energy_hBN} and figure \ref{fig:Exciton_binding-energy-1}. 

The A1s intralayer exciton binding energies in both environments are more than 2X larger than the A1s interlayer exciton binding energies. From A2s to A5s, their difference gradually decreases to $\sim$ 8 meV and 0 meV, in the enviroments of vacuum and hBN, respectively. Furthermore, the intralayer and interlayer exciton binding energies in the environment of vacuum are $\sim$ 3.7X and $\sim$ 3.4X larger than in the enviroment of hBN.

In addition, the similarity between the hBN-encapsulated twisted InSe bilayer and InSe monolayer can be seen through the same analysis as fig. \ref{fig:bandstructure_twisted_InSe-bilayer-AA-stacking_ONETEP_QE}(a). The VBM of InSe monolayer from the primitive cell calculation match well with the effective band structure of the hBN-encapsulated A-type and B-type twisted InSe bilayers from the supercell calculations (fig. \ref{fig:bandstructure_InSe-hBN-InSe_ONETEP_QE}). The calculated intralayer exciton binding energies in the environment of vacuum and hBN are shown in table \ref{Exciton_binding-energy_InSe-hBN-InSe}. In the calculation, $U^{\mathrm{ML}}$ (eq. \eqref{eq:U_ML} in the supporting information) was used as the Coulomb potential because the trilayer heterostructure system can be regarded as two separated InSe layers plus a hBN layer. The result in the environment of vacuum is 1.7X larger than in the environemnt of hBN for A1s intralayer exciton binding energy.

\end{multicols}

\begin{table}[h!]
\centering
\begin{tabularx}{\textwidth} { 
  | >{\centering\arraybackslash}X 
  | >{\centering\arraybackslash}X 
  | >{\centering\arraybackslash}X 
  | >{\centering\arraybackslash}X 
  | >{\centering\arraybackslash}X
  | >{\centering\arraybackslash}X | }
  \hline
   & A1s & A2s & A3s & A4s & A5s \\
 \hline
Intralayer exciton (meV) & $\sim$441.1 & $\sim$169.6 & $\sim$97.5 & $\sim$66.1 & $\sim$48.7 \\
 \hline
Interlayer exciton (meV) & $\sim$183.2 & $\sim$110.1 & $\sim$73.4 & $\sim$52.8 & $\sim$40.4 \\
 \hline
\end{tabularx}
 \caption{ Calculated exciton binding energies of twisted InSe bilayers based on all twist angles considered in this work for the A-type and B-type systems in the environment of vacuum. Some parameters used are according to previous literature: effective mass for electrons ($m_{\mathrm{e}}$)= 0.16 m$_{\mathrm{0}}$ \cite{PBE-GW_consistent}, $\tilde{\epsilon}= \epsilon_{\mathrm{vacuum}}$= 1, screening length $r_{*}= 7.7/\epsilon_{\mathrm{hBN}}*\epsilon_{\mathrm{vacuum}}$= 38.9 \AA \, \cite{screening-length_InSe, In-plane_polarisability, interlayer-distance_InSe, dielectric_constant-hBN} ($\epsilon_{\mathrm{hBN}}= \sqrt{\epsilon_{\parallel}*\epsilon_{\perp}}= \sqrt{6.9*3.7}$= 5.1 \cite{dielectric_constant-hBN} and $\epsilon_{\mathrm{vacuum}}$= 1). Some parameters used are according to this work: the effective mass for holes ($m_{\mathrm{h}}$)= [1.39, 1.75] m$_{\mathrm{0}}$ and interlayer distance ($d$)= [8.81, 8.89] \AA.} 
\label{Exciton_binding-energy_vacuum}
\end{table}

\begin{table}[h!]
\centering
\begin{tabularx}{\textwidth} { 
  | >{\centering\arraybackslash}X 
  | >{\centering\arraybackslash}X 
  | >{\centering\arraybackslash}X 
  | >{\centering\arraybackslash}X 
  | >{\centering\arraybackslash}X
  | >{\centering\arraybackslash}X | }
 \hline
   & A1s & A2s & A3s & A4s & A5s \\
 \hline
Intralayer exciton (meV) & $\sim$119.1 & $\sim$23.2 & $\sim$9.7 & $\sim$5.3 & $\sim$3.3 \\
 \hline
Interlayer exciton (meV) & $\sim$54.5 & $\sim$16.8 & $\sim$7.9 & $\sim$4.6 & $\sim$3.0 \\
 \hline
\end{tabularx}
 \caption{ Calculated exciton binding energies of twisted InSe bilayers based on all twist angles considered in this work for the A-type and B-type systems in the environment of hBN. Some parameters used are according to previous literature: effective mass for electrons ($m_{\mathrm{e}}$)= 0.16 m$_{\mathrm{0}}$ \cite{PBE-GW_consistent}, screening length ($r_{*}$)= 7.7 \AA \, \cite{screening-length_InSe, In-plane_polarisability, interlayer-distance_InSe}, $\tilde{\epsilon}= \epsilon_{\mathrm{hBN}}= \sqrt{\epsilon_{\parallel}*\epsilon_{\perp}}= \sqrt{6.9*3.7}$= 5.1 \cite{dielectric_constant-hBN}. Some parameters used are according to this work: the effective mass for holes ($m_{\mathrm{h}}$)= [1.39, 1.75] m$_{\mathrm{0}}$ and interlayer distance ($d$)= [8.81, 8.89] \AA.} 
\label{Exciton_binding-energy_hBN}
\end{table}

\begin{table}[h!]
\centering
\begin{tabularx}{\textwidth} { 
  | >{\centering\arraybackslash}X 
  | >{\centering\arraybackslash}X 
  | >{\centering\arraybackslash}X 
  | >{\centering\arraybackslash}X 
  | >{\centering\arraybackslash}X
  | >{\centering\arraybackslash}X | }
 \hline
   & A1s & A2s & A3s & A4s & A5s \\
   \hline
Intralayer exciton (meV) (vacuum) & 232.9 & 60.9 & 27.4 & 15.5 & 9.9 \\
 \hline
Intralayer exciton (meV) (hBN) & 138.8 & 27.8 & 11.5 & 6.2 & 3.9 \\
 \hline
\end{tabularx}
 \caption{ Calculated exciton binding energies of twisted InSe/hBN/InSe heterostructure of the A-type and B-type systems in the environments of vacuum and hBN. Some parameters used are according to previous literature: effective mass for electrons ($m_{\mathrm{e}}$)= 0.18 m$_{\mathrm{0}}$ \cite{PBE-GW_consistent}, screening length $r_{*}$= 7.7/$\epsilon_{\mathrm{hBN}}*\tilde{\epsilon}$= 12.86 \AA \, \cite{screening-length_InSe, In-plane_polarisability, interlayer-distance_InSe, dielectric_constant-hBN} ($\epsilon_{\mathrm{hBN}}=\sqrt{\epsilon_{\parallel}*\epsilon_{\perp}}= \sqrt{6.9*3.7}$= 5.1  \cite{dielectric_constant-hBN}), $\tilde{\epsilon}=(\epsilon_{\mathrm{vacuum}}+\epsilon_{\mathrm{hBN}})/2= \sqrt{1.0+5.1}/2$= 3.0. Some parameters used are according to this work: the effective mass for holes of InSe monolayer ($m_{\mathrm{h}}$)= 2.22 m$_{\mathrm{0}}$.}
\label{Exciton_binding-energy_InSe-hBN-InSe}
\end{table}

\begin{multicols}{2}
\section{Conclusions}
The atomic and electronic parameters of InSe monolayer and InSe bilayers with different stacking configurations are discussed in this work. The binding energy, band gap and effective mass for holes increase as the interlayer distance increases. The most stable stacking configuration is the B-3. The stacking configuration with the largest band gap and the effective mass for holes are the A-1. The band gap and the effective mass for holes in InSe monolayer are larger than InSe bilayer with different stacking configurations because there is no interlayer hybridisation in the monolayer form. 

Only the band structure of the A-2 and B-1 show the band crossings for the two bands below the Fermi level. The $p_{\mathrm{z}}$ orbitals of Se atoms and $s$ orbitals of In atoms give the largest contributions to the VBM and CBM, respectively, for InSe monolayer and InSe bilayer with different stacking configurations. The variation of the VBM in InSe bilayer with different stacking configurations can be explained by the variation of the $p_{\mathrm{z}}$ orbital contribution of each Se atom. 

The residual force in each innermost atom displays a similar distribution with the distribution of interlayer distance according to the regions corresponding to each stacking configuration in twisted InSe bilayer. The residual forces in the outermost atoms are smaller than the innermost atoms because there is a direct interaction between the  innermost atoms in the two different layers. The same reason also leads to the larger residual forces for innermost Se atoms than In atoms. The innermost atoms mostly do the vertical movements when forming twisted system (for at least the twist angle considered here) from the analysis of residual force.  

The spin-orbit coupling does not change significantly the VBM, especially around the highest energy kpoint of the VBM where we concern. The variation of the VBM near $\textbf{K}$ and around the kpoint we concern becomes larger and remains the same as the twist angle increases, respecstively. The effective band structure of the A-type (B-type) twisted InSe bilayer with acceptable lattice mismatch can be well described by the A-type (B-type) InSe bilayer with different stacking configurations from the primitive cell calculations. It is necessary to include the bands from both the top and bottom InSe layers from the primitive cell calculations along the same kpoint path (especially at large twist angles) for getting a better mapping on the twisted InSe bilayer.

The effective mass for holes will increase if the interlayer distance in twisted InSe bilayer increases. However, the interlayer distance of A-type and B-type twisted InSe bilayers with different twist angles considered in this work are 8.85$\pm$0.04 \AA. The differences of effective mass of holes and band gap among different twist angles are very small, they are calculated to be 1.57$\pm$0.18 m$_{\mathrm{o}}$ and 0.95$\pm$0.03 eV, respectively.

The curvature of the VBM becomes less obvious owing to the increase of band folding as the twist angle decreases and not owing to the changes of the VBM at different twist angles. The effective mass for holes of InSe bilayer at different twist angles are approximately the same and the exciton binding energies were calculated to be $\sim$ 441.1 meV and 119.1 meV for A1s intralayer excitons in the environments of vacuum and hBN, respectively. While the A1s interlayer exciton binding energies are $\sim$ 183.2 meV and $\sim$ 54.5 meV in the environments of vacuum and hBN, respectively. 

In addition, the hBN layer in the twisted InSe/hBN/InSe heterostructuere only decreases the interlayer hybridisation between two InSe layers. A good match between the effective band structure of twisted InSe/hBN/InSe heterostructure and the bands of InSe monolayer from the primitive cell calculation can also be seen. This gives A1s intralayer exciton binding energies of $\sim$ 232.9 meV and $\sim$ 138.8 meV in the environments of vacuum and hBN, respectively.

\section{Acknowledments}
The Engineering and Physical Sciences Research Council supported N.D.M.H through grant EP/P01139X/1.
S.M.L were supported by University of Warwick
Chancellor’s Scholarships. Computing resources
were provided by the Scientific Computing Research
Technology Platform of the University of Warwick, and
the UK national high performance computing service, ARCHER and ARCHER2, via the UKCP consortium (EP/P022561/1). We acknowledge the use of Athena at HPC Midlands+, which was funded by the EPSRC through Grant No. EP/P020232/1.

\bibliographystyle{unsrt}
\bibliography{references}
\end{multicols}

\newpage

\section{Supporting information}
\renewcommand{\thefigure}{S\arabic{figure}}
\renewcommand{\thetable}{S\arabic{table}}

Fig. \ref{fig:InSe-monolayer_pdos} shows the projected band structure of InSe monolayer. Table \ref{PDOS_InSe-monolayer-VBM} and table \ref{PDOS_InSe-monolayer-CBM} show the projected density of states for the valence band maximum (VBM) and the conduction band minimum (CBM) in InSe monolayer, respectively. There is no difference of orbital contribution from the same type of atom in InSe monolayer. The $p_{\mathrm{z}}$ orbitals of Se and In atoms predominate the VBM before and after 0.74 $\Gamma$-$\textbf{K}$, respectively. The CBM is predominated by the $s$ orbitals of In atoms. The decrease of $p_{\mathrm{z}}$ orbitals of Se atoms leads to the decrease of separation between the VBM and the band below it as approaching $\textbf{K}$ because the interaction between two Se atoms in the two different layers is the strongest. Due to the structural symmetry, the $p_{\mathrm{x}}$ and $p_{\mathrm{y}}$ contribute equally. The projected band structures for the A-type and B-type InSe bilayers in fig. \ref{fig:InSe-bilayer_pdos} looks similar to InSe monolayer, however, there are some  differences from InSe monolayer in details.

\setcounter{figure}{0}
\begin{figure} [H]
    \centering
    \includegraphics[width=\textwidth]{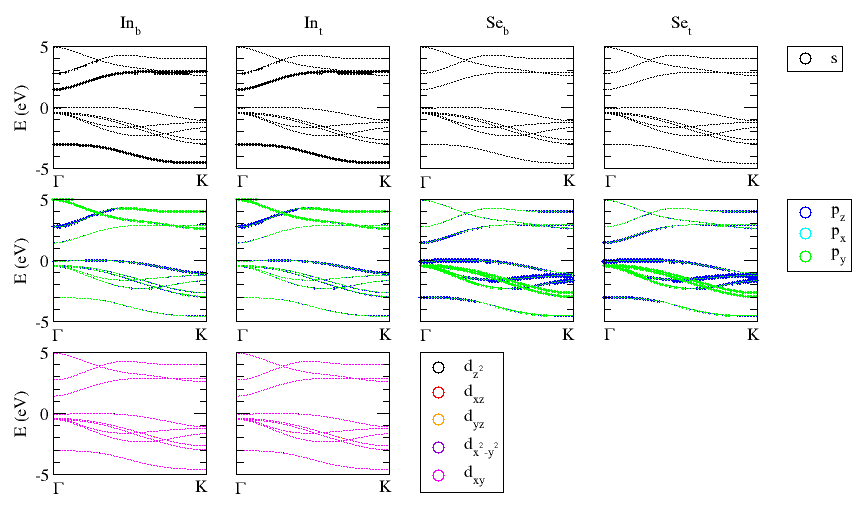}
 \caption{ Projected band structures for InSe monolayer. The subscript with `t' and `b' refer to the top and bottom InSe layers, respectively.} 
\label{fig:InSe-monolayer_pdos}
\end{figure}

\setcounter{table}{0}
\begin{table}[h!]
\centering
\begin{tabularx}{\textwidth} { 
  | >{\centering\arraybackslash}X 
  | >{\centering\arraybackslash}X 
  | >{\centering\arraybackslash}X | }
 \hline
 K-length & In & Se \\
 \hline
 $\Gamma$ & s: 0.048, $p_{\mathrm{z}}$: 0.062, $d_{{\mathrm{z}}^{2}}$: 0.003 & $p_{\mathrm{z}}$: 0.375  \\
 \hline
 0.1 & $s$: 0.040, $p_{\mathrm{z}}$: 0.077, $d_{{\mathrm{\mathrm{z}}}^{2}}$: 0.003 & $p_{\mathrm{z}}$:0.344, $p_{\mathrm{x}}$: 0.012, $p_{\mathrm{y}}$: 0.012   \\
 \hline
 0.2 & $s$: 0.035, $p_{\mathrm{z}}$: 0.097, $d_{{\mathrm{z}}^{2}}$: 0.003 & $s$: 0.001, $p_{\mathrm{z}}$: 0.329, $p_{\mathrm{x}}$: 0.013, $p_{\mathrm{y}}$: 0.013   \\
 \hline
 0.3 & $s$: 0.035, $p_{\mathrm{z}}$: 0.110, $d_{{\mathrm{z}}^{2}}$: 0.003 & $s$: 0.002, $p_{\mathrm{z}}$: 0.317, $p_{\mathrm{x}}$: 0.013, $p_{\mathrm{y}}$: 0.013   \\
 \hline
 0.4 & $s$: 0.036, $p_{\mathrm{z}}$: 0.121, $d_{{\mathrm{z}}^{2}}$: 0.002 & $s$: 0.003, $p_{\mathrm{z}}$: 0.298, $p_{\mathrm{x}}$: 0.016, $p_{\mathrm{y}}$: 0.016   \\
 \hline
 0.5 & $s$: 0.039, $p_{\mathrm{z}}$: 0.131, $p_{\mathrm{x}}$: 0.003, $p_{\mathrm{y}}$: 0.003, $d_{{\mathrm{z}}^{2}}$: 0.002 & $s$: 0.004, $p_{\mathrm{z}}$: 0.268, $p_{\mathrm{x}}$: 0.022, $p_{\mathrm{y}}$: 0.022   \\
 \hline
 0.6 & $s$: 0.042, $p_{\mathrm{z}}$: 0.143, $p_{\mathrm{x}}$: 0.006, $p_{\mathrm{y}}$: 0.006, $d_{{\mathrm{z}}^{2}}$: 0.002 & $s$: 0.005, $p_{\mathrm{z}}$: 0.230, $p_{\mathrm{x}}$: 0.031, $p_{\mathrm{y}}$: 0.031   \\
 \hline
 0.7 & $s$: 0.045, $p_{\mathrm{z}}$: 0.160, $p_{\mathrm{x}}$: 0.009, $p_{\mathrm{y}}$: 0.009, $d_{{\mathrm{z}}^{2}}$: 0.001 & $s$: 0.006, $p_{\mathrm{z}}$: 0.186, $p_{\mathrm{x}}$: 0.040, $p_{\mathrm{y}}$: 0.040   \\
 \hline
 0.8 & $s$: 0.049, $p_{\mathrm{z}}$: 0.188, $p_{\mathrm{x}}$: 0.010, $p_{\mathrm{y}}$: 0.010, $d_{{\mathrm{z}}^{2}}$: 0.001 & $s$: 0.005, $p_{\mathrm{z}}$: 0.131, $p_{\mathrm{x}}$: 0.052, $p_{\mathrm{y}}$: 0.052   \\
 \hline
 0.9 & $s$: 0.057, $p_{\mathrm{z}}$: 0.236, $p_{\mathrm{x}}$: 0.005, $p_{\mathrm{y}}$: 0.005, $d_{{\mathrm{z}}^{2}}$: 0.001 & $s$: 0.002, $p_{\mathrm{z}}$: 0.053, $p_{\mathrm{x}}$: 0.069, $p_{\mathrm{y}}$: 0.069   \\
 \hline
 $\textbf{K}$ & $s$: 0.064, $p_{\mathrm{z}}$: 0.271, $d_{{\mathrm{z}}^{2}}$: 0.001 & $p_{\mathrm{x}}$: 0.081, $p_{\mathrm{y}}$: 0.081   \\
 \hline
\end{tabularx}
 \caption{ Details of the valance band maximum for the projected band structure in fig. \ref{fig:InSe-monolayer_pdos}.} 
\label{PDOS_InSe-monolayer-VBM}
\end{table}

\begin{table}[h!] 
\centering
\begin{tabularx}{\textwidth} { 
  | >{\centering\arraybackslash}X 
  | >{\centering\arraybackslash}X 
  | >{\centering\arraybackslash}X | }
 \hline
 K-length & In & Se \\
 \hline
 $\Gamma$ & $s$: 0.196, $p_{\mathrm{z}}$: 0.049 & $s$: 0.043, $p_{\mathrm{z}}$: 0.193  \\
 \hline
 0.1 & $s$: 0.191, $p_{\mathrm{z}}$: 0.038, $p_{\mathrm{x}}$: 0.002, $p_{\mathrm{y}}$: 0.002 & $s$: 0.037, $p_{\mathrm{z}}$: 0.172, $p_{\mathrm{x}}$: 0.019, $p_{\mathrm{y}}$: 0.019   \\
 \hline
 0.2 & $s$: 0.194, $p_{\mathrm{z}}$: 0.022, $p_{\mathrm{x}}$: 0.005, $p_{\mathrm{y}}$: 0.005 & $s$:  0.029, $p_{\mathrm{z}}$: 0.147, $p_{\mathrm{x}}$: 0.040, $p_{\mathrm{y}}$: 0.040   \\
 \hline
 0.3 & $s$: 0.208, $p_{\mathrm{z}}$: 0.009, $p_{\mathrm{x}}$: 0.006, $p_{\mathrm{y}}$: 0.006 & $s$:  0.021, $p_{\mathrm{z}}$: 0.129, $p_{\mathrm{x}}$: 0.052, $p_{\mathrm{y}}$: 0.052   \\
 \hline
 0.4 & $s$: 0.229, $p_{\mathrm{z}}$: 0.002, $p_{\mathrm{x}}$: 0.004, $p_{\mathrm{y}}$: 0.004 & $s$:  0.014, $p_{\mathrm{z}}$: 0.110, $p_{\mathrm{x}}$: 0.062, $p_{\mathrm{y}}$: 0.062   \\
 \hline
 0.5 & $s$: 0.249, $p_{\mathrm{x}}$: 0.001, $p_{\mathrm{y}}$: 0.001 & $s$: 0.007, $p_{\mathrm{z}}$: 0.086, $p_{\mathrm{x}}$: 0.072, $p_{\mathrm{y}}$: 0.072   \\
 \hline
 0.6 & $s$: 0.262, $p_{\mathrm{z}}$: 0.002 & $s$: 0.003, $p_{\mathrm{z}}$: 0.059, $p_{\mathrm{x}}$: 0.080, $p_{\mathrm{y}}$: 0.080   \\
 \hline
 0.7 & $s$: 0.271, $p_{\mathrm{z}}$: 0.006 & $s$: 0.001, $p_{\mathrm{z}}$: 0.034, $p_{\mathrm{x}}$: 0.086, $p_{\mathrm{y}}$: 0.086   \\
 \hline
 0.8 & $s$: 0.278, $p_{\mathrm{z}}$: 0.011 & $p_{\mathrm{z}}$: 0.015, $p_{\mathrm{x}}$: 0.089, $p_{\mathrm{y}}$: 0.089   \\
 \hline
 0.9 & $s$: 0.283, $p_{\mathrm{z}}$: 0.015 & $p_{\mathrm{z}}$: 0.003, $p_{\mathrm{x}}$: 0.091, $p_{\mathrm{y}}$: 0.091   \\
 \hline
 $\textbf{K}$ & $s$: 0.285, $p_{\mathrm{z}}$: 0.016 & $p_{\mathrm{x}}$: 0.091, $p_{\mathrm{y}}$: 0.091   \\
 \hline
\end{tabularx}
 \caption{ Details of the conduction band minimum for the projected band structure in fig. \ref{fig:InSe-monolayer_pdos}. } 
\label{PDOS_InSe-monolayer-CBM}
\end{table}

\newpage

\begin{figure} [H]
    \centering
    \includegraphics[width=0.72\textwidth]{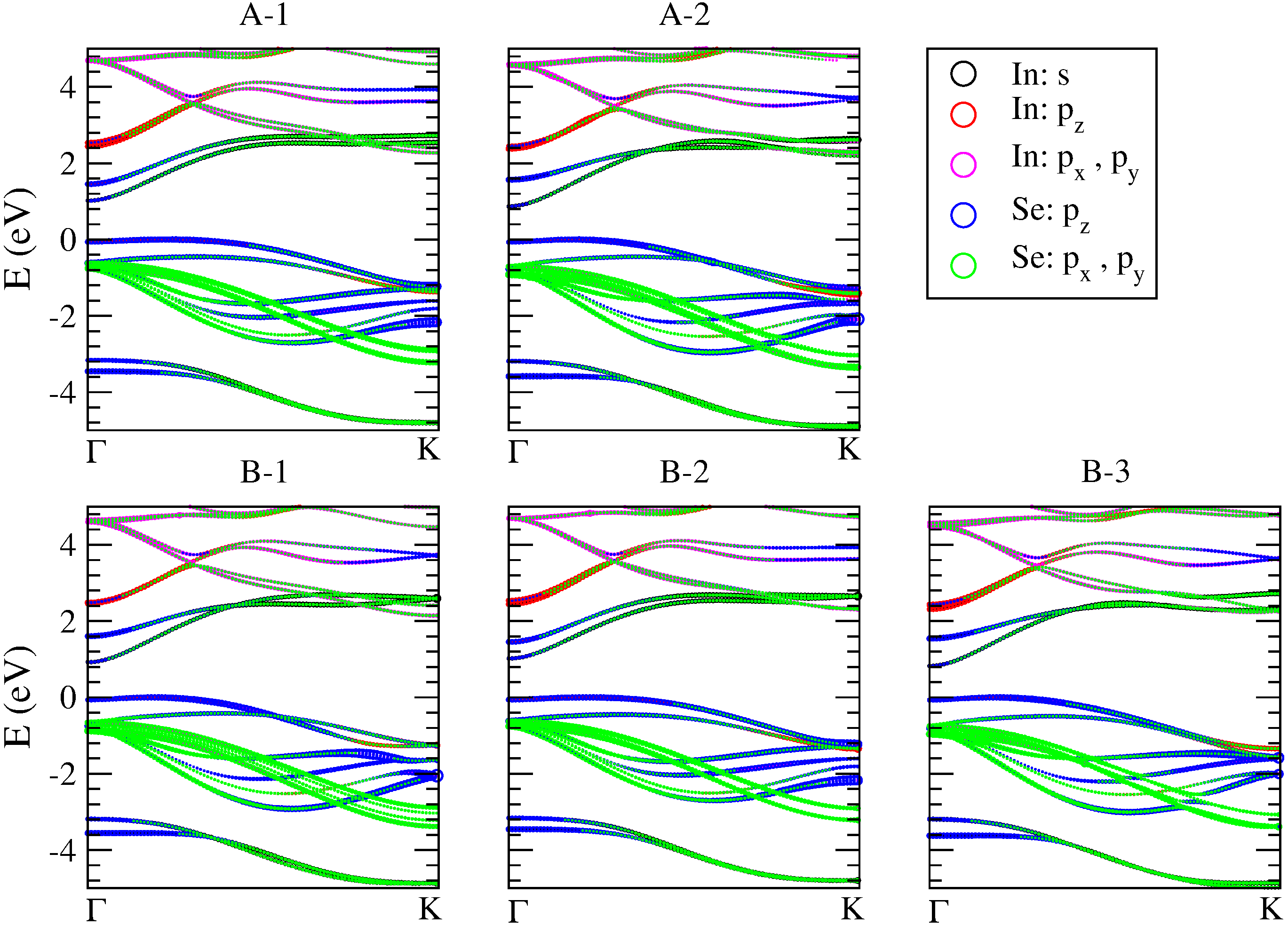}    
 \caption{ Projected band structures for the A-type and B-type InSe bilayers. Black: $s$ orbital of In atom, red: $p_{\mathrm{z}}$ orbital of In atom, magenta: $p_{\mathrm{x}}$ and $p_{\mathrm{y}}$ orbitals of In atom, blue: $p_{\mathrm{z}}$ orbital of Se atom, green: $p_{\mathrm{x}}$ and $p_{\mathrm{y}}$ orbitals of Se atom.} 
\label{fig:InSe-bilayer_pdos}
\end{figure}

\begin{figure} [H]
    \centering
    \includegraphics[width=0.65\textwidth]{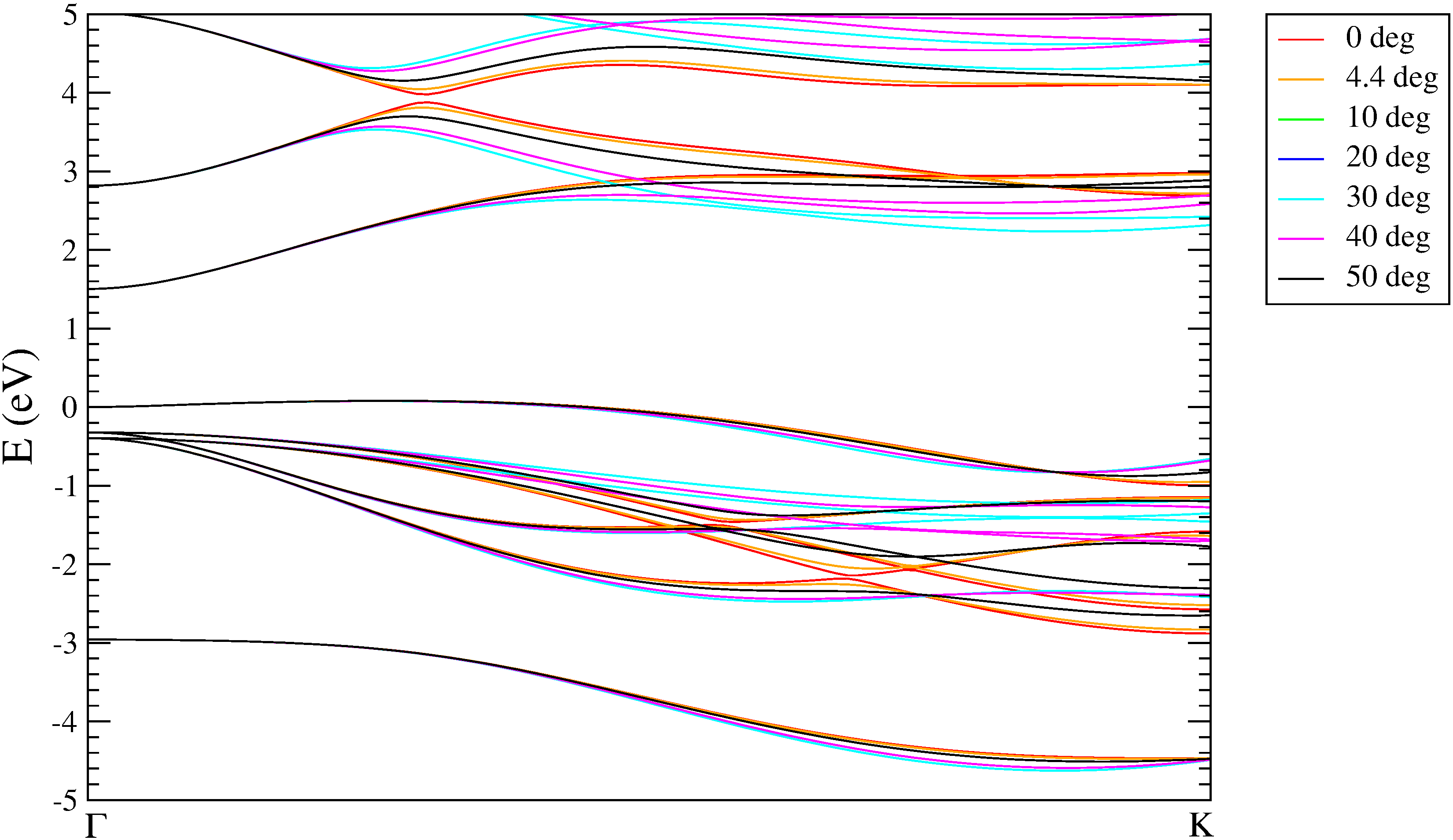}    
 \caption{ Various band structures of twisted InSe monolayers along $\Gamma$ to $\textbf{K}$ of the untwisted (0$^{\mathrm{o}}$) InSe monolayer. The Fermi level is shifted to the energy of the valence band maximum at $\Gamma$.} 
\label{fig:bandstructure_InSe-monolayer_rotation}
\end{figure}

Fig. \ref{fig:bandstructure_InSe-monolayer_rotation} shows the various band structures of twisted InSe monolayers along $\Gamma$ to $\textbf{K}$ of the untwisted (0$^{\mathrm{o}}$) InSe monolayer. The band structures remain almost the same from $\Gamma$ to the highest energy kpoint of the VBM (the kpoint we concern) for various twist angles. It suggests that the  band gap and the effective mass for holes are not affected by the variation of kpoint path at different twist angles.

\begin{figure} [H]
\captionsetup[subfigure]{justification=centering}
\begin{subfigure}{0.45\textwidth}
    \centering
    \includegraphics[width=0.62\textwidth]{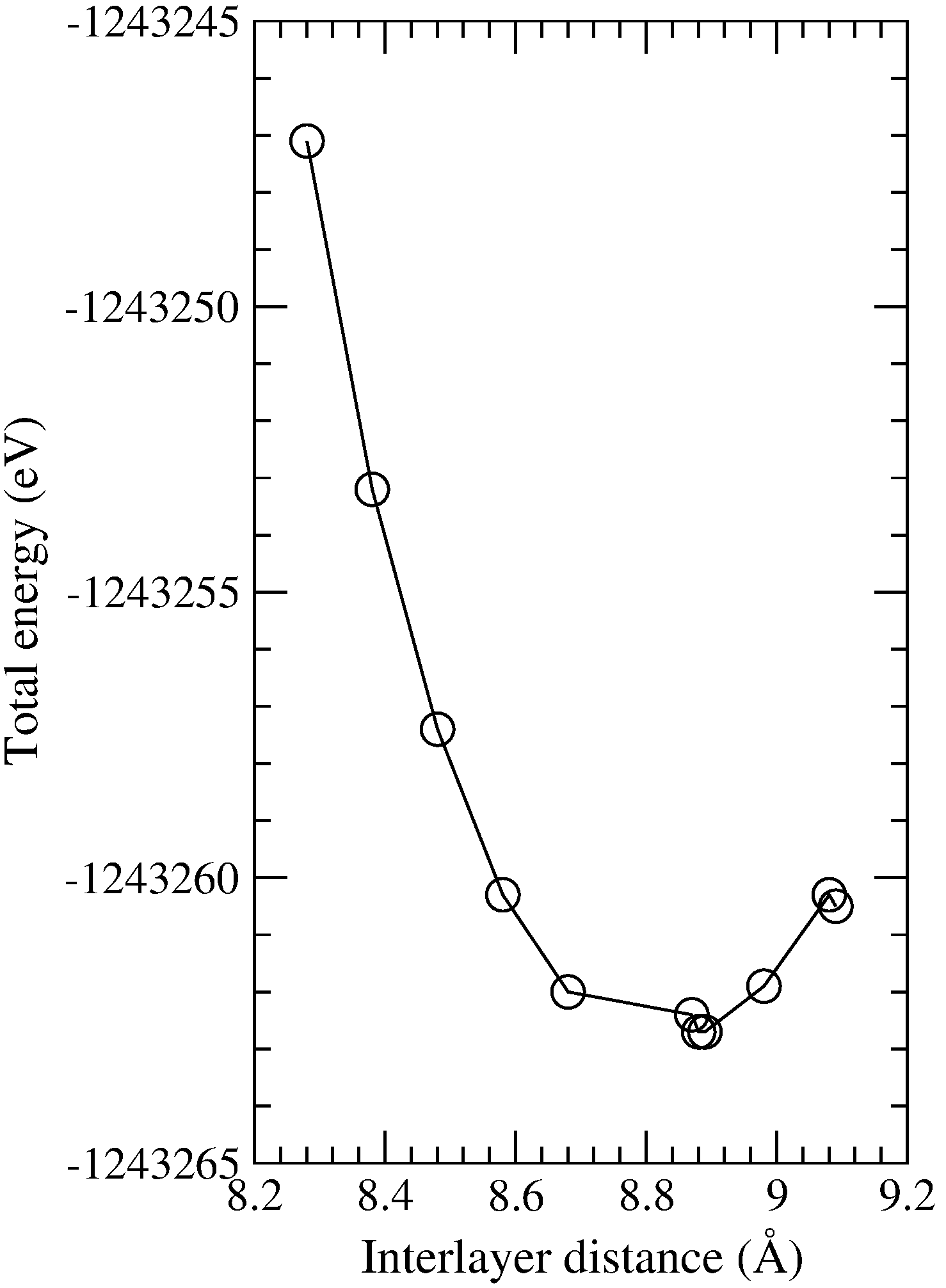}
    \caption{}
\end{subfigure}
\begin{subfigure}{0.45\textwidth}
  \centering
  \includegraphics[width=0.62\textwidth]{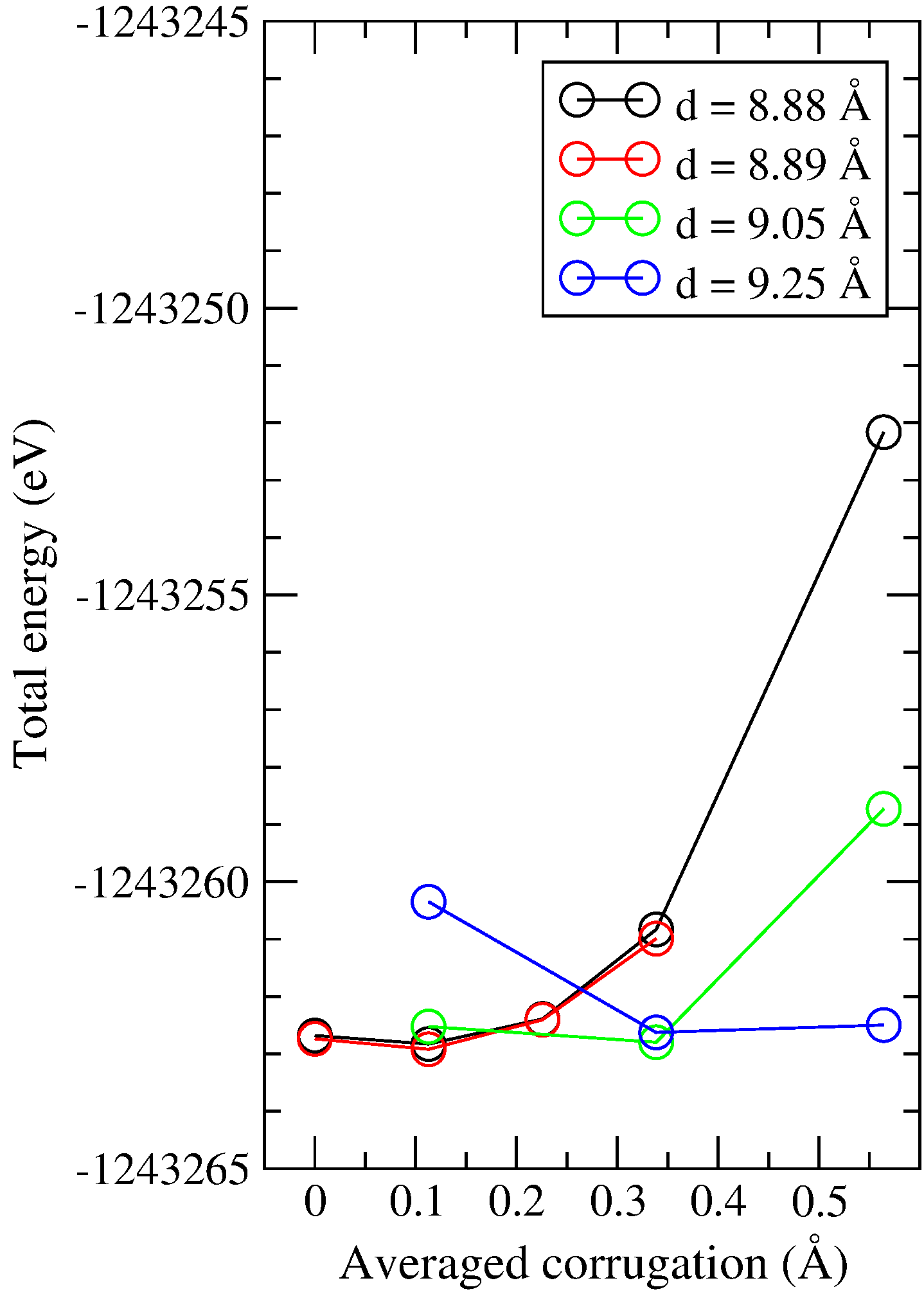}
  \caption{}
\end{subfigure}
 \caption{ Total energy versus (a) interlayer distance (corrugation was not considered in the calculation) (b) averanged corrugation for the A-type twisted InSe bilayer with a twist angle of 4.4$^{\mathrm{o}}$.} 
\label{fig:interlayer_distance-corrugation}
\end{figure}

Fig. \ref{fig:interlayer_distance-corrugation}(a) shows the total energy versus interlayer distance for the A-type twisted InSe bilayer with a twist angle of 4.4$^{\mathrm{o}}$. The interlayer distance is optimised to be 8.89 \AA. The small corrugation ($\simeq$ 0.1 \AA) is obtained at the optimised interlayer distance (d= 8.88 \AA \, and d= 8.89 \AA ). In contrast, a larger corrugation ($\simeq$ 0.3 \AA) is obtained if deviating from the optimised interlayer distance (d= 9.05 \AA \, and d= 9.25 \AA). According to the previous literature \cite{Twisted_InSe_bilayer}, the averaged corrugation in this work is defined as 

\begin{equation}
\mathrm{Averaged \: \: corrugation}= \frac{1}{4}\sum_{n=1}^{4}| d^{n}-d |
\end{equation}

\noindent where $d^{n}$ and $d$ are the averaged interlayer distance calculated from each In and Se sublayers in the top layer to the same sublayer in the bottom layer when including out-of-plane corrugation and the optimised interlayer distance without including out-of-plane corrugation, respectively. 

\newpage

The effective band structure with the consideration of the corrugation looks similar to the effective band structure without consideration of the corrugation. The separation between neighbouring bands such as the VBM and the band below it slightly increases by the corrugation. 

\begin{figure} [H]
\captionsetup[subfigure]{justification=centering}
\begin{subfigure}{0.5\textwidth}
  \centering
  \includegraphics[width=\textwidth]{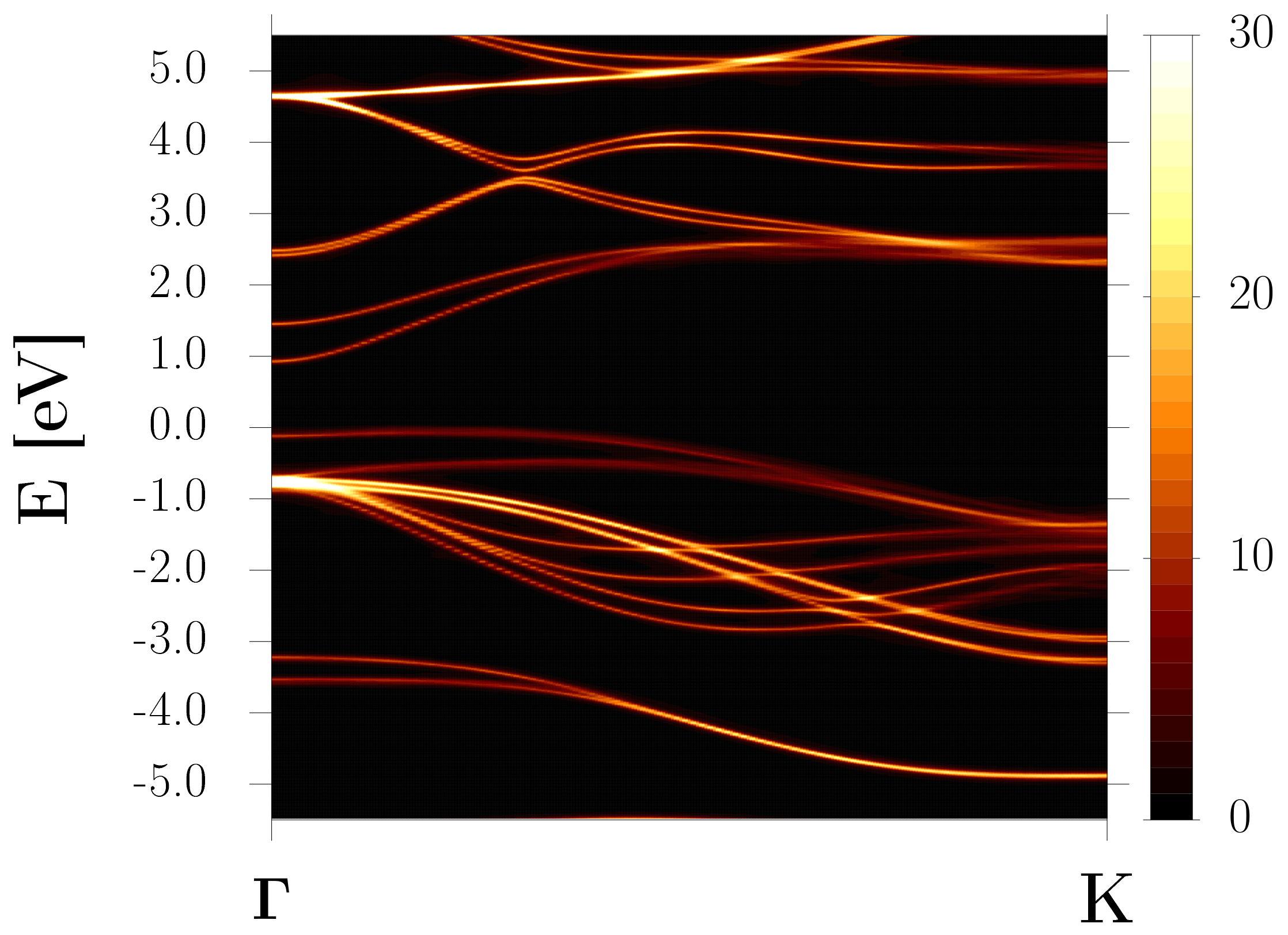}
  \caption{Without the corrugation}
\end{subfigure}
\begin{subfigure}{0.5\textwidth}
  \centering
  \includegraphics[width=\textwidth]{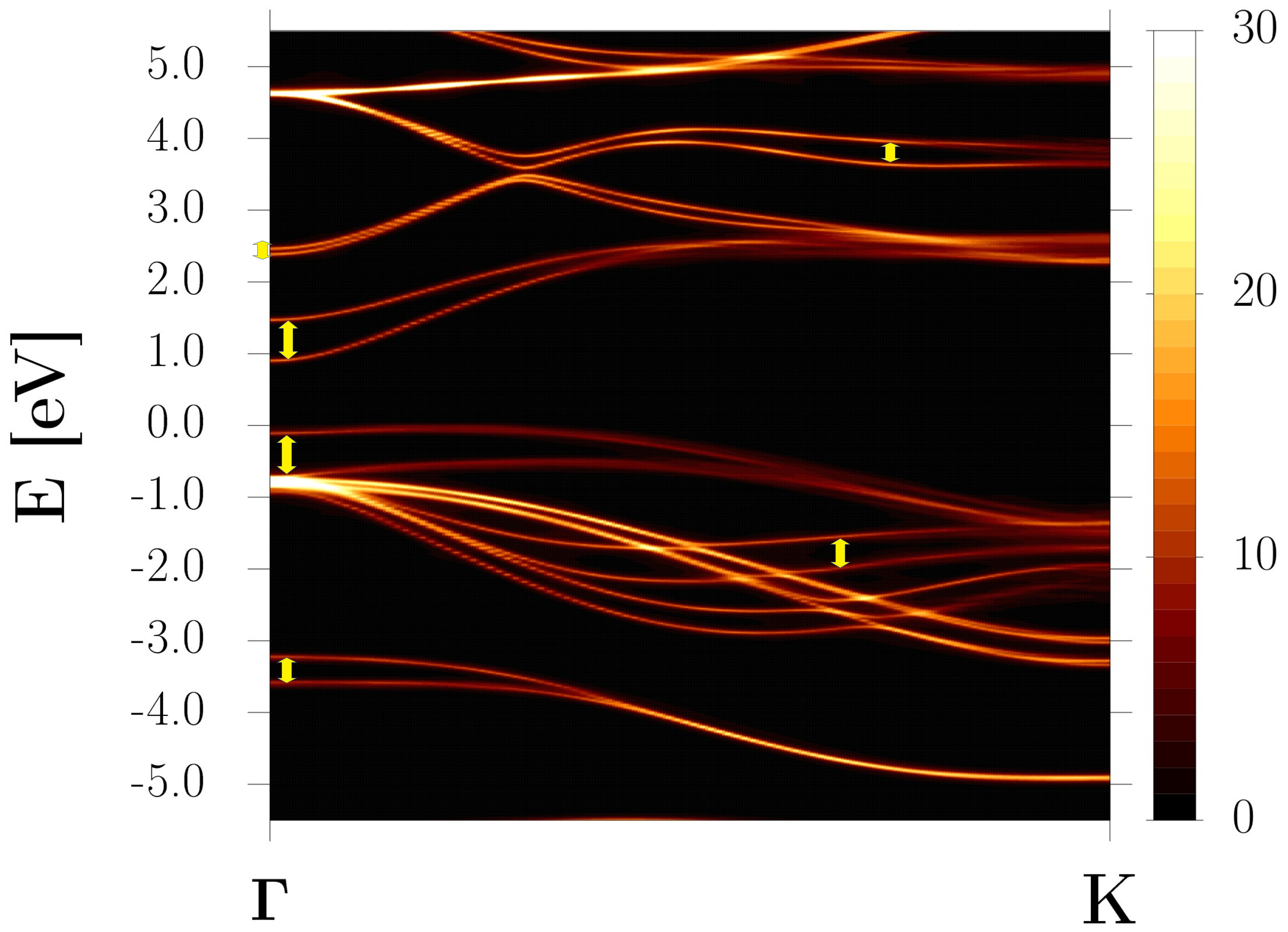}
  \caption{With the corrugation}
\end{subfigure}
 \caption{ Comparison of the effective band structure of the A-type twisted InSe bilayer with a twist angle of 4.4$^{\mathrm{o}}$ (a) before (b) after geometry optimisation. The double-headed arrows show the increase of separation between some neighbouring bands.}
\label{fig:compare_with-without_corrugation}
\end{figure}

\newpage

\begin{figure} [H]
\captionsetup[subfigure]{justification=centering}
\begin{subfigure}{0.5\textwidth}
  \centering
  \includegraphics[width=\textwidth]{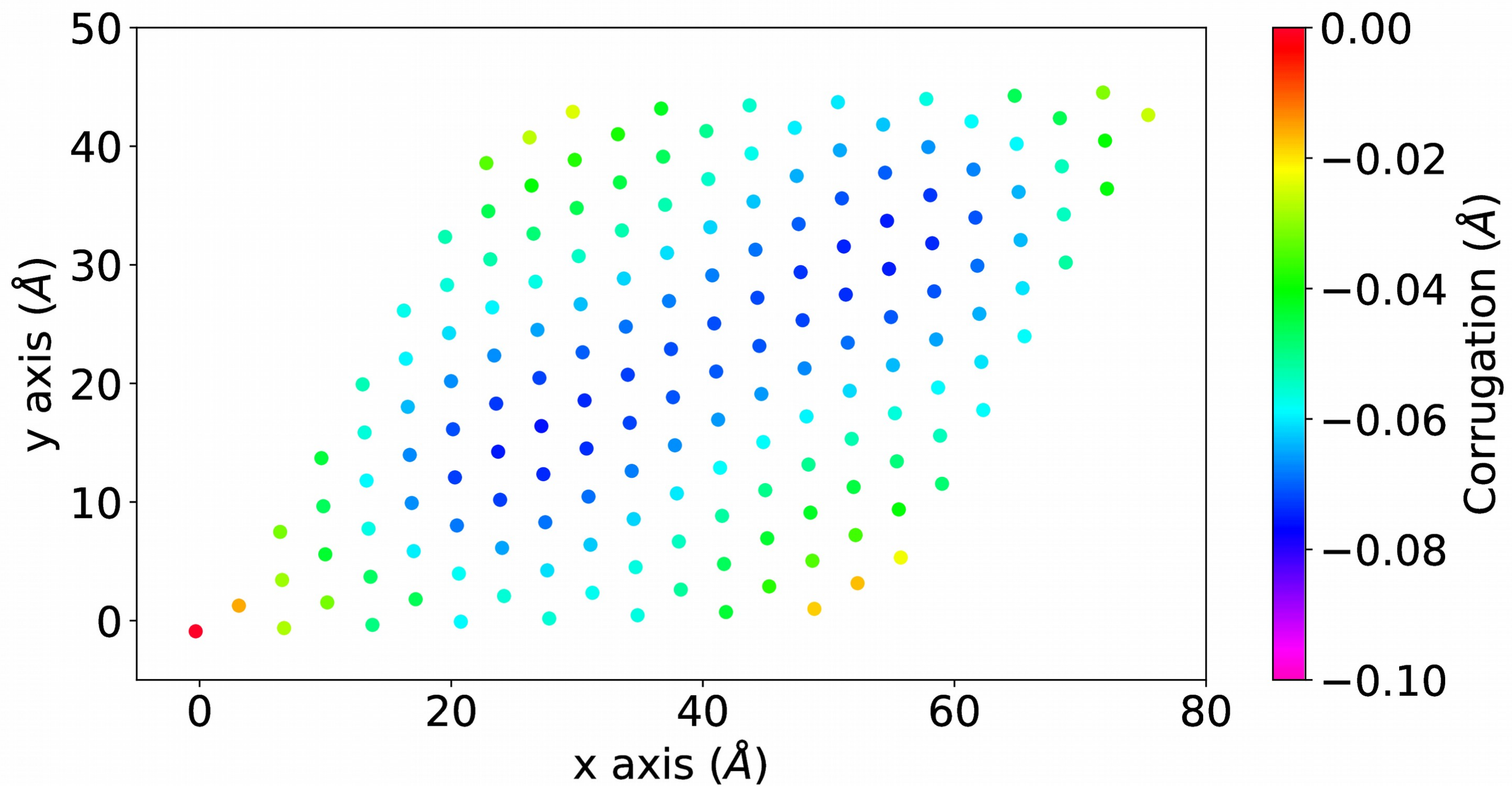}
  \caption{In$_{\mathrm{t}}$}
\end{subfigure}
\begin{subfigure}{0.5\textwidth}
  \centering
  \includegraphics[width=\textwidth]{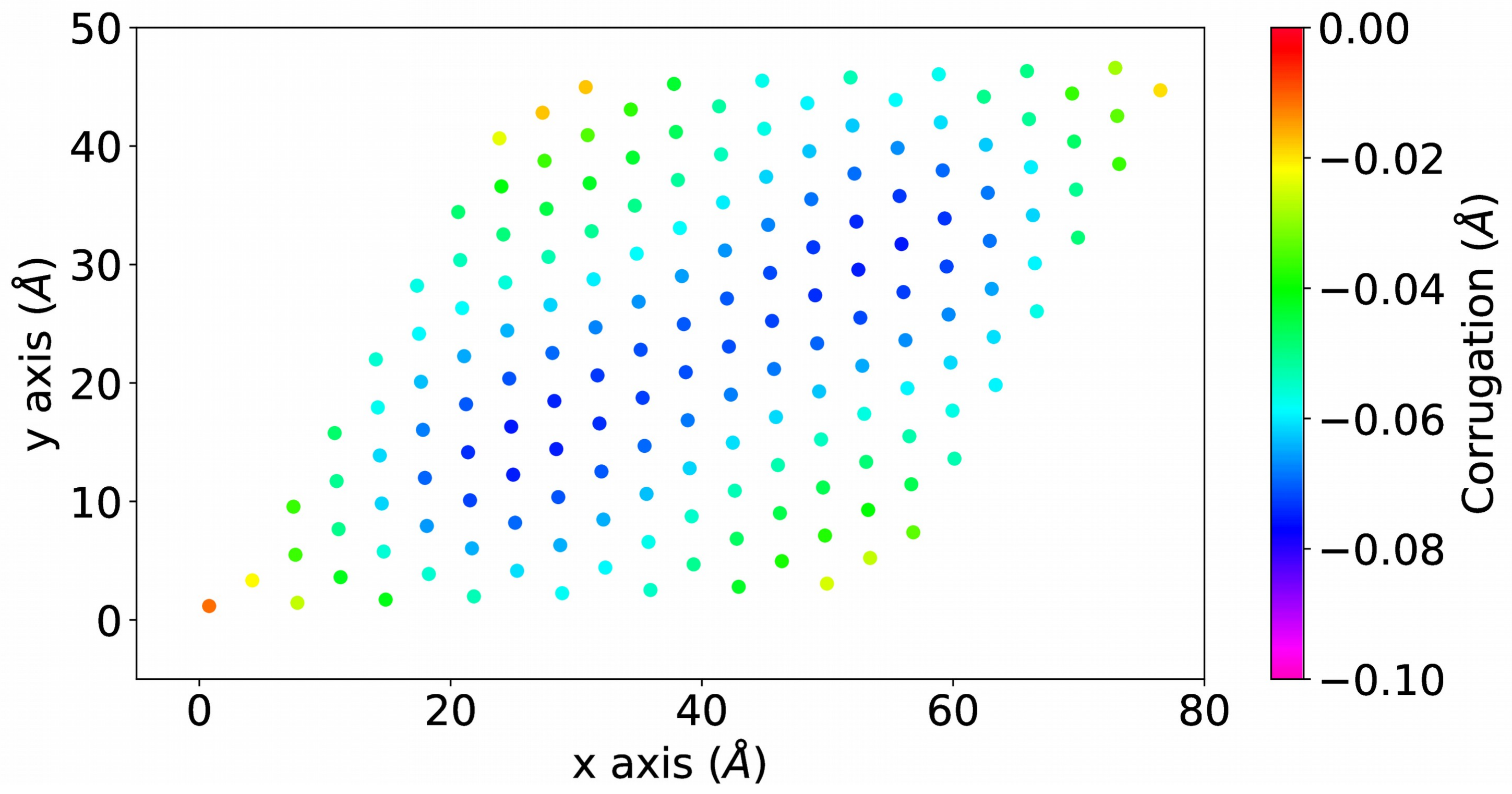}
  \caption{Se$_{\mathrm{t}}$}
\end{subfigure}
\begin{subfigure}{0.5\textwidth}
  \centering
  \includegraphics[width=\textwidth]{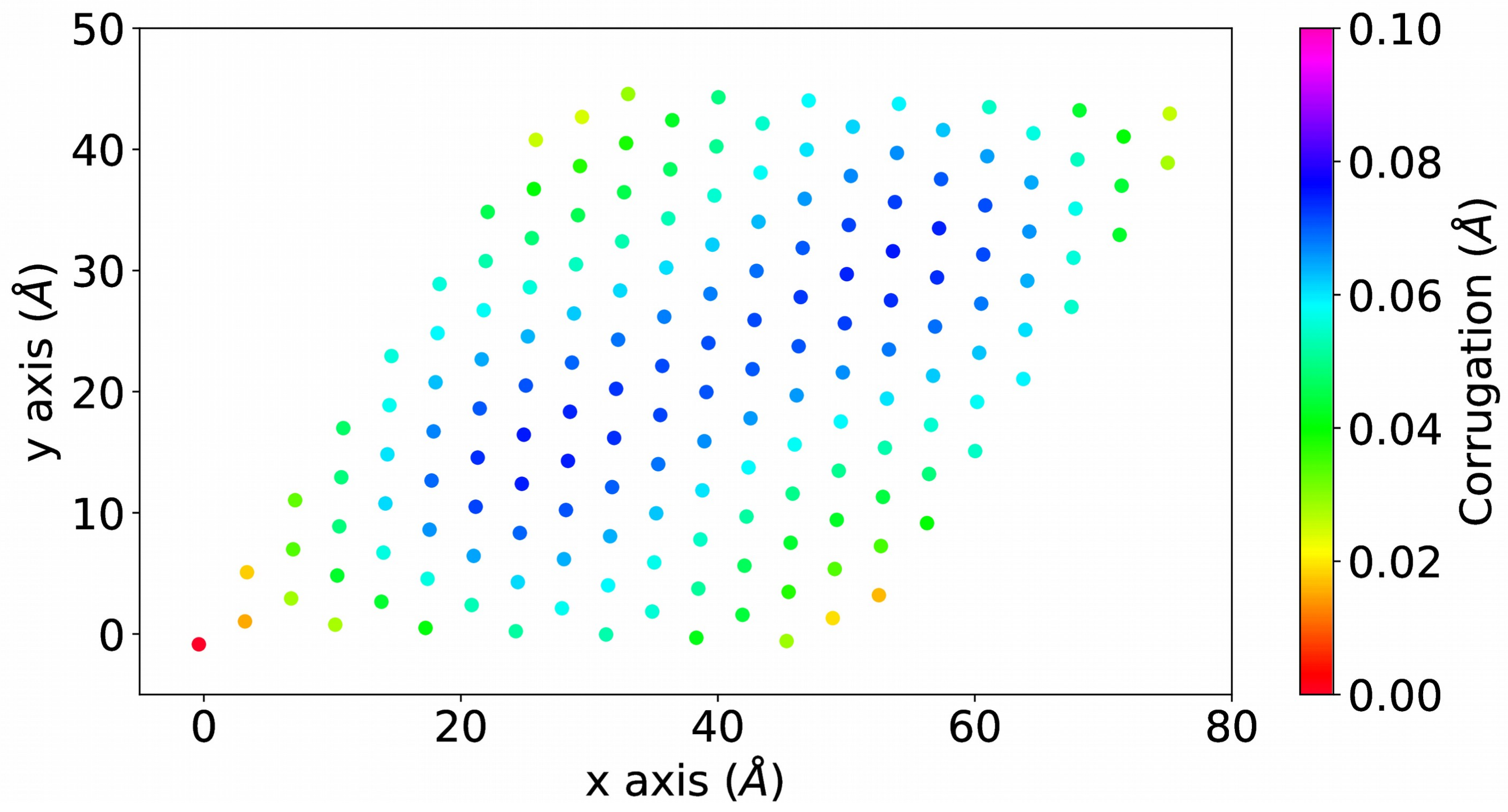}
  \caption{In$_{\mathrm{b}}$}
\end{subfigure}
\begin{subfigure}{0.5\textwidth}
  \centering
  \includegraphics[width=\textwidth]{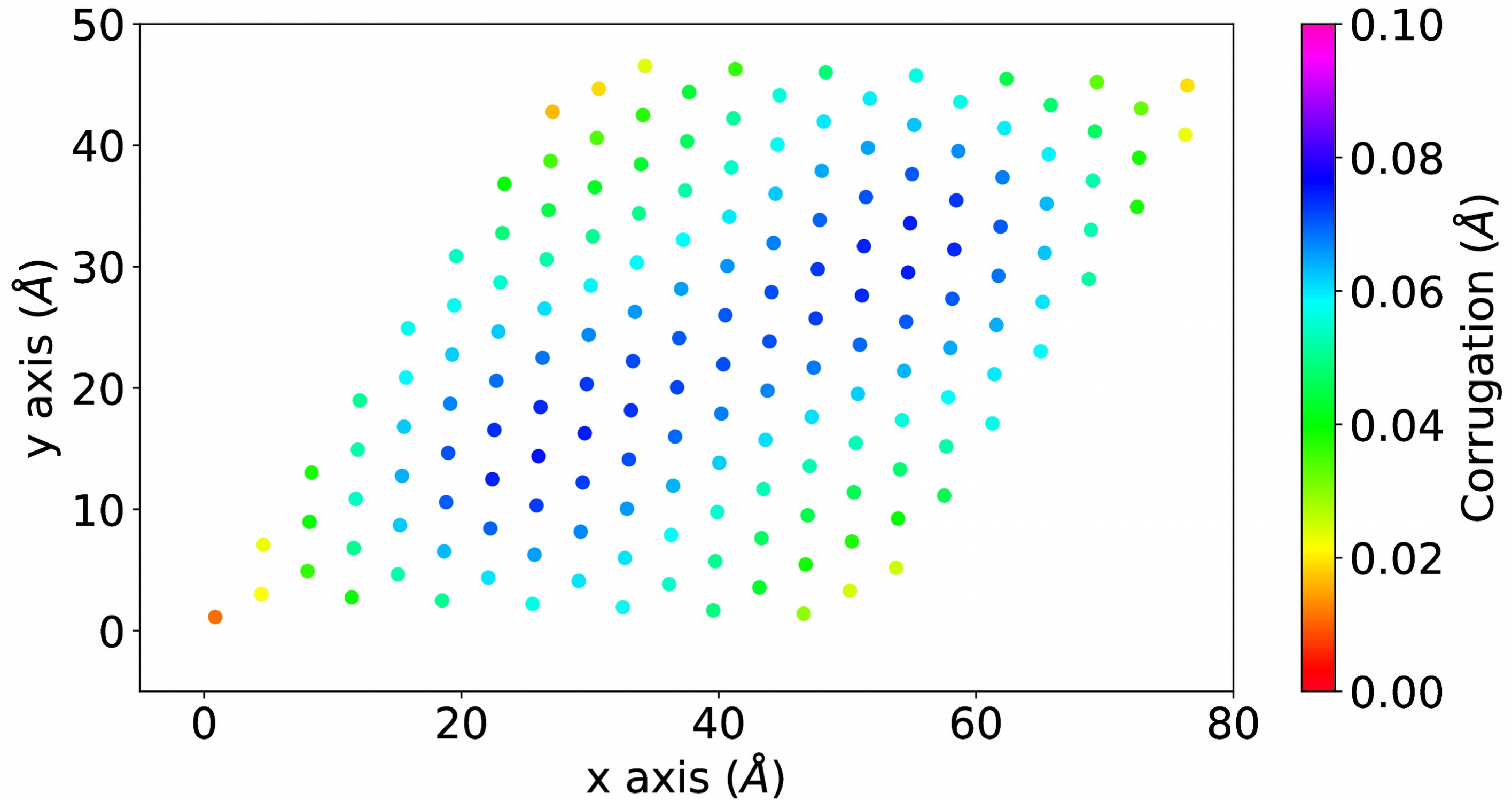}
  \caption{Se$_{\mathrm{b}}$}
\end{subfigure}
 \caption{ An initial atomic structure with the averaged corrugation ($\simeq$ 0.1 \AA) included. The subscript `t' and `b' refer to the top and bottom layers, respectively.}
\label{fig:Corrugation_distribution}
\end{figure}

An initial atomic structure was given according to the distribution of the interlayer distance corresponding to different stacking configurations seen in some regions of twisted InSe bilayer (see table \ref{InSe-bilayer_parameters}). Fig. \ref{fig:Corrugation_distribution} shows an initial atomic structure with the consideration of the averaged corrugation ($\simeq$ 0.1 \AA) for the A-type twisted InSe bilayer with a twist angle of 4.4$^{\mathrm{o}}$. 

\newpage

Before structural relaxation, the innermost In atoms (In$_{\mathrm{bt}}$ and In$_{\mathrm{tb}}$) do not show a similar distribution of the residual total force to the neighbouring Se atoms (Se$_{\mathrm{bt}}$ and Se$_{\mathrm{tb}}$).

\begin{figure} [H]
\captionsetup[subfigure]{justification=centering}
\begin{subfigure}{0.5\textwidth}
  \centering
  \includegraphics[width=\textwidth]{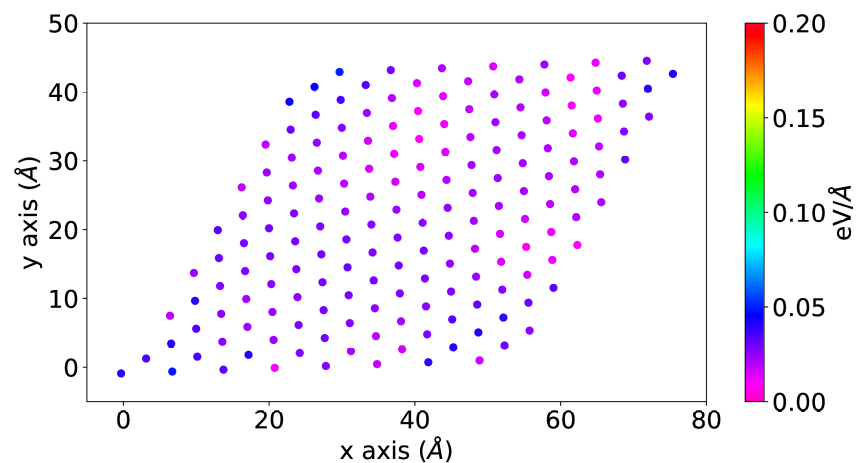}
  \caption{In$_{\mathrm{bt}}$}
\end{subfigure}
\begin{subfigure}{0.5\textwidth}
  \centering
  \includegraphics[width=\textwidth]{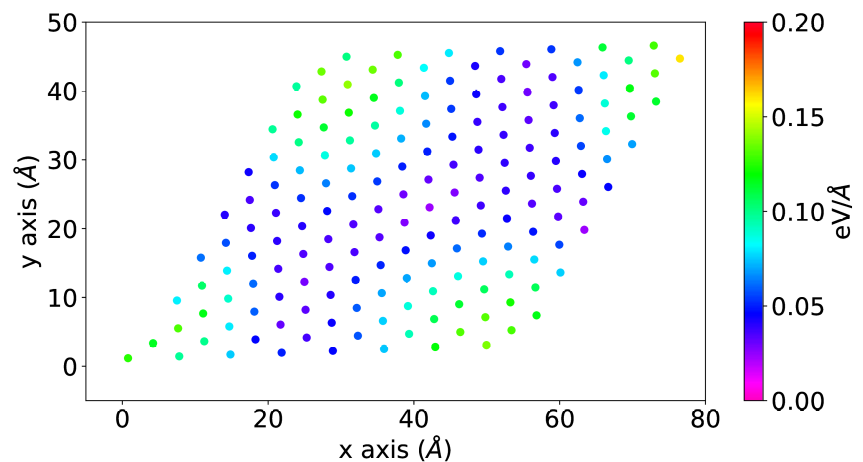}
  \caption{Se$_{\mathrm{bt}}$}
\end{subfigure}
\begin{subfigure}{0.5\textwidth}
  \centering
  \includegraphics[width=\textwidth]{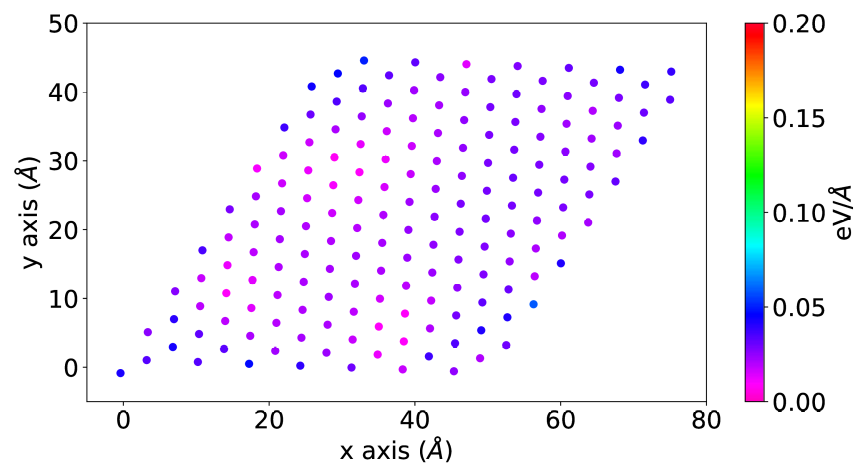}
  \caption{In$_{\mathrm{tb}}$}
\end{subfigure}
\begin{subfigure}{0.5\textwidth}
  \centering
  \includegraphics[width=\textwidth]{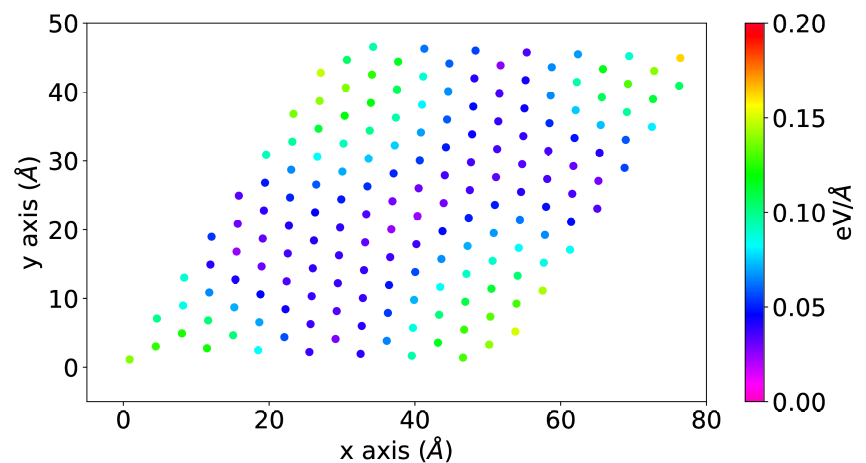}
  \caption{Se$_{\mathrm{tb}}$}
\end{subfigure}
 \caption{ Distribution of the residual total force in the innermost In and Se atoms for the A-type twisted InSe bilayer with a twist angle of 4.4$^{\mathrm{o}}$ before geometry optimisation.}
\label{fig:Force_distribution_before-relaxation}
\end{figure}

\newpage

After the structural relaxation, the similarity between fig. \ref{fig:Force_distribution_after-relaxation} and fig. \ref{fig:Corrugation_distribution} is remained for the residual force in the z direction. The distribution is similar for the neigbouring In and Se atoms, whereas the residual force is larger in Se atoms. The z direction of the residual force is opposite in the top and bottom InSe layers. The total residual forces mostly originate from the residual forces along the z direction.

\begin{figure} [H]
\captionsetup[subfigure]{justification=centering}
\begin{subfigure}{0.5\textwidth}
  \centering
  \includegraphics[width=\textwidth]{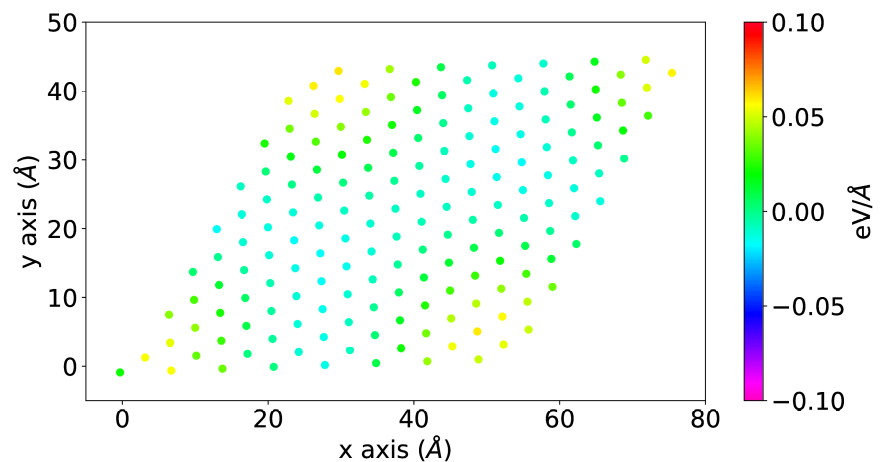}
  \caption{In$_{\mathrm{bt}}$}
\end{subfigure}
\begin{subfigure}{0.5\textwidth}
  \centering
  \includegraphics[width=\textwidth]{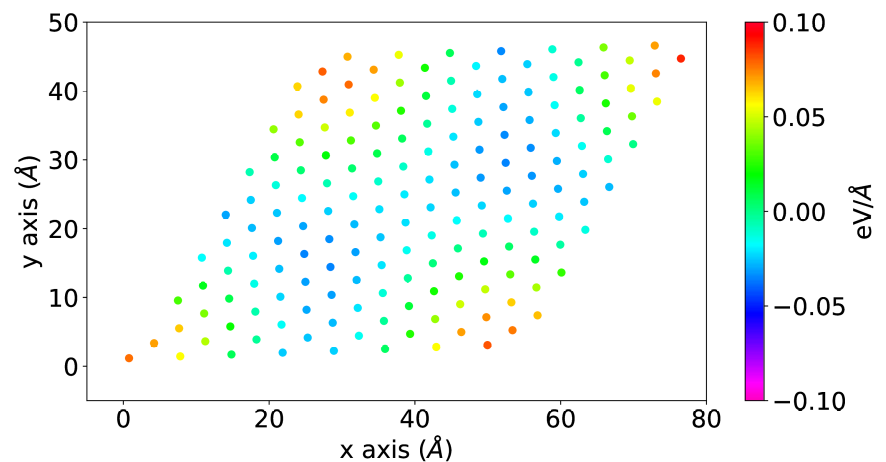}
  \caption{Se$_{\mathrm{bt}}$}
\end{subfigure}
\begin{subfigure}{0.5\textwidth}
  \centering
  \includegraphics[width=\textwidth]{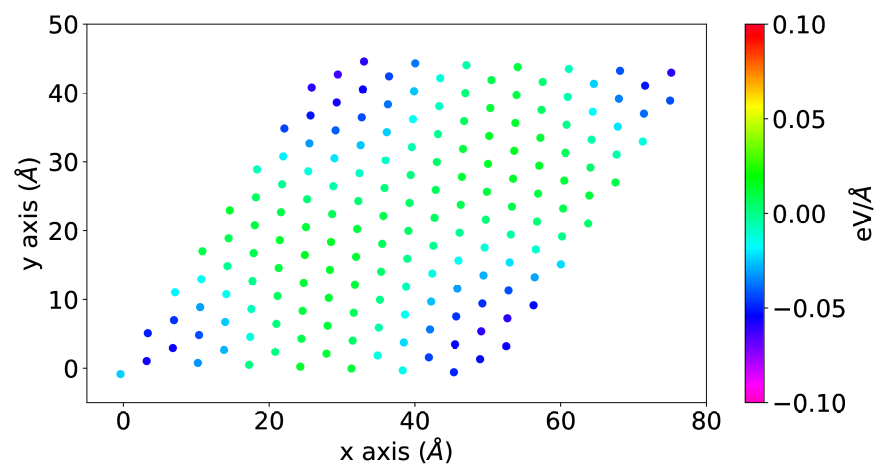}
  \caption{In$_{\mathrm{tb}}$}
\end{subfigure}
\begin{subfigure}{0.5\textwidth}
  \centering
  \includegraphics[width=\textwidth]{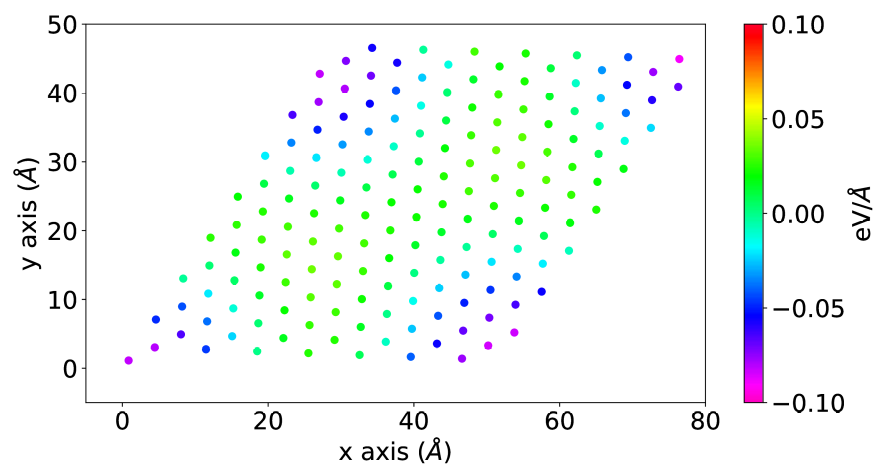}
  \caption{Se$_{\mathrm{tb}}$}
\end{subfigure}
 \caption{ Distribution of the residual force along the z direction in the innermost In and Se atoms for the A-type twisted InSe bilayer with a twist angle of 4.4$^{\mathrm{o}}$ after geometry optimisation.}
\label{fig:Force_distribution_z}
\end{figure}  

\newpage

After the structural relaxation, the similarity between fig. \ref{fig:Force_distribution_after-relaxation} and fig. \ref{fig:Corrugation_distribution} is remained for the residual force in xy plane. The distribution is similar for the neigbouring In and Se atoms, whereas the residual force in the xy plane is larger in Se atoms. 

\begin{figure} [H]
\captionsetup[subfigure]{justification=centering} 
\begin{subfigure}{0.5\textwidth}
  \centering 
  \includegraphics[width=\textwidth]{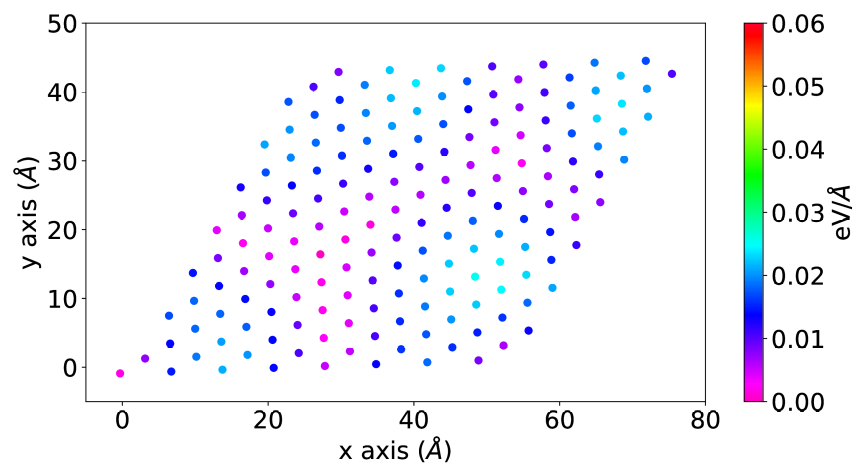}
  \caption{In$_{\mathrm{bt}}$}
\end{subfigure}
\begin{subfigure}{0.5\textwidth}
  \centering
  \includegraphics[width=\textwidth]{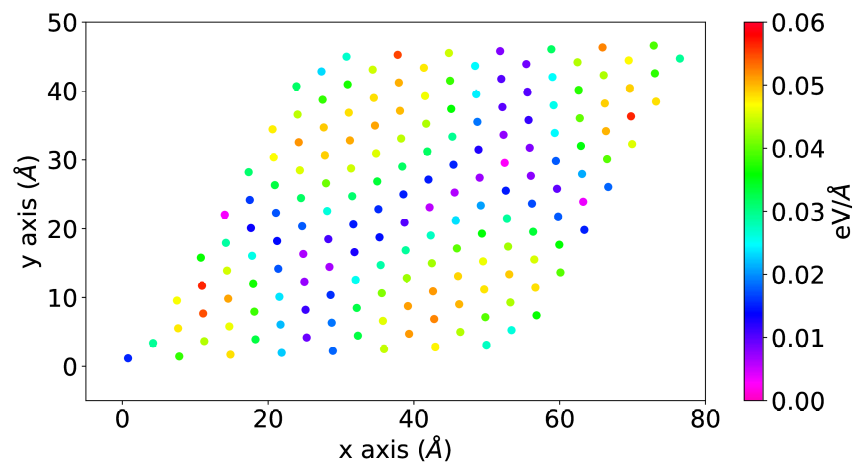}
  \caption{Se$_{\mathrm{bt}}$}
\end{subfigure}
\begin{subfigure}{0.5\textwidth}
  \centering
  \includegraphics[width=\textwidth]{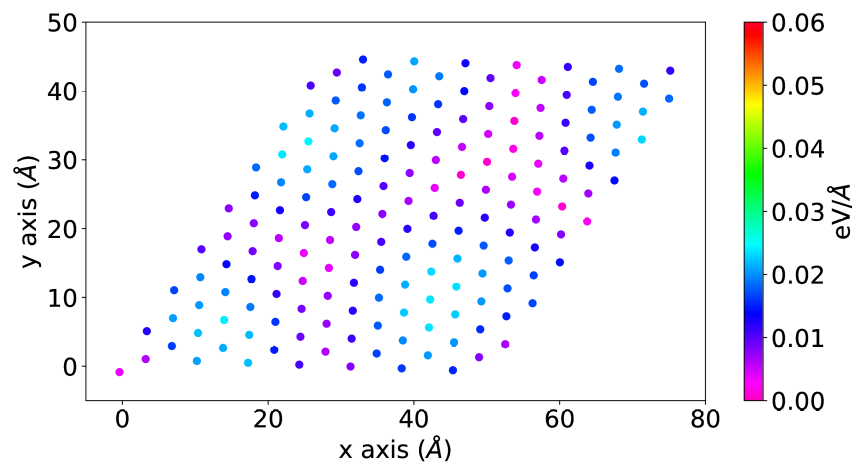}
  \caption{In$_{\mathrm{tb}}$}
\end{subfigure}
\begin{subfigure}{0.5\textwidth}
  \centering
  \includegraphics[width=\textwidth]{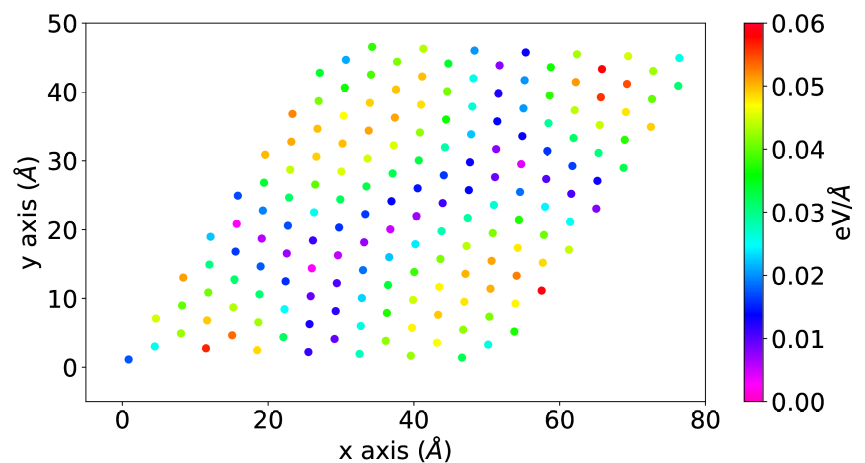}
  \caption{Se$_{\mathrm{tb}}$}
\end{subfigure}
 \caption{ Distribution of the residual force in the xy plane in the innermost In and Se atoms for the A-type twisted InSe bilayer with a twist angle of 4.4$^{\mathrm{o}}$ after geometry optimisation.}
\label{fig:Force_distribution_xy}
\end{figure}

\begin{figure} [H]
\captionsetup[subfigure]{justification=centering}
\begin{subfigure}{0.498\textwidth}
  \centering
  \includegraphics[width=0.73\textwidth]{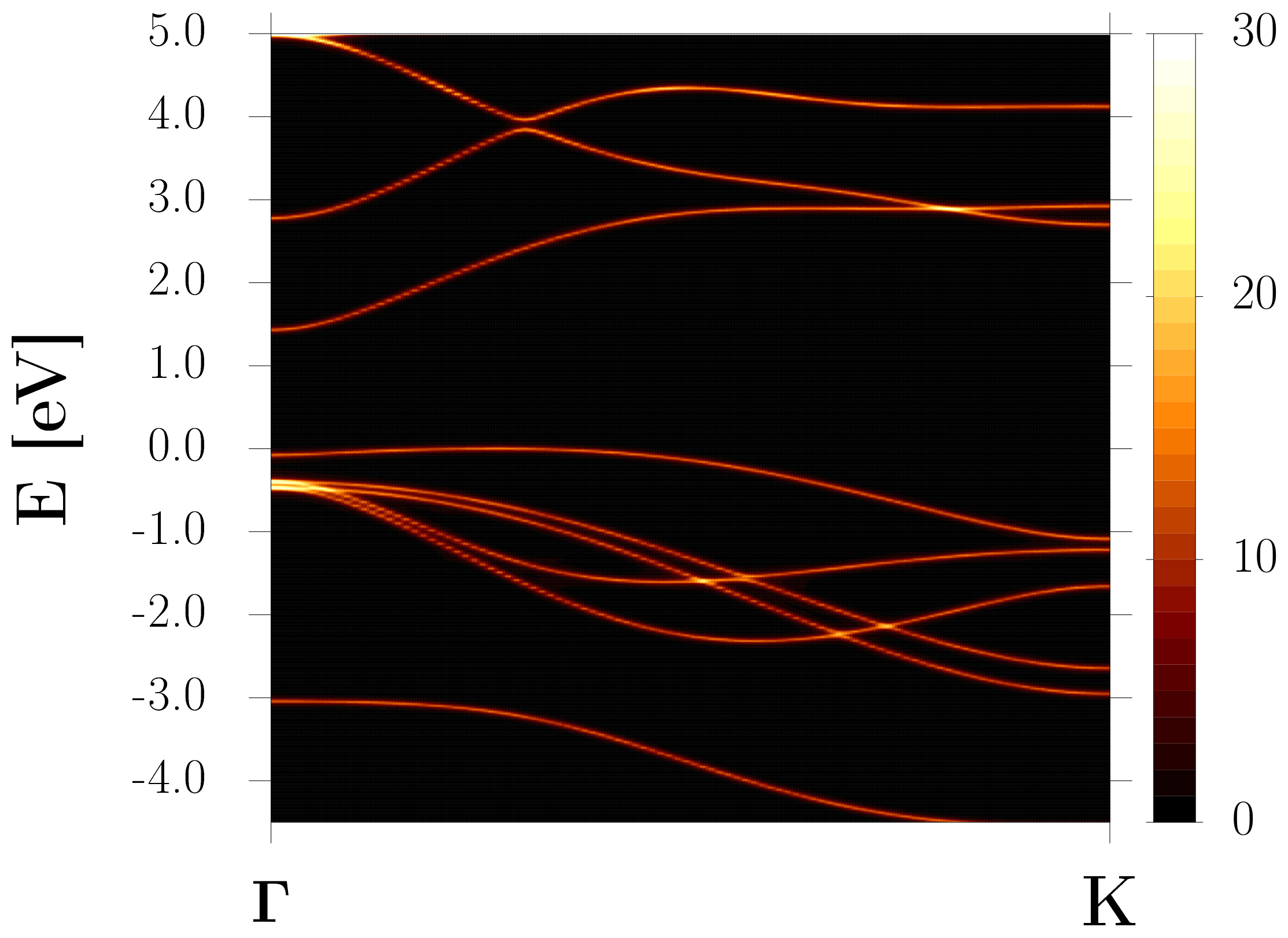}
  \caption{InSe monolayer (without SOC)}
\end{subfigure}
\begin{subfigure}{0.5\textwidth}
  \centering
  \includegraphics[width=0.73\textwidth]{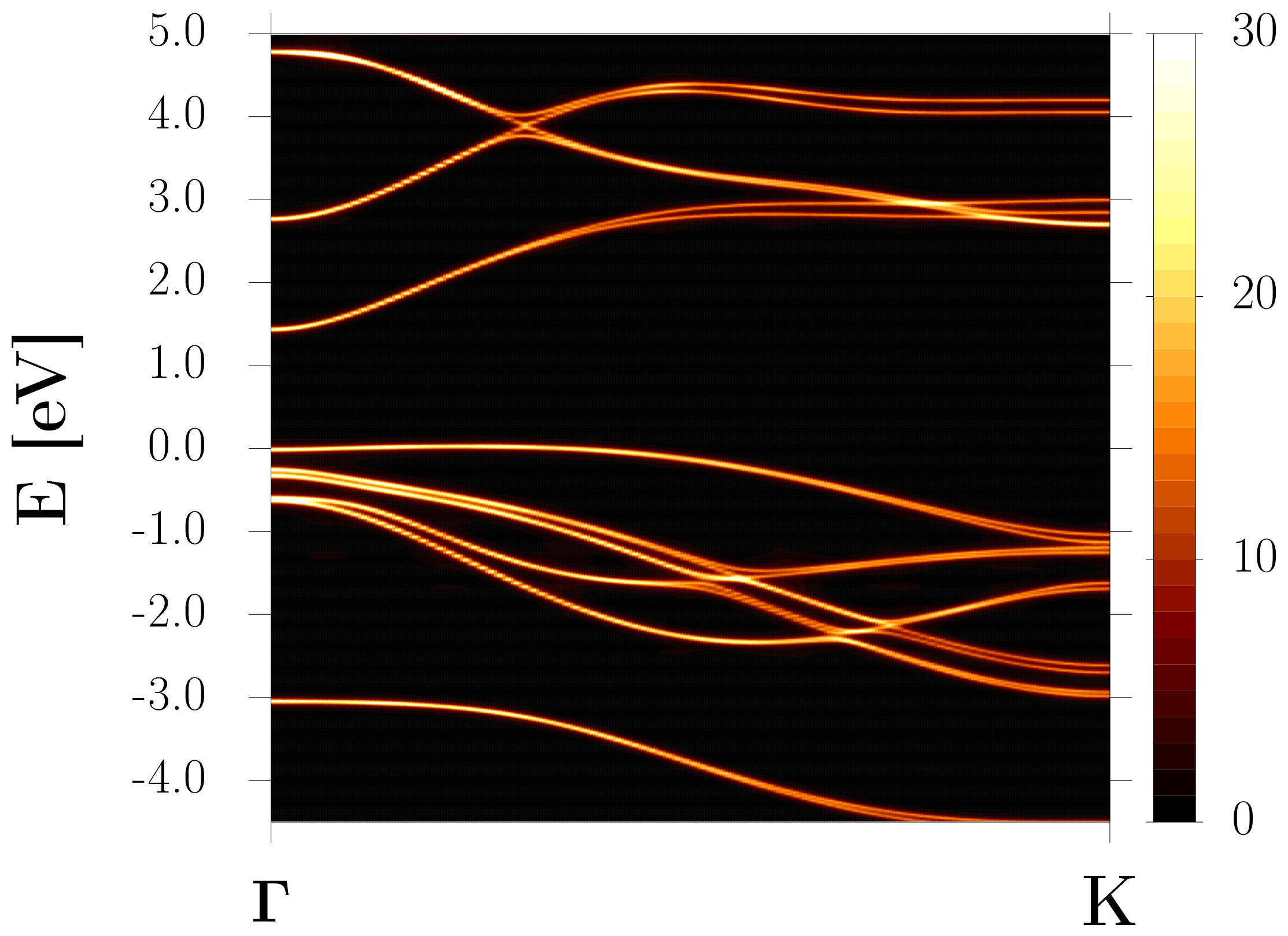}
  \caption{InSe monolayer (with SOC)}
\end{subfigure}
\begin{subfigure}{0.498\textwidth}
  \centering
  \includegraphics[width=0.73\textwidth]{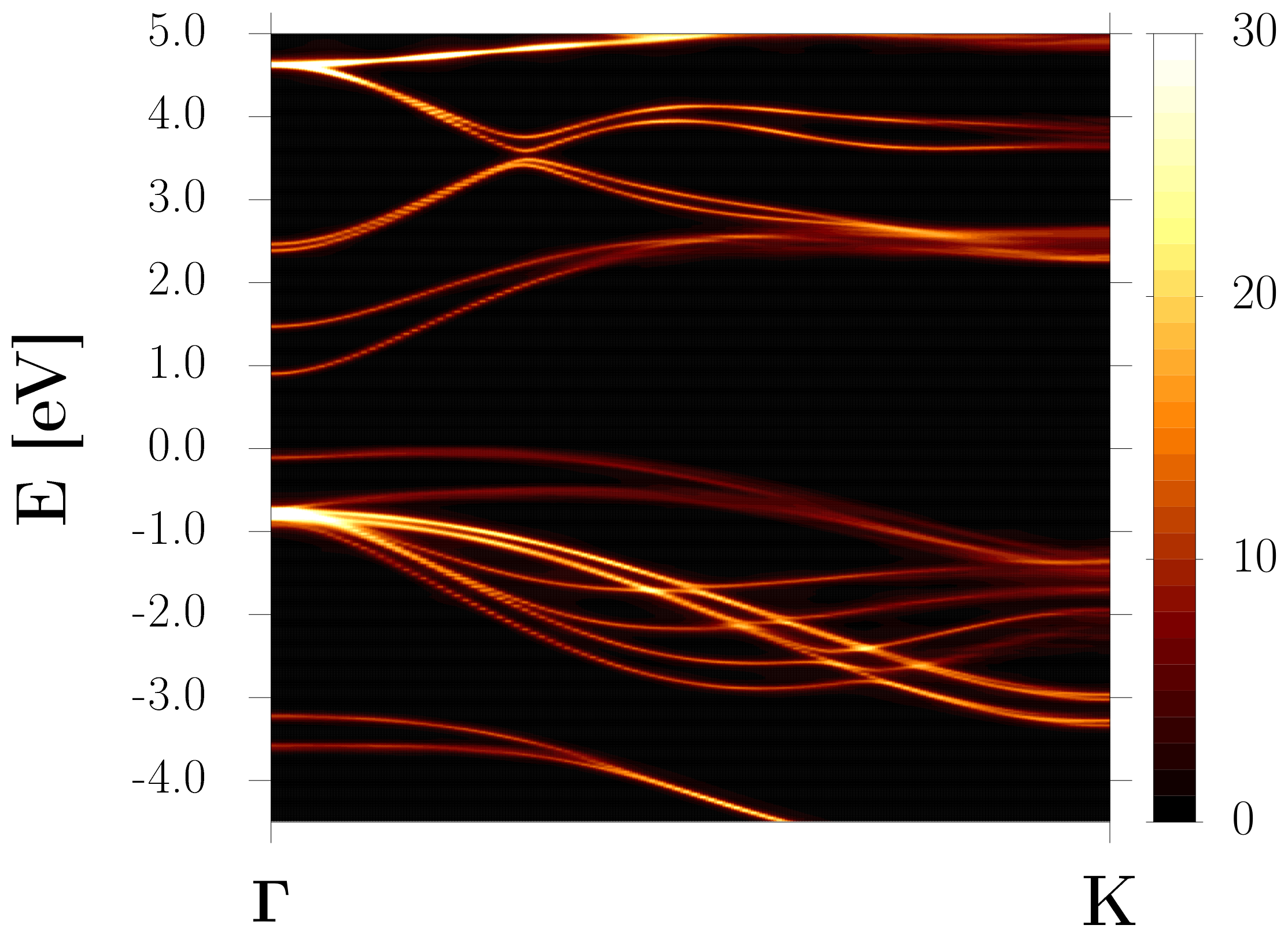}
  \caption{A-type twisted InSe bilayer (without SOC)}
\end{subfigure}
\begin{subfigure}{0.5\textwidth}
  \centering
  \includegraphics[width=0.73\textwidth]{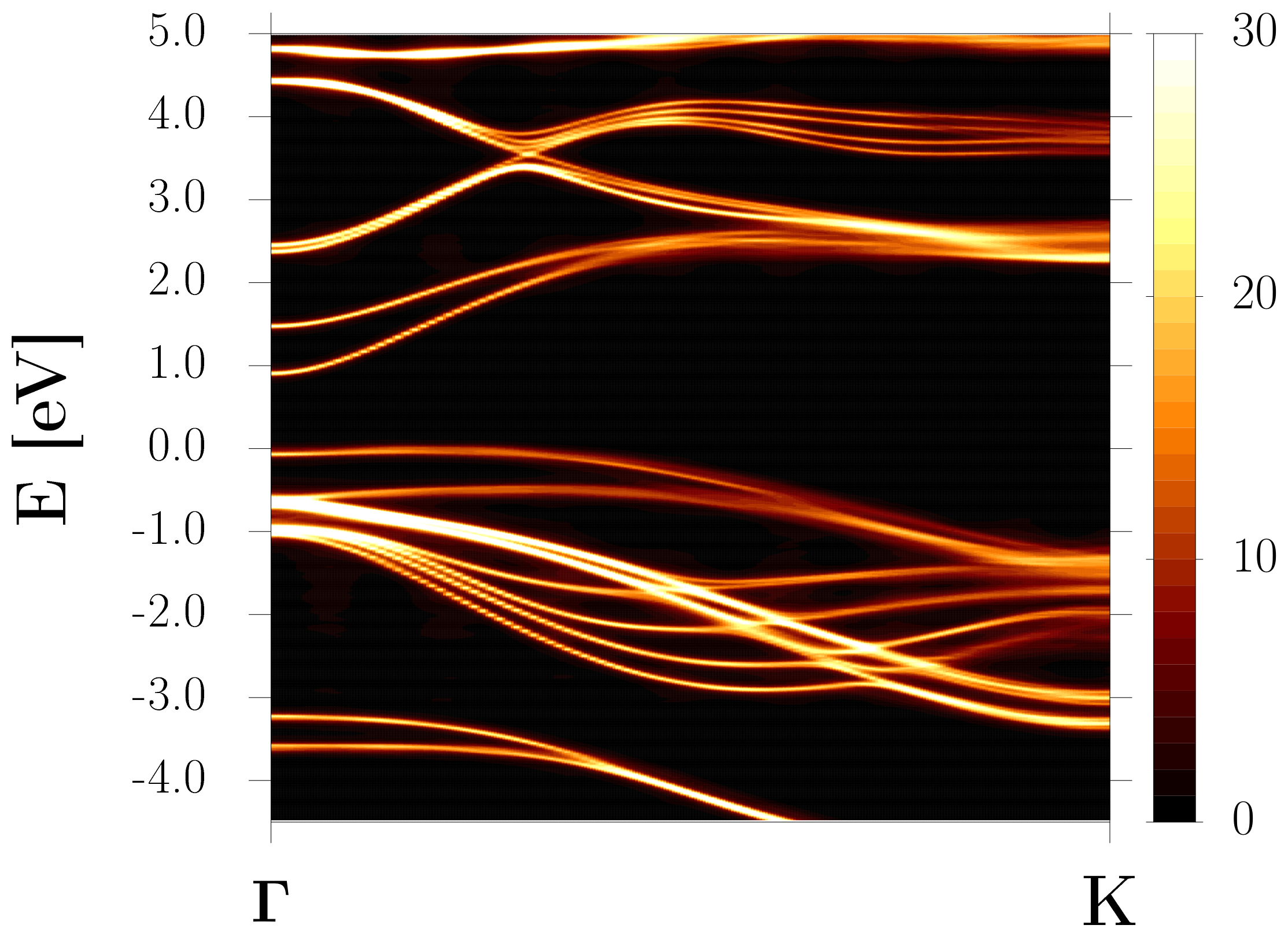}
  \caption{A-type twisted InSe bilayer (with SOC)}
\end{subfigure}
\begin{subfigure}{0.498\textwidth}
  \centering
  \includegraphics[width=0.73\textwidth]{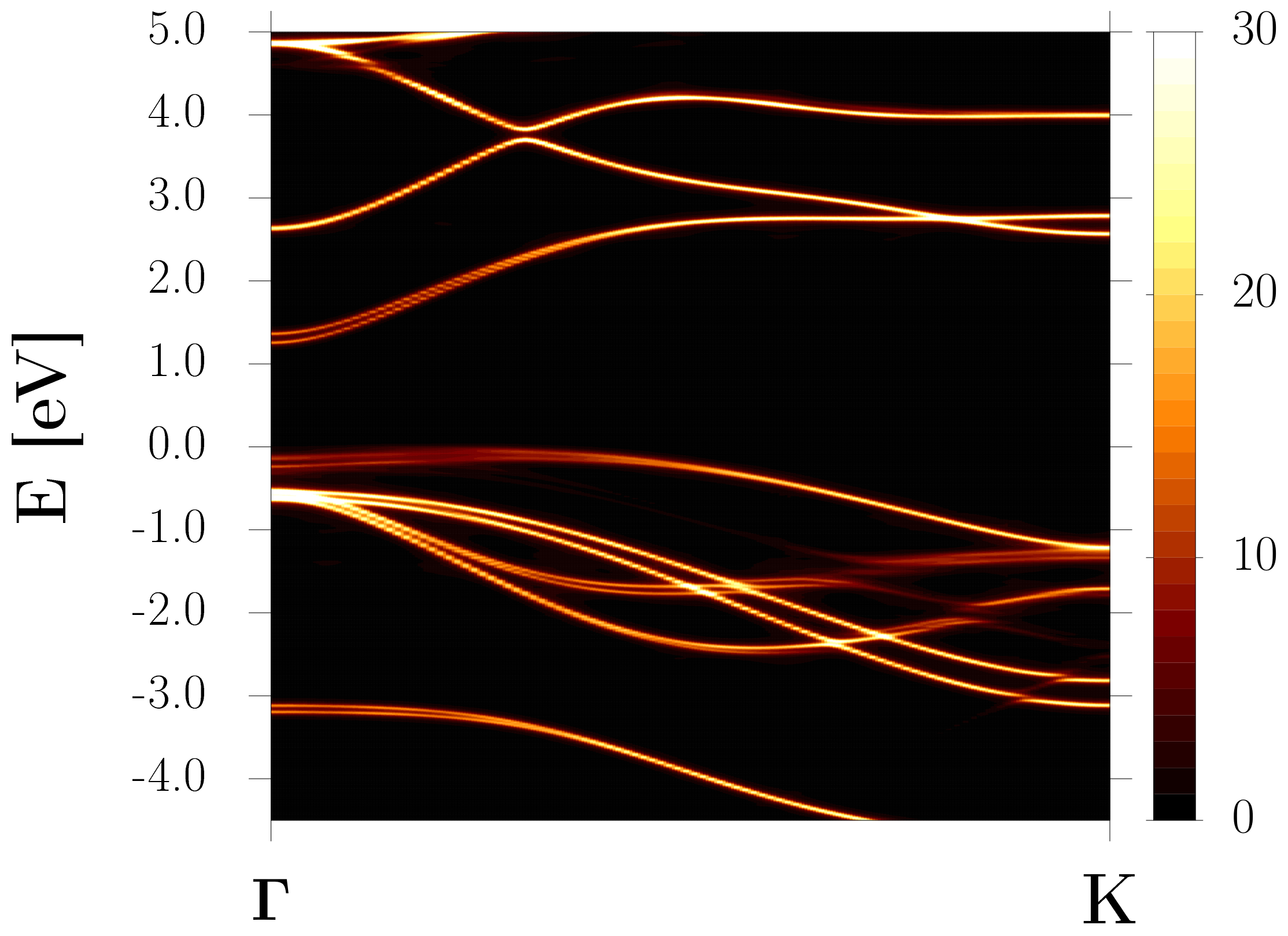}
  \caption{HBN-encapsulated A-type twisted InSe bilayer (without SOC)}
\end{subfigure}
\begin{subfigure}{0.5\textwidth}
  \centering
  \includegraphics[width=0.73\textwidth]{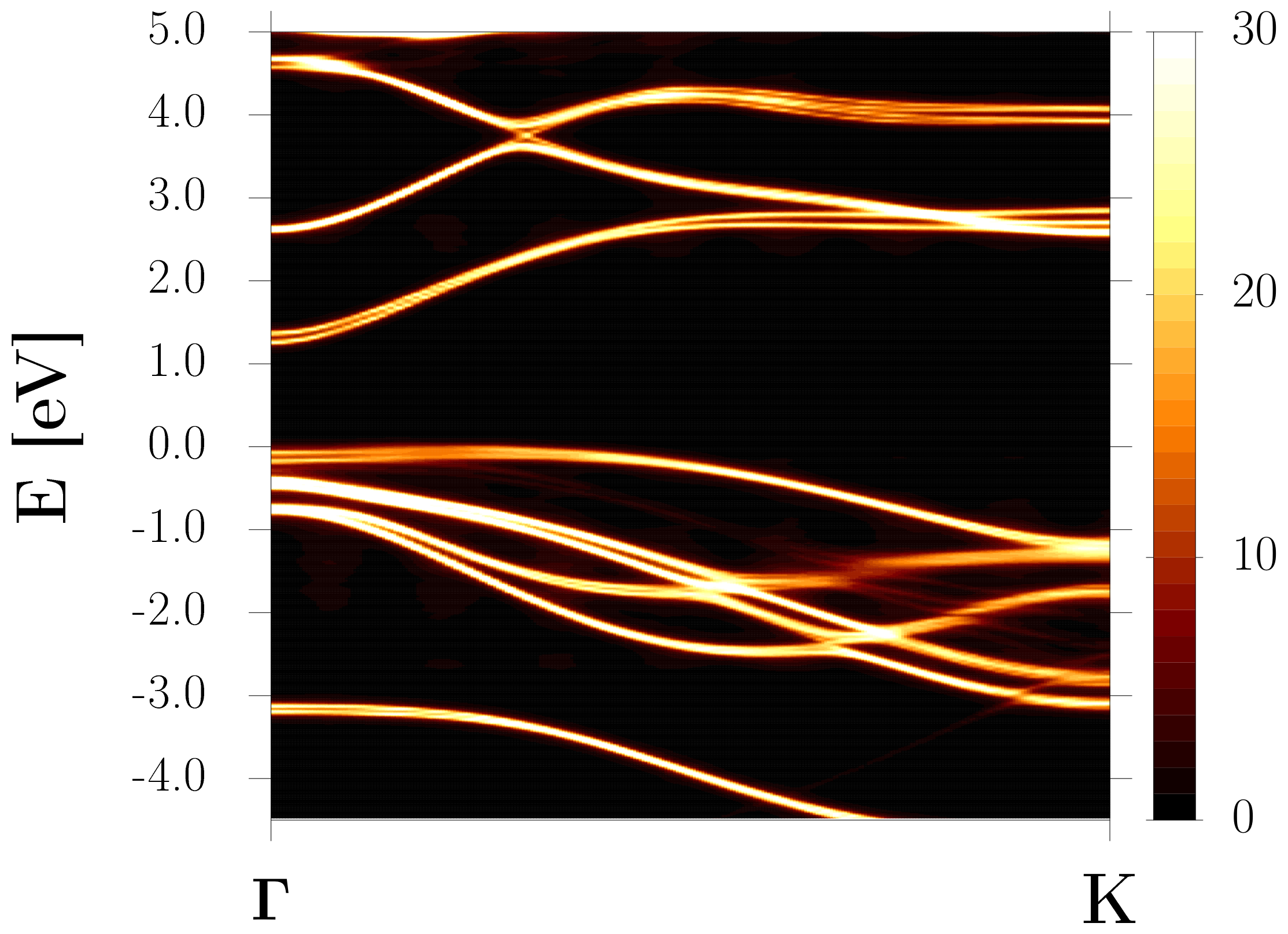}
  \caption{HBN-encapsulated A-type twisted InSe bilayer (with SOC)}
\end{subfigure}
 \caption{ Effective band structure of (a)(b) InSe monolayer (c)(d) A-type twisted InSe bilayer (e)(f) hBN-encapsulated A-type twisted InSe bilayer. The spin-orbit coupling was considered in (b)(d)(f).}
\label{fig:bandstructure_AA-stacking_soc_no-soc}
\end{figure}

The effective band structures of InSe monolayer, the A-type twisted InSe bilayer and the hBN-encapsulated A-type twisted InSe bilayer with a twist angle of 4.4$^{\mathrm{o}}$ are shown in fig. \ref{fig:bandstructure_AA-stacking_soc_no-soc}. The inclusion of spin-orbit coupling in the calculation leads to the band splittings. The splitting of the VBM is only slightly pronounced near $\textbf{K}$ for these three different systems.

\newpage

\begin{figure} [H]
\captionsetup[subfigure]{justification=centering}
\begin{subfigure}{0.5\textwidth}
  \centering
  \includegraphics[width=0.73\textwidth]{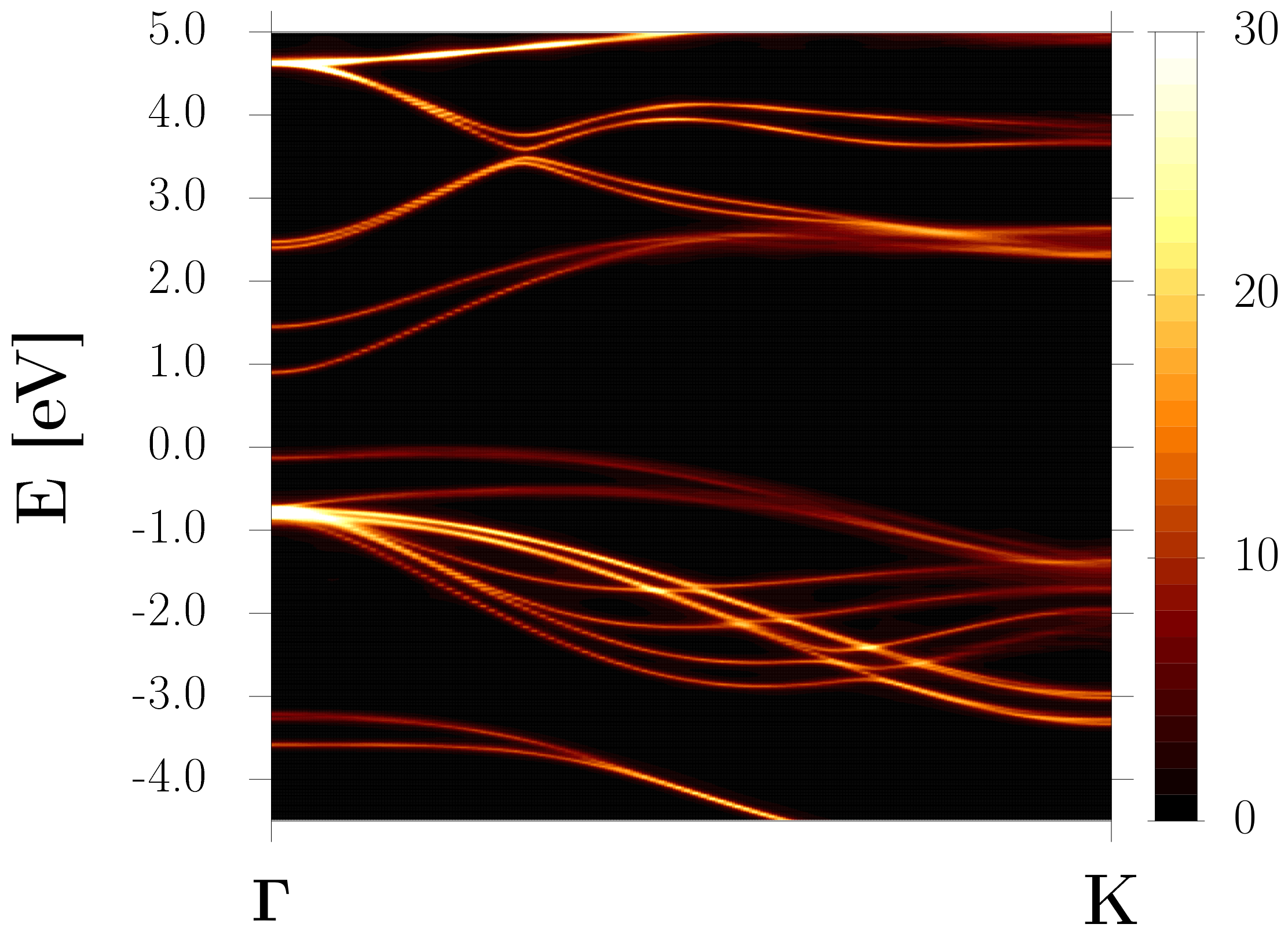}
  \caption{$\theta$= 4.4$^{\mathrm{o}}$ (without SOC)}
\end{subfigure}
\begin{subfigure}{0.5\textwidth}
  \centering
  \includegraphics[width=0.73\textwidth]{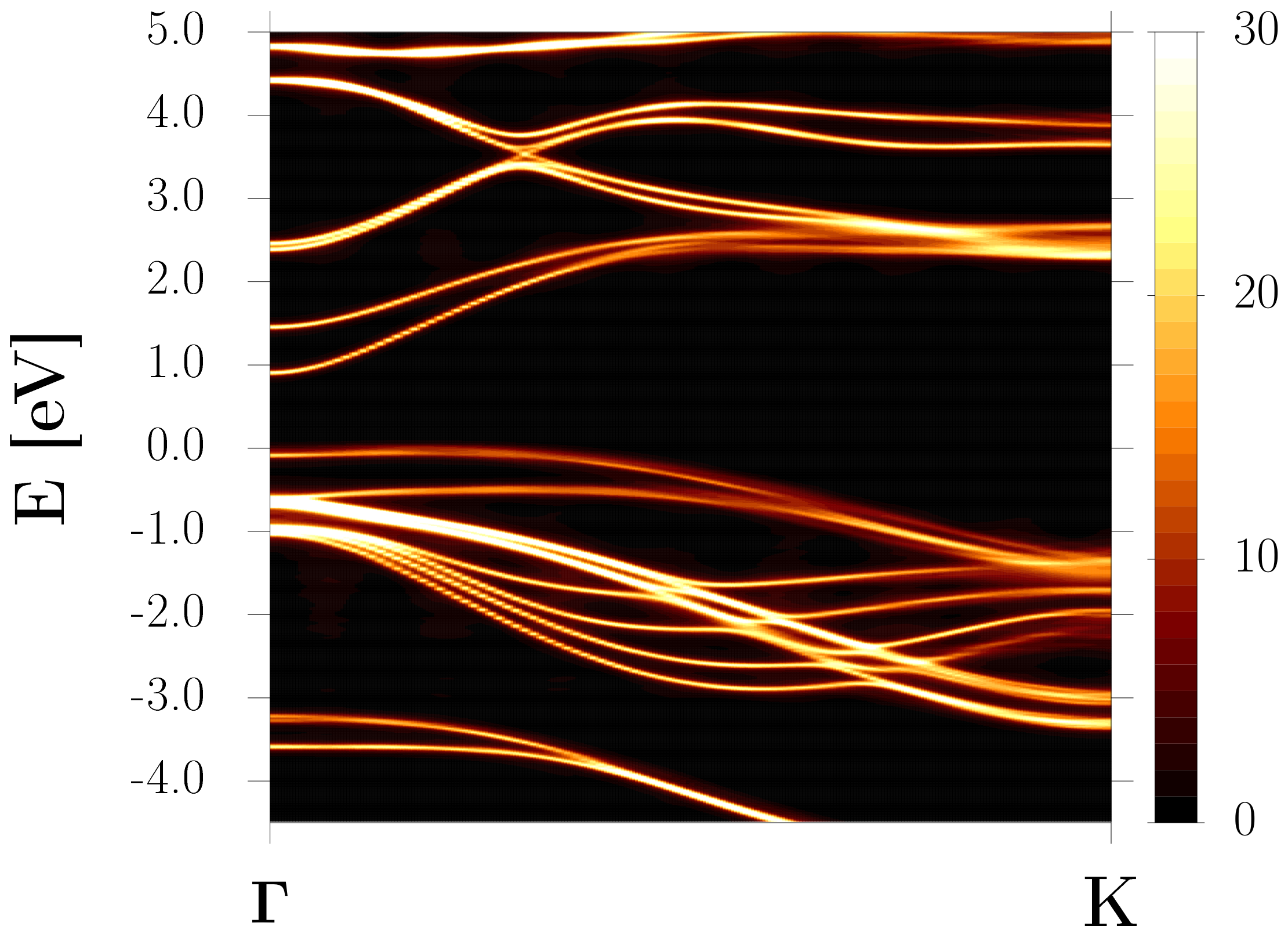}
  \caption{$\theta$= 4.4$^{\mathrm{o}}$ (with SOC)}
\end{subfigure}
\begin{subfigure}{0.5\textwidth}
  \centering
  \includegraphics[width=0.73\textwidth]{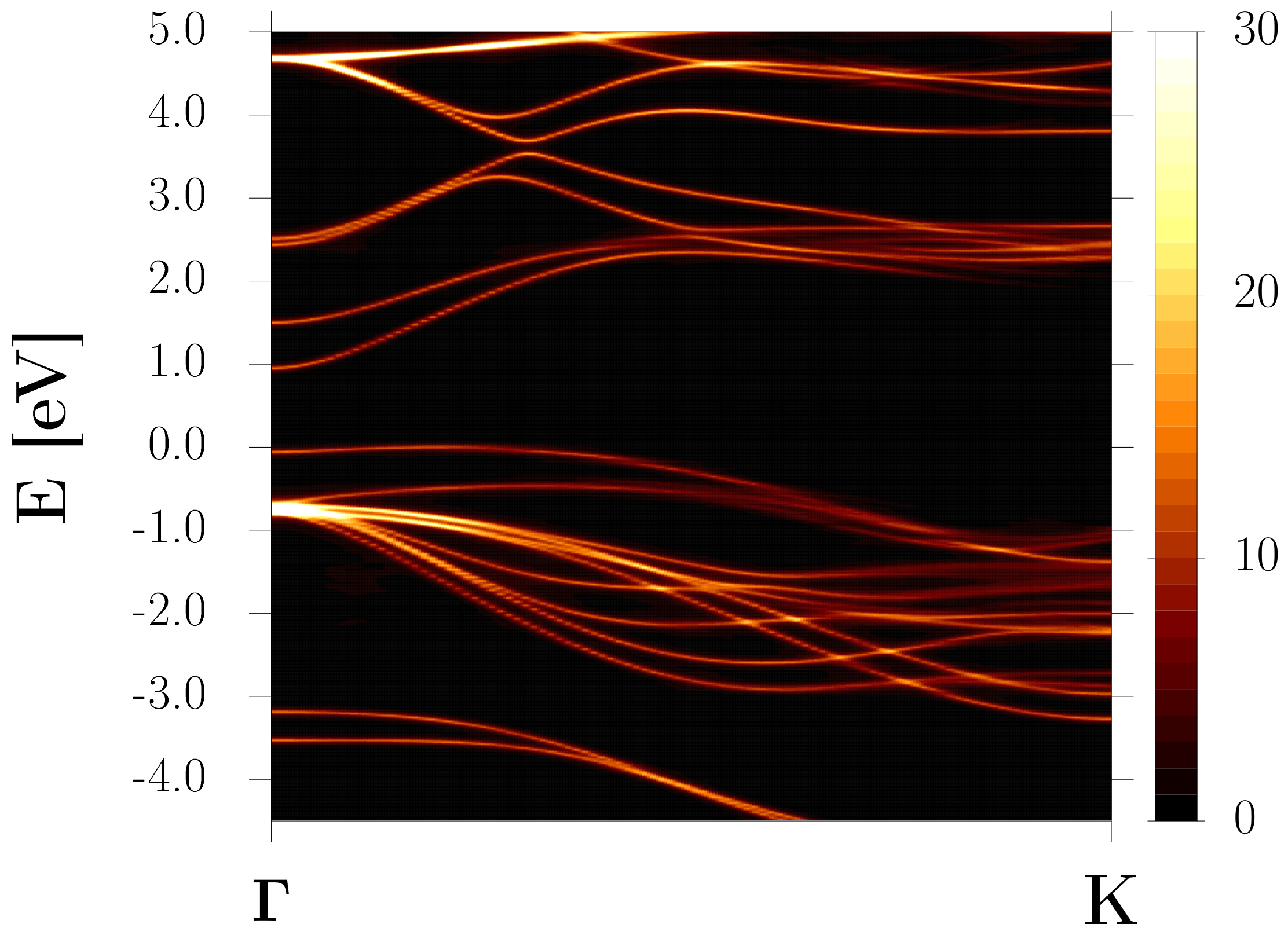}
  \caption{$\theta$= 17.9$^{\mathrm{o}}$ (without SOC)}
\end{subfigure}
\begin{subfigure}{0.5\textwidth}
  \centering
  \includegraphics[width=0.73\textwidth]{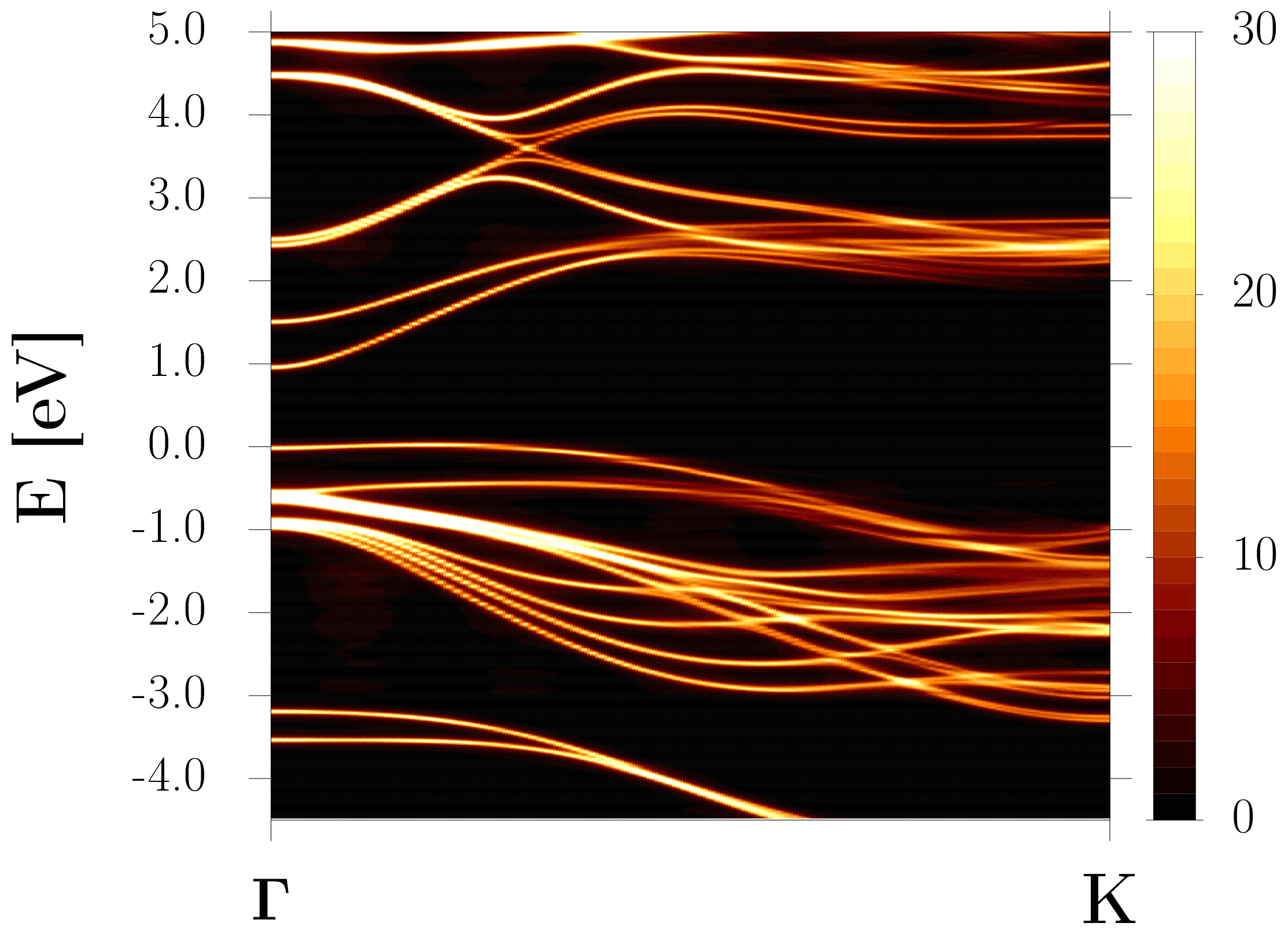}
  \caption{$\theta$= 17.9$^{\mathrm{o}}$ (with SOC)}
\end{subfigure}
\begin{subfigure}{0.5\textwidth}
  \centering
  \includegraphics[width=0.73\textwidth]{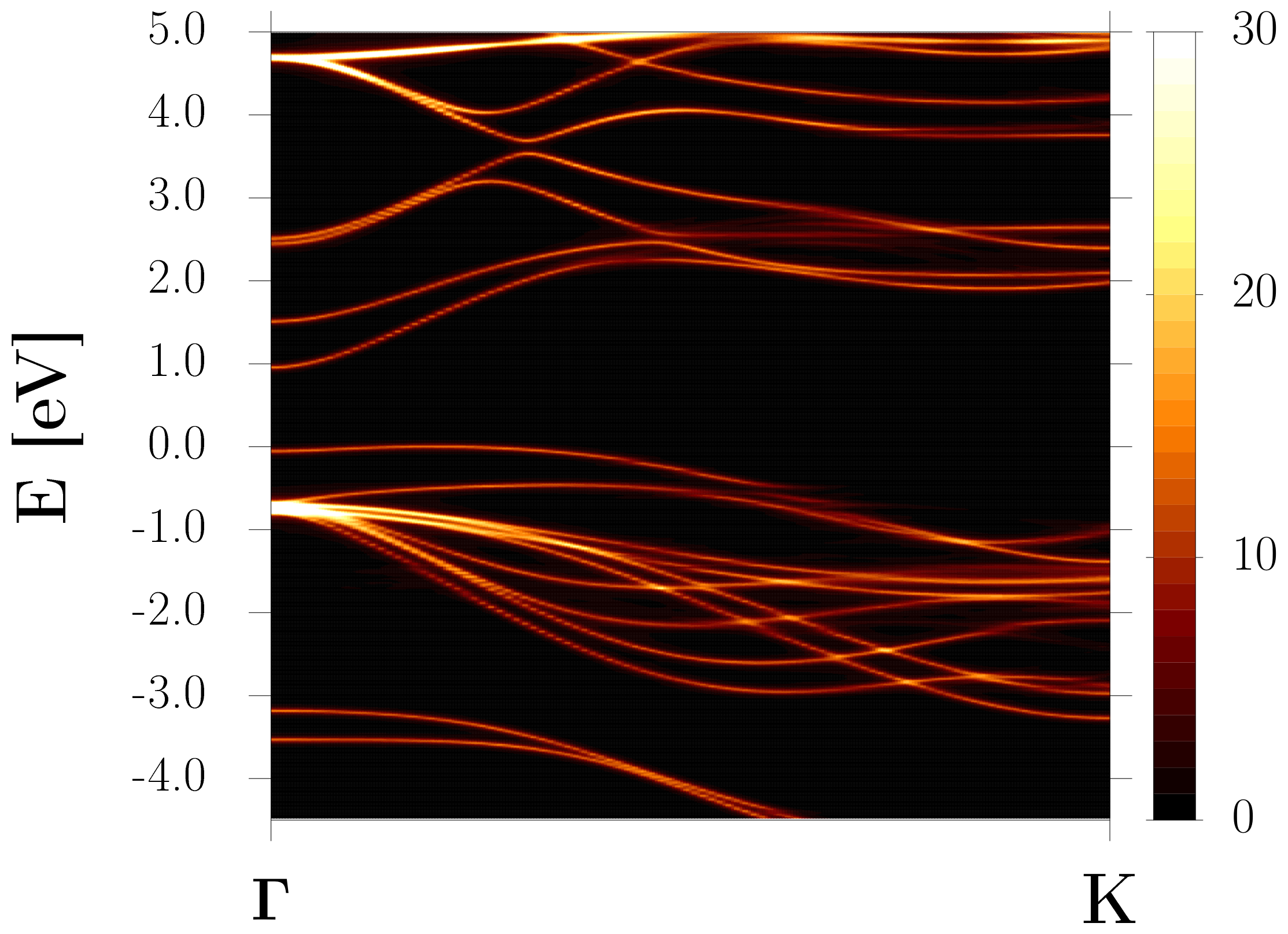}
  \caption{$\theta$= 27.8$^{\mathrm{o}}$ (without SOC)}
\end{subfigure}
\begin{subfigure}{0.5\textwidth}
  \centering
  \includegraphics[width=0.73\textwidth]{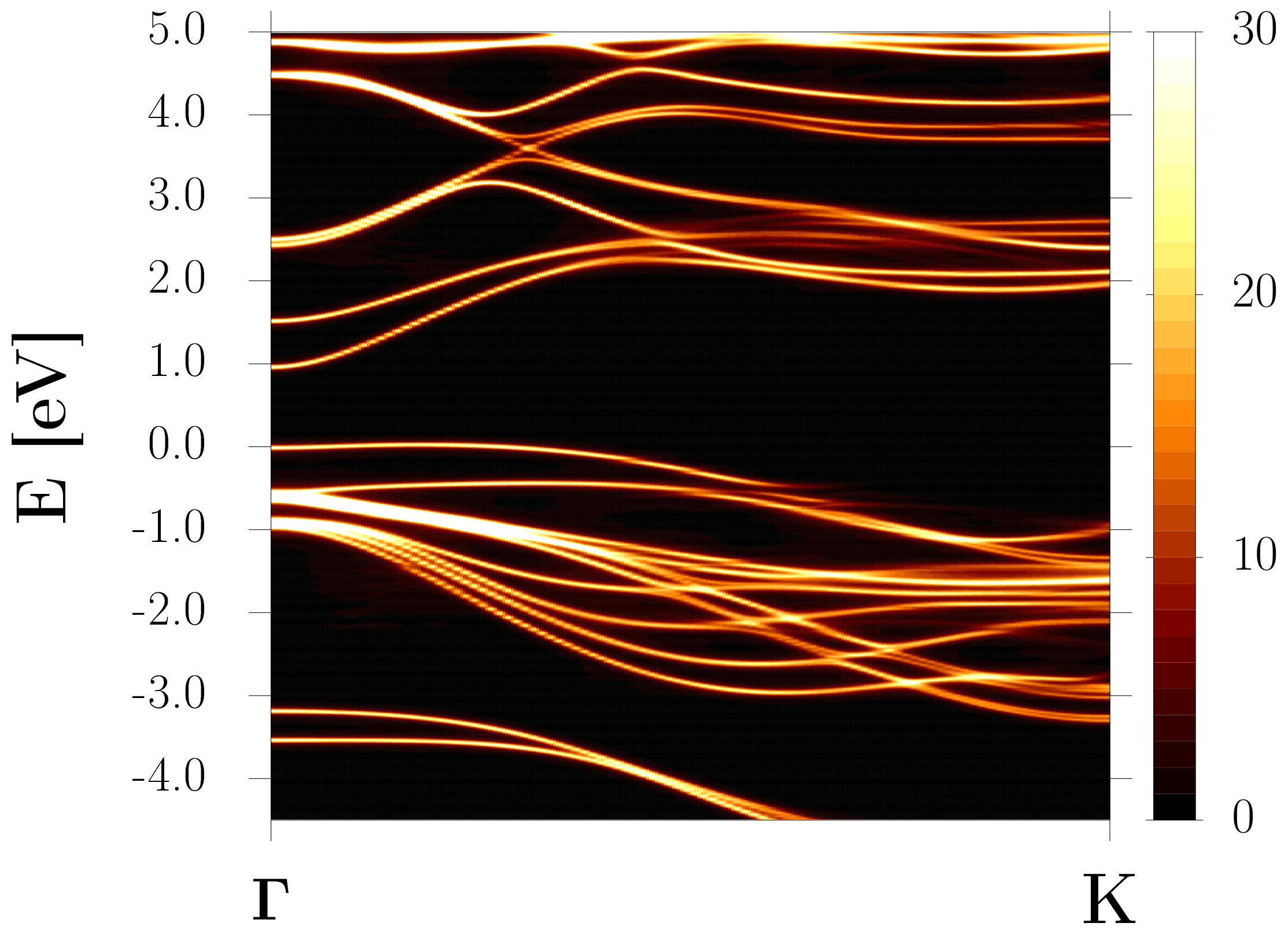}
  \caption{$\theta$= 27.8$^{\mathrm{o}}$ (with SOC)}
\end{subfigure}
 \caption{ Effective band structure of the B-type twisted InSe bilayer with a twist angle of (a)(b) $\theta$= 4.4$^{\mathrm{o}}$ (c)(d) $\theta$= 17.9$^{\mathrm{o}}$ (e)(f) $\theta$= 27.8$^{\mathrm{o}}$. The spin-orbit coupling was considered in (b)(d)(f). }
\label{fig:bandstructure_AB-stacking_soc_no-soc}
\end{figure}

The effective band structures of the B-type twisted InSe bilayers with twist angles of 4.4$^{\mathrm{o}}$, 17.9$^{\mathrm{o}}$ and 27.8$^{\mathrm{o}}$ are shown in fig. \ref{fig:bandstructure_AB-stacking_soc_no-soc}. The inclusion of spin-orbit coupling in the calculation leads to the band splittings. The splitting of the VBM is only slightly pronounced near $\textbf{K}$ for these three different systems.

\newpage

\begin{figure} [H]
    \centering
    \includegraphics[width=0.5\textwidth]{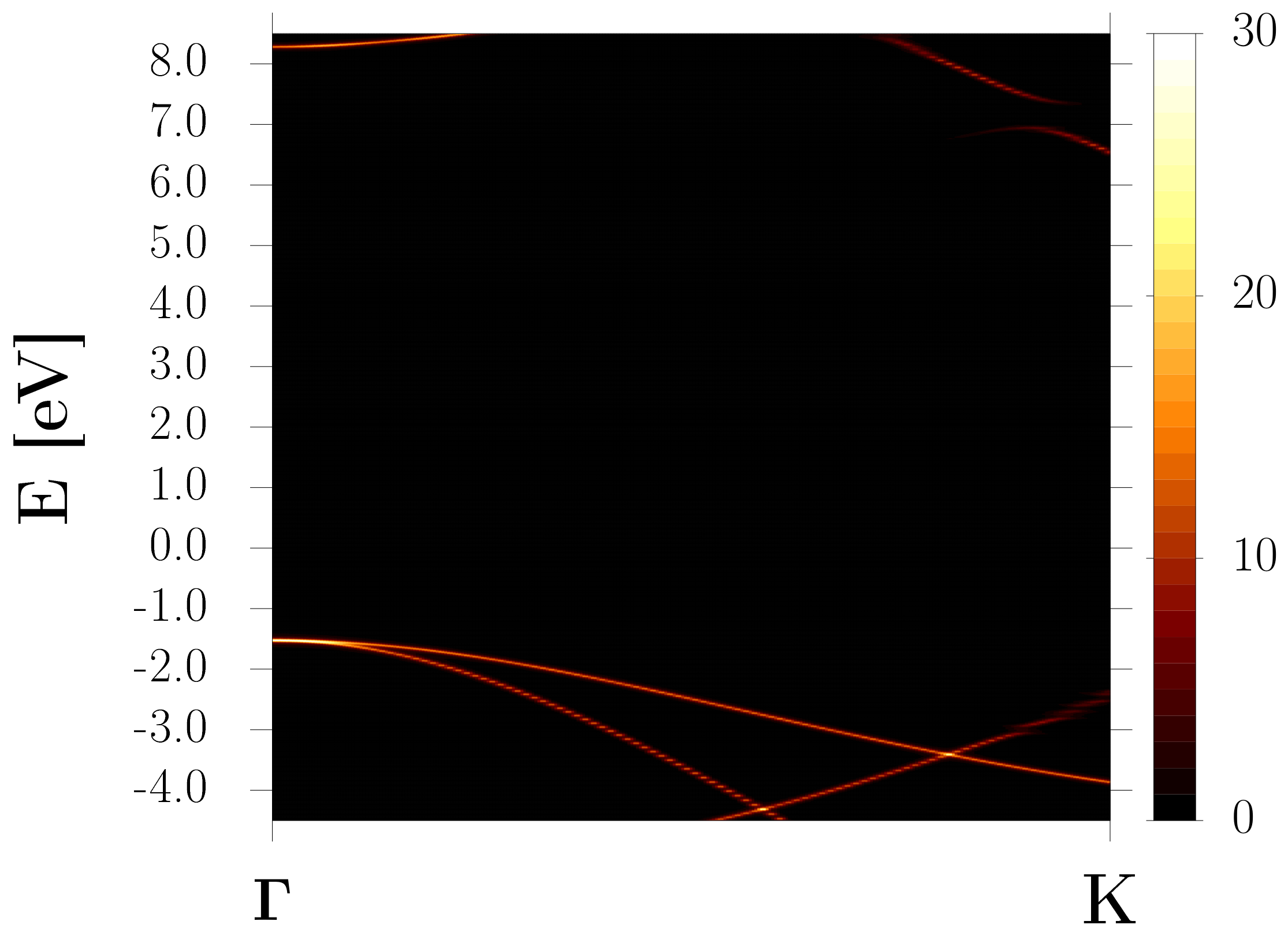}
 \caption{ Effective band structure of the hBN-encapsulated A-type twisted InSe bilayer ($\theta$= 4.4$^{\mathrm{o}}$) with a projection on the hBN layer.} 
\label{fig:InSe-hBN-InSe_projected_on_hBN}
\end{figure}

\begin{figure} [H]
    \centering
    \includegraphics[width=0.65\textwidth]{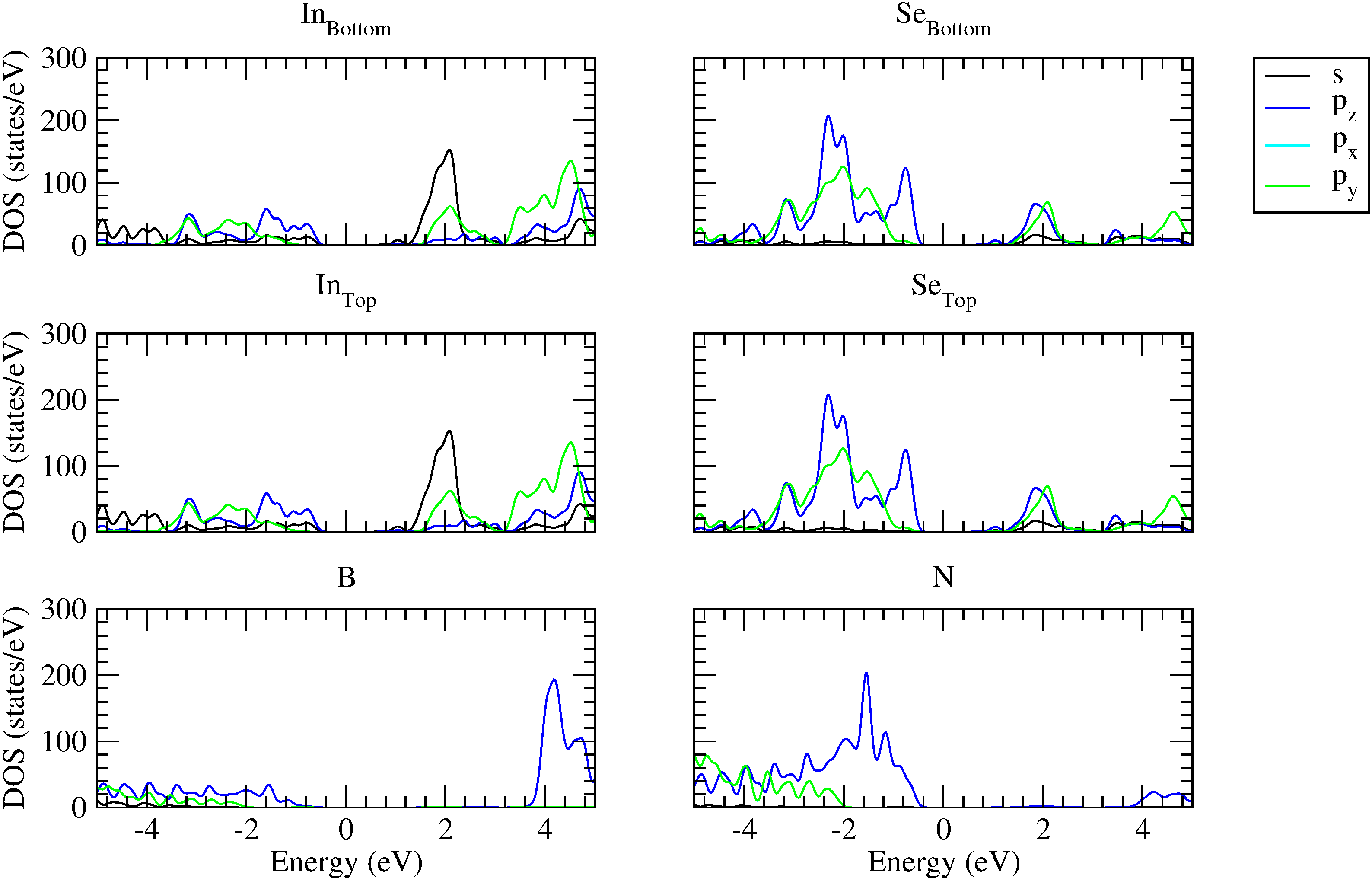}
 \caption{ Projected density of states of the hBN-encapsulated A-type twisted InSe bilayer ($\theta$= 4.4$^{\mathrm{o}}$). The subsripts ‘t’ and ‘b’ refer to the top or bottom InSe layers.} 
\label{fig:InSe-hBN-InSe_pdos}
\end{figure}

The hBN layer does not affect the VBM of twisted InSe layers in the hBN-encapsulated twisted InSe bilayer because the energy difference between the VBM of the hBN layer and the VBM of twisted InSe bilayer is at least 1.5 eV (see fig. \ref{fig:InSe-hBN-InSe_projected_on_hBN} and fig. \ref{fig:InSe-hBN-InSe_pdos}). 

\begin{figure} [H]
\captionsetup[subfigure]{justification=centering}
\begin{subfigure}{0.5\textwidth}
  \centering
  \includegraphics[width=0.8\textwidth]{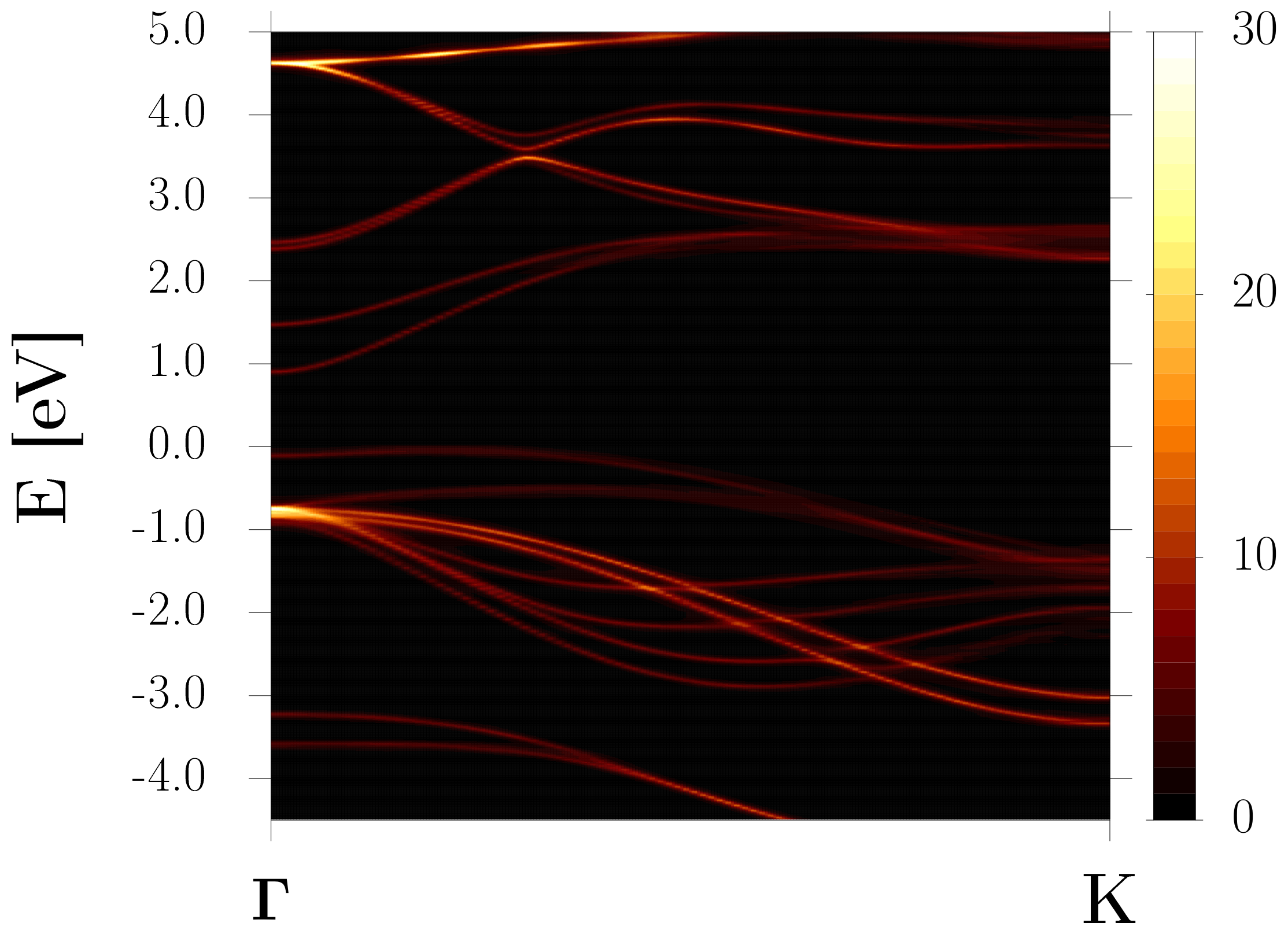}
  \caption{Twisted InSe bilayer (projected on \\ the bottom layer) ($\theta$= 4.4$^{\mathrm{o}}$)}
\end{subfigure}
\begin{subfigure}{0.5\textwidth}
  \centering
  \includegraphics[width=0.8\textwidth]{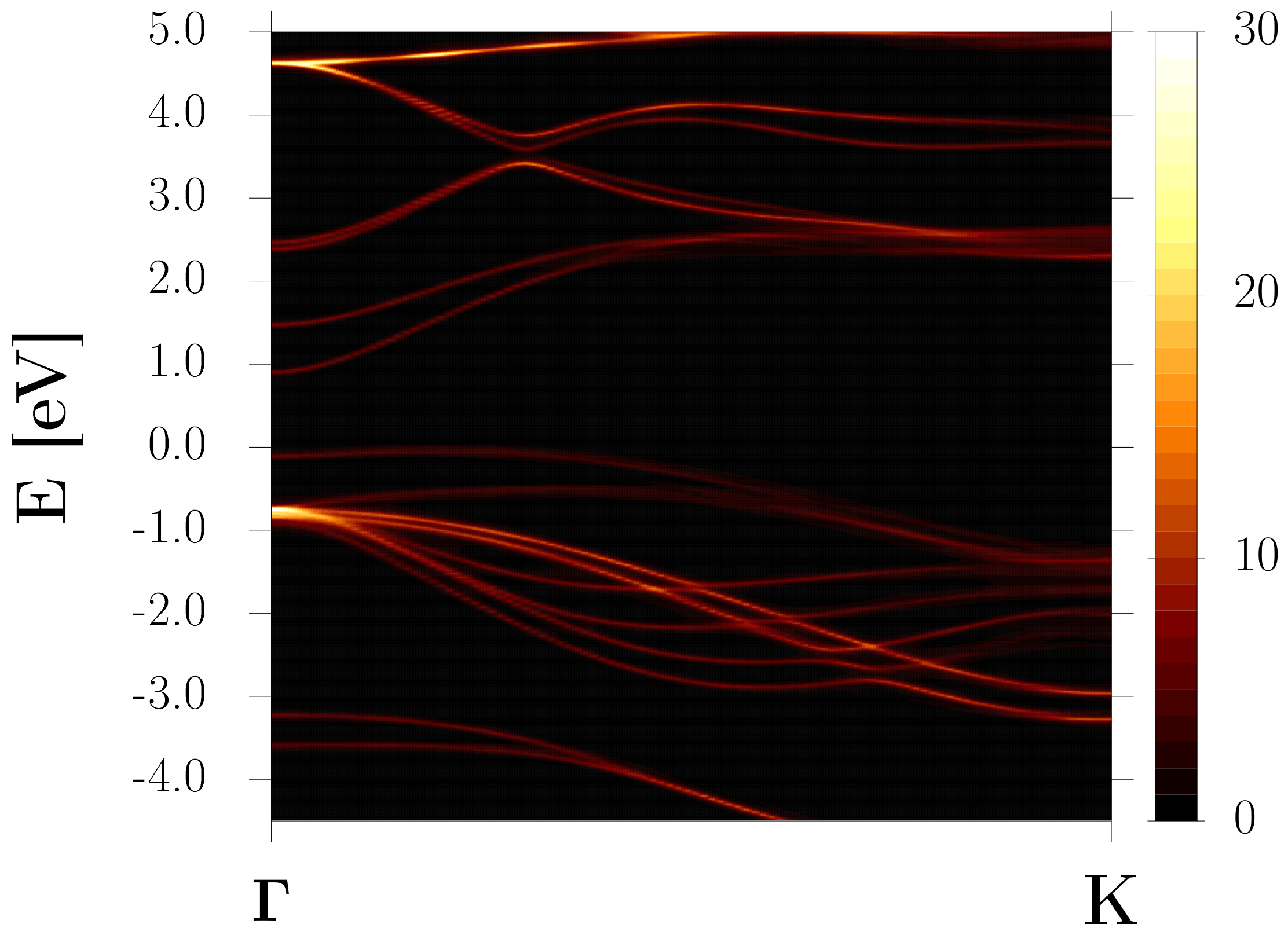}
  \caption{Twisted InSe bilayer (projected on \\ the top layer) ($\theta$= 4.4$^{\mathrm{o}}$)}
\end{subfigure}
\begin{subfigure}{0.5\textwidth}
  \centering
  \includegraphics[width=0.8\textwidth]{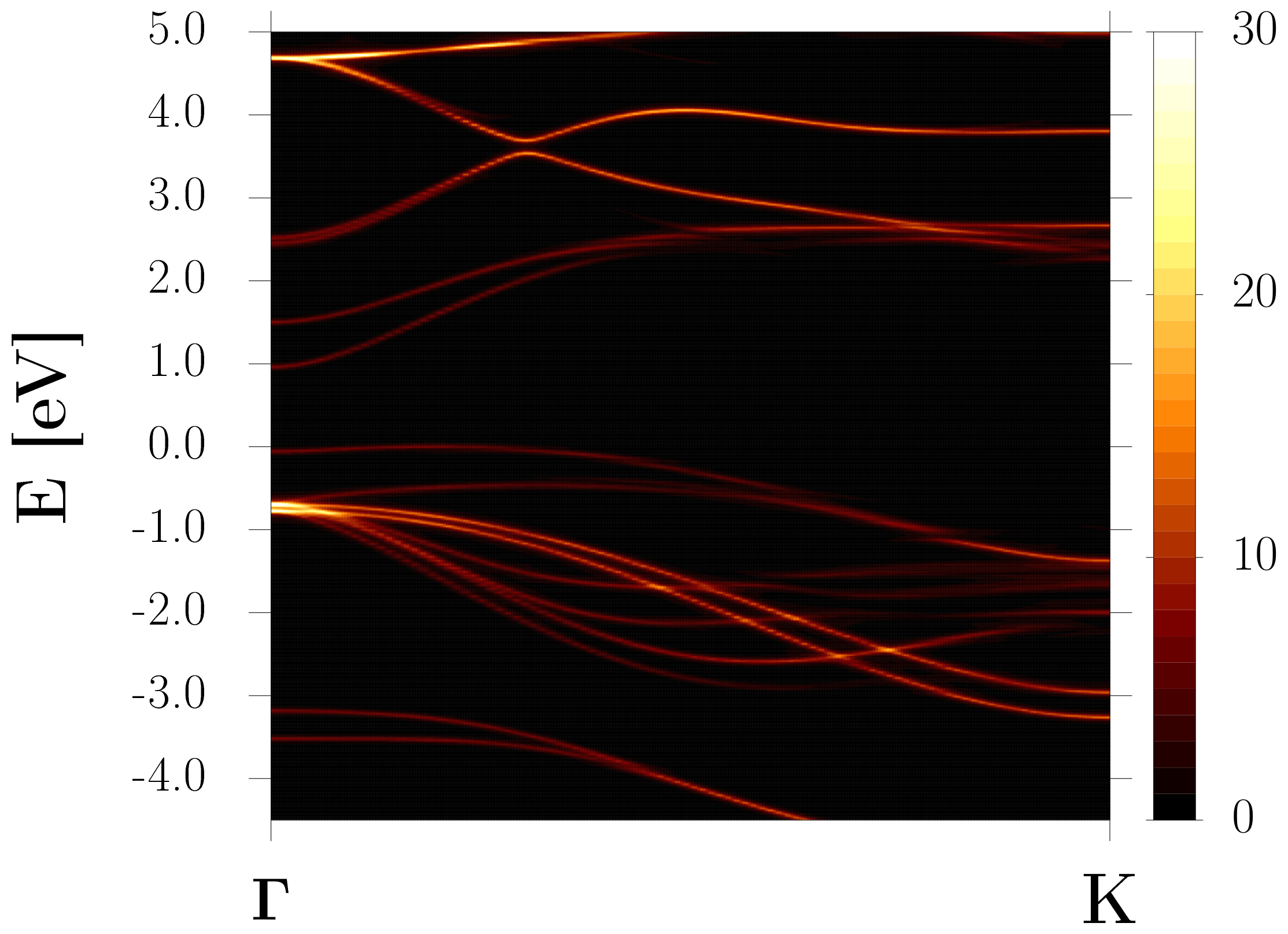}
  \caption{Twisted InSe bilayer (projected on \\ the bottom layer) ($\theta$= 17.9$^{\mathrm{o}}$)}
\end{subfigure}
\begin{subfigure}{0.5\textwidth}
  \centering
  \includegraphics[width=0.8\textwidth]{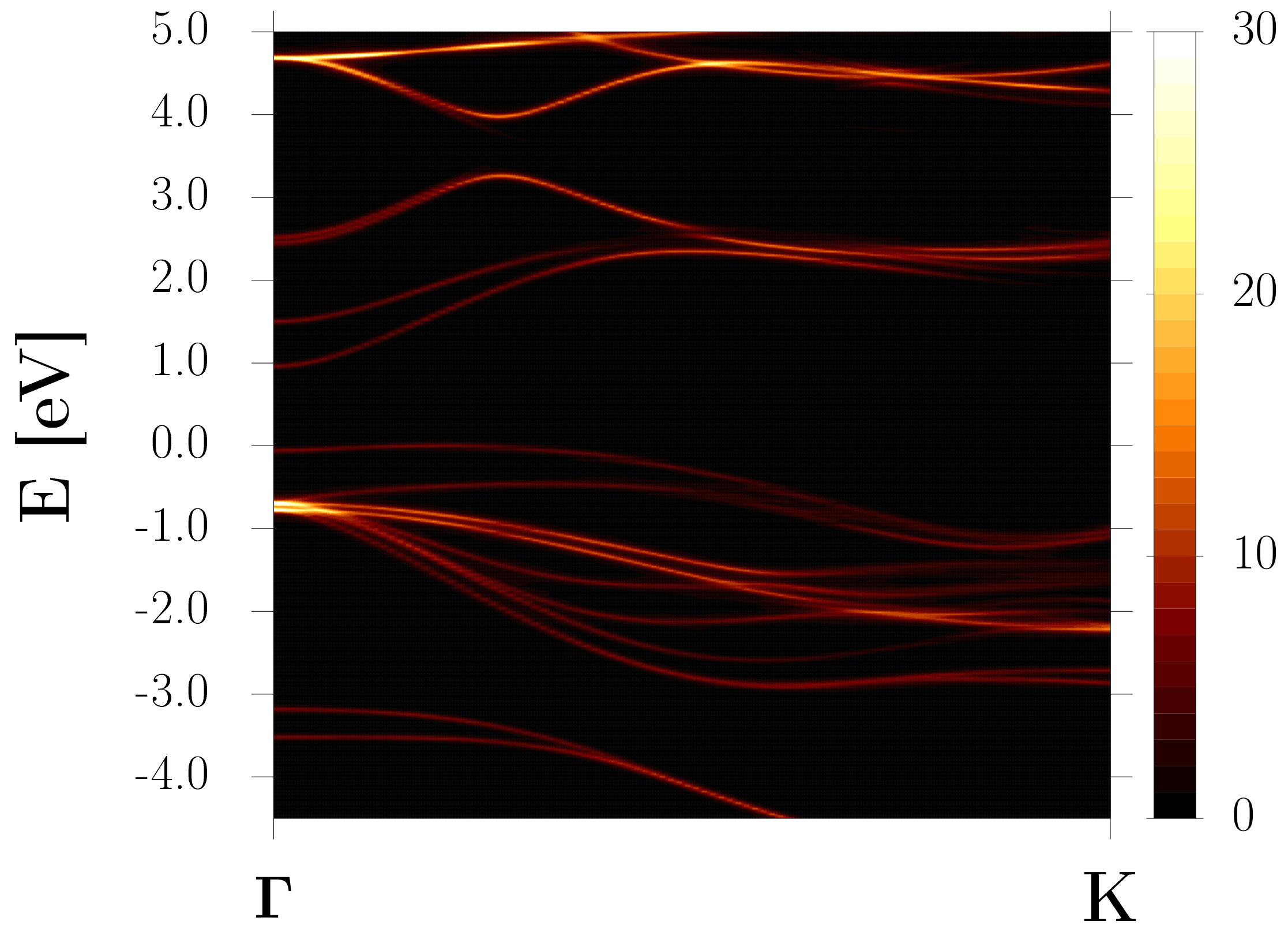}
  \caption{Twisted InSe bilayer (projected on \\ the top layer) ($\theta$= 17.9$^{\mathrm{o}}$)}
\end{subfigure}
\begin{subfigure}{0.5\textwidth}
  \centering
  \includegraphics[width=0.8\textwidth]{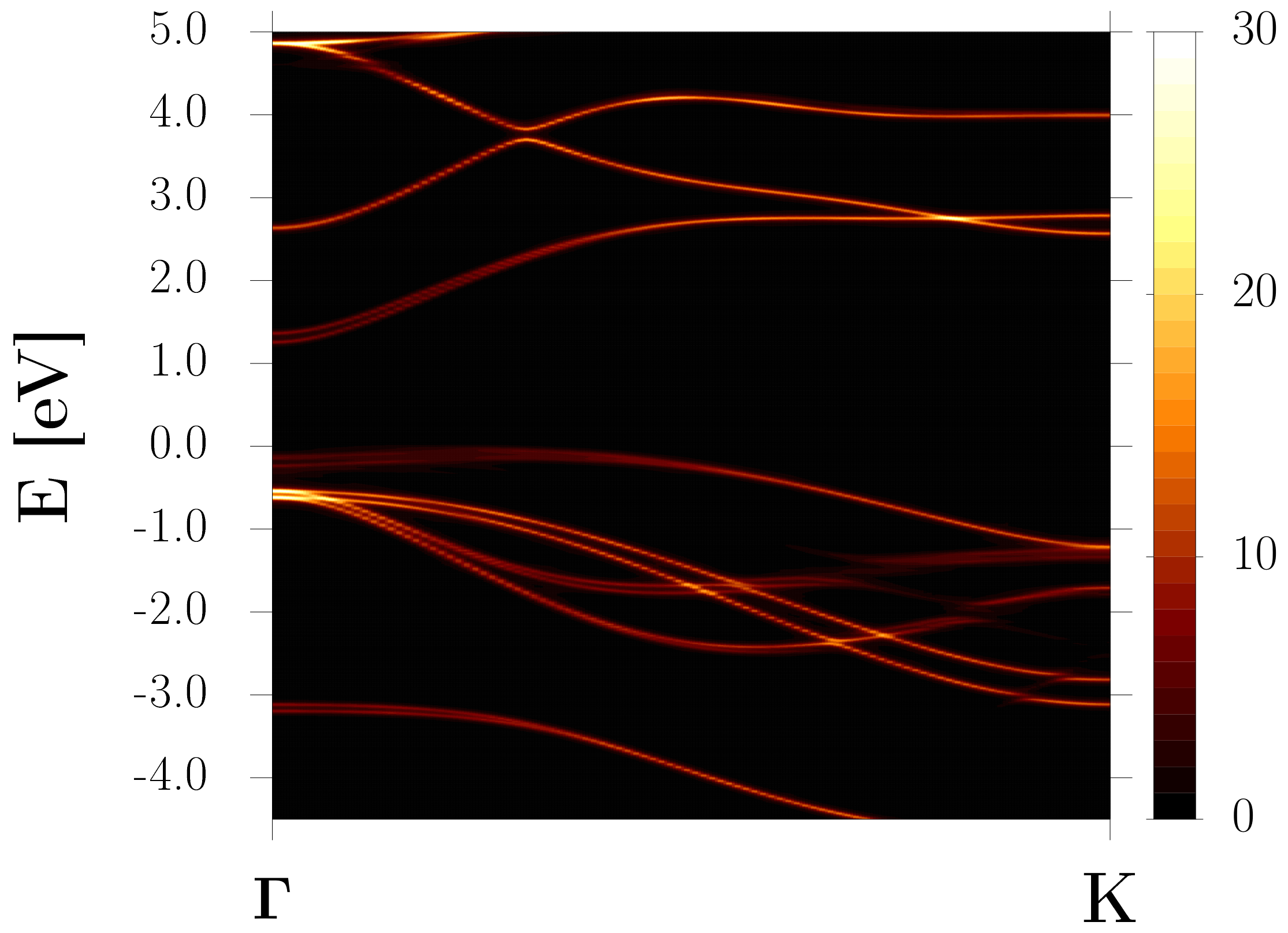}
  \caption{HBN-encapsulated twisted InSe bilayer (projected on the bottom layer) ($\theta$= 4.4$^{\mathrm{o}}$)}
\end{subfigure}
\begin{subfigure}{0.5\textwidth}
  \centering
  \includegraphics[width=0.8\textwidth]{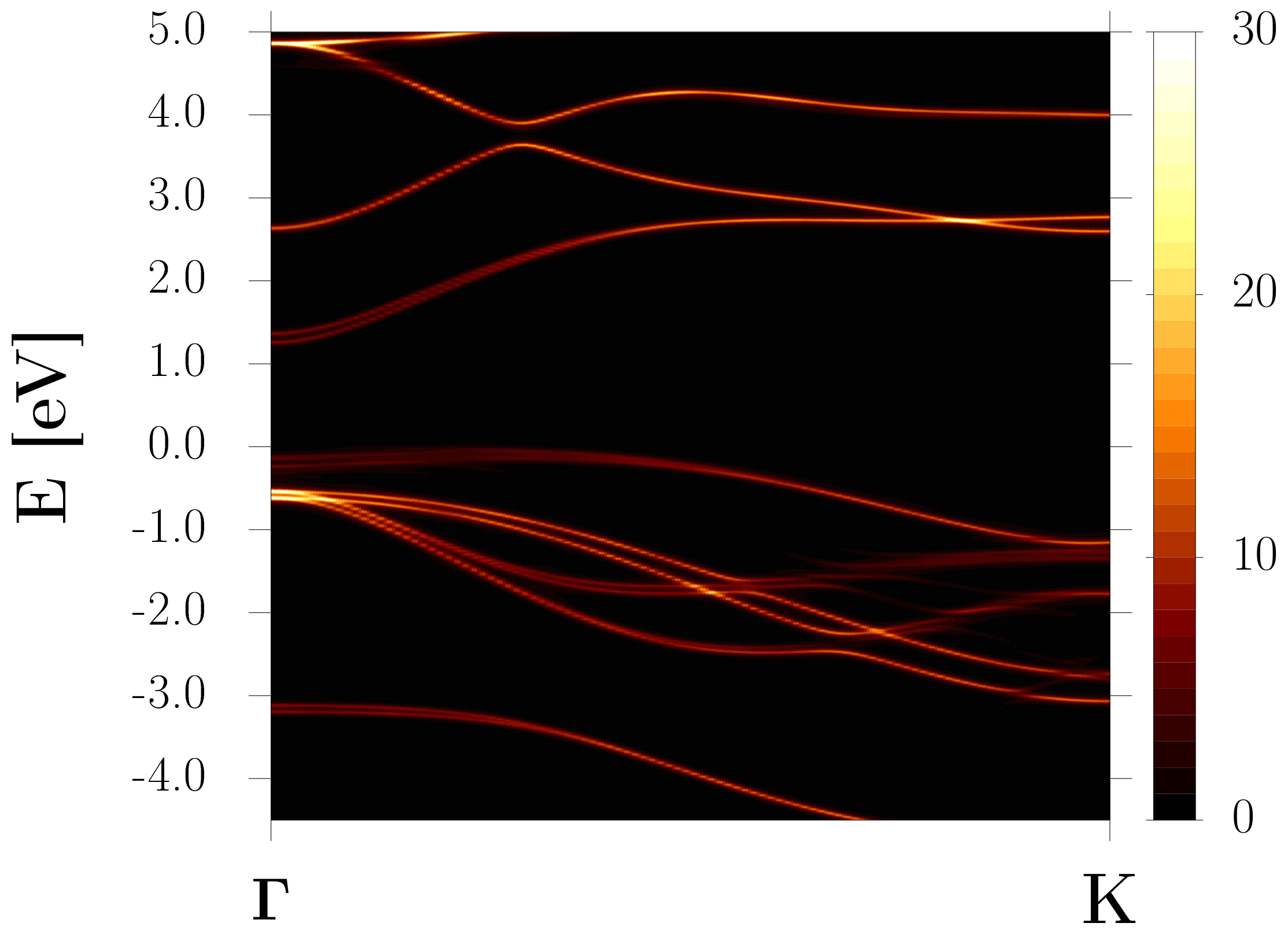}
  \caption{HBN-encapsulated twisted InSe bilayer (projected on the top layer) ($\theta$= 4.4$^{\mathrm{o}}$)}
\end{subfigure}
 \caption{ Effective band structure projected on the (a)(c)(e) bottom  and (b)(d)(f) top InSe layer of the A-type twisted InSe bilayer and hBN encapsulated A-type twisted InSe bilayer. Twisted InSe bilayers with a twist angle of (a)(b) 4.4$^{\mathrm{o}}$ and (c)(d) 17.9$^{\mathrm{o}}$. HBN-encapsulated A-type twisted InSe bilayer with a twist angle of (e)(f) 4.4$^{\mathrm{o}}$. The kpoint path is according to $\Gamma$ to $\textbf{K}$ of the bottom layer. Spin-orbit coupling was not included.}
\label{fig:bandstructure_AA-stacking_top_bottom}
\end{figure}

Fig. \ref{fig:bandstructure_AA-stacking_top_bottom} shows the effective band structures of the A-type twisted InSe bilayer and the hBN-encapsulated A-type twisted InSe bilayer with a  twist angle of 4.4$^{\mathrm{o}}$ projected on the bottom and top InSe layers along $\Gamma$ to $\textbf{K}$ of the bottom layer. The difference of the VBM between the bottom and top layers increases near $\textbf{K}$ as the twist angle increases (see also fig. \ref{fig:bandstructure_InSe-monolayer_rotation}).   

\newpage

\begin{figure} [H]
\captionsetup[subfigure]{justification=centering}
\begin{subfigure}{0.5\textwidth}
  \centering
  \includegraphics[width=0.7\textwidth]{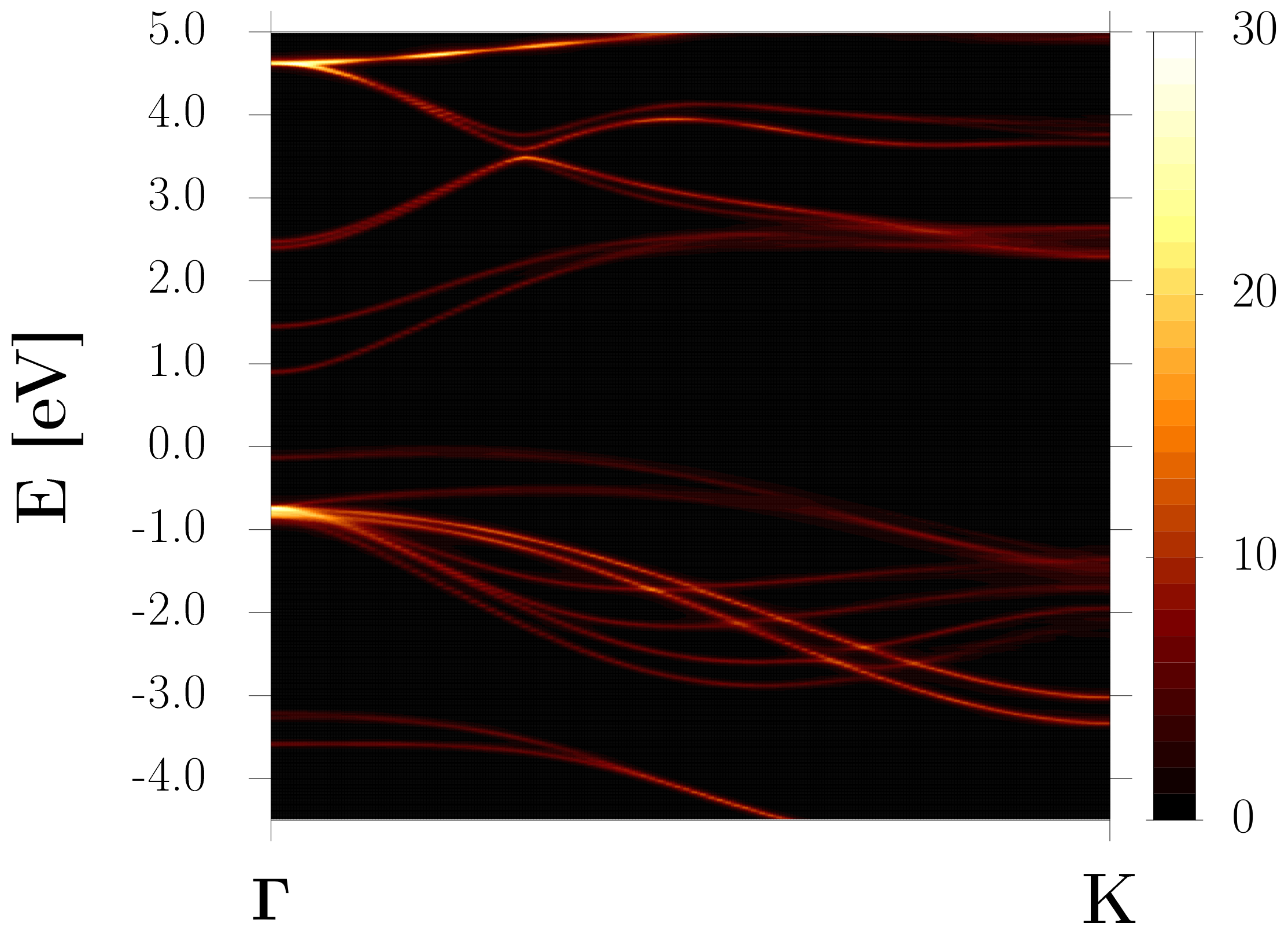}
  \caption{ $\theta$= 4.4$^{\mathrm{o}}$ (projected on \\ the bottom layer)}
\end{subfigure}
\begin{subfigure}{0.5\textwidth}
  \centering
  \includegraphics[width=0.7\textwidth]{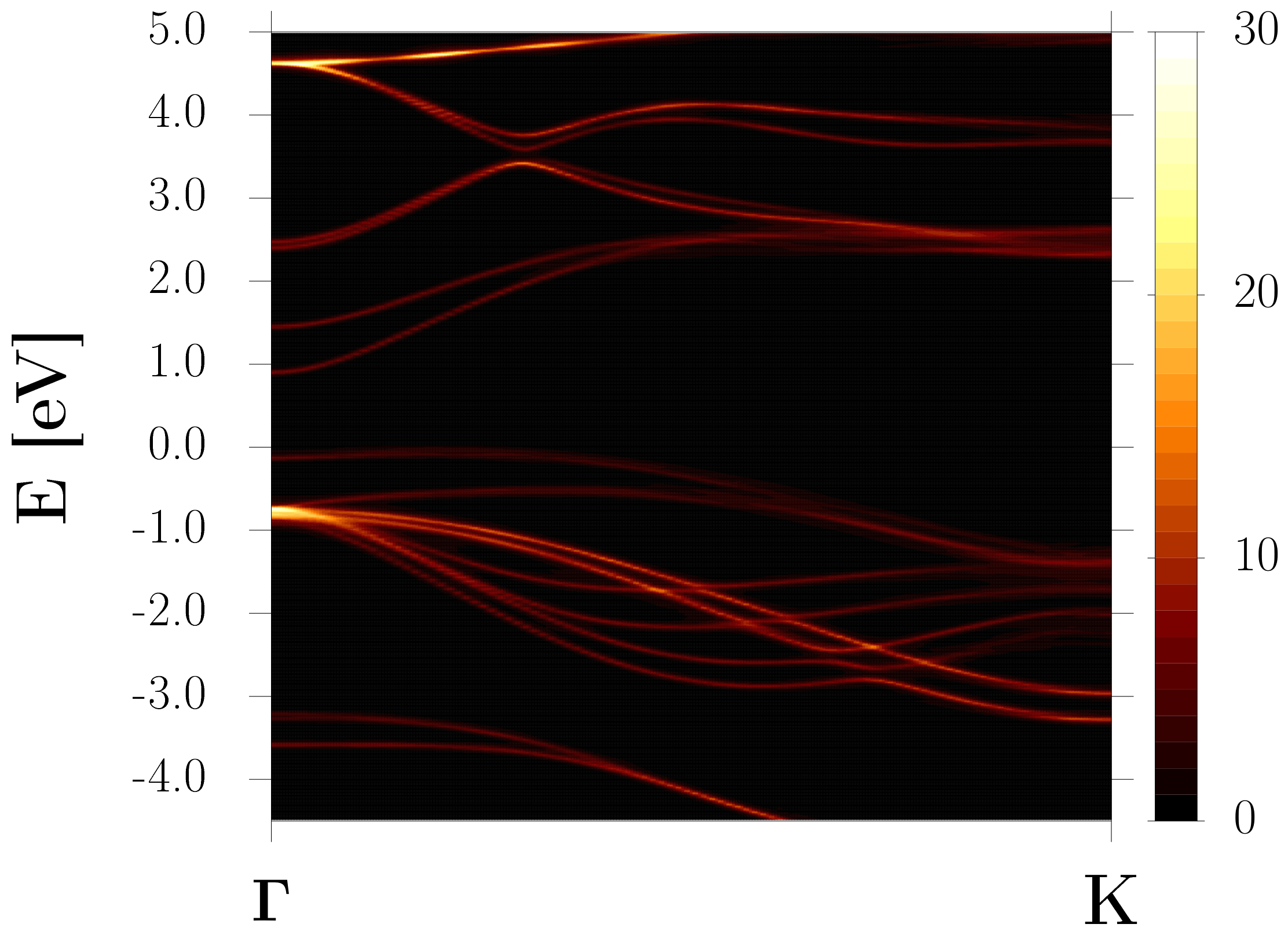}
  \caption{$\theta$= 4.4$^{\mathrm{o}}$ (projected on \\ the top layer)}
\end{subfigure}
\begin{subfigure}{0.5\textwidth}
  \centering
  \includegraphics[width=0.7\textwidth]{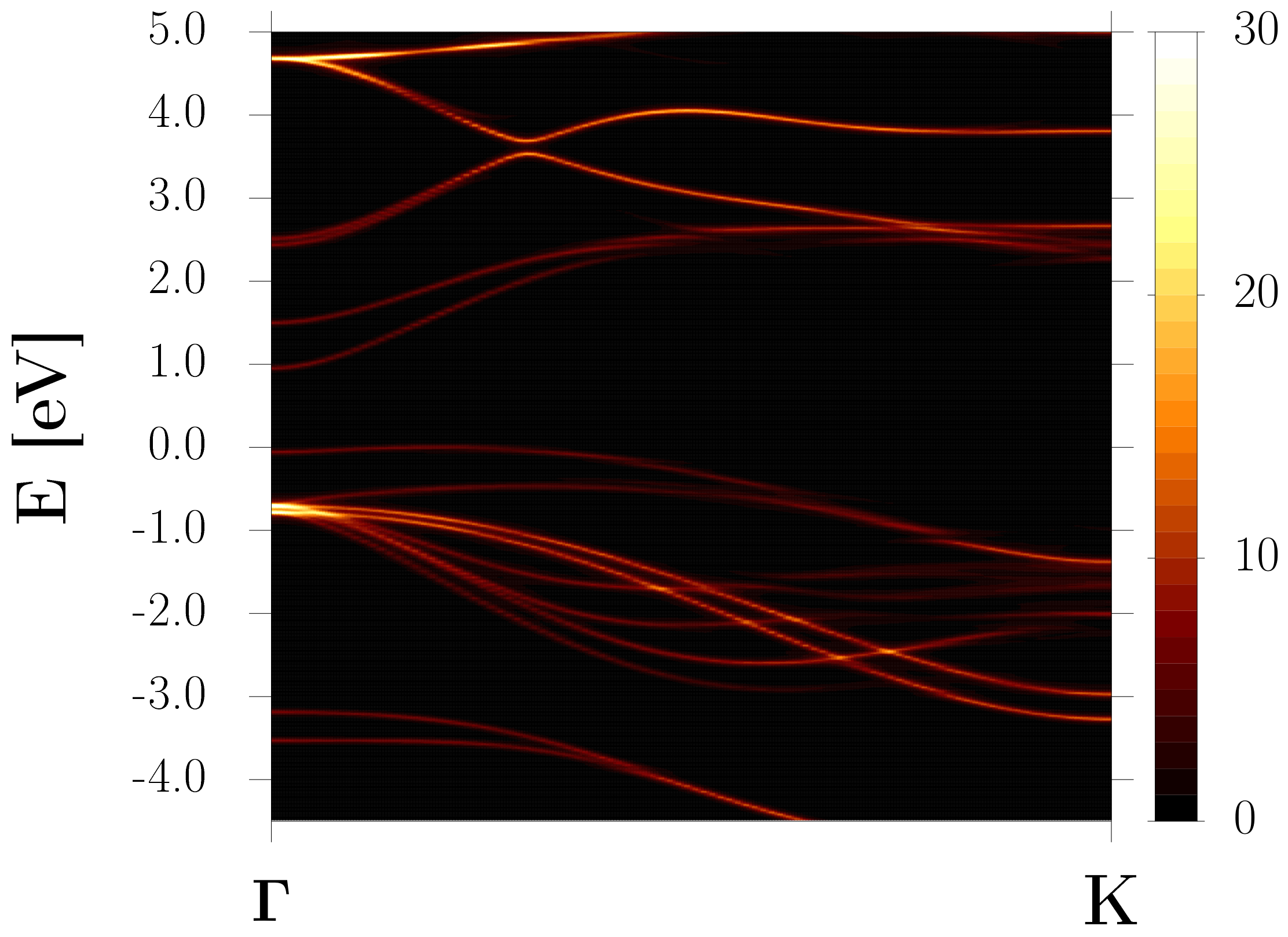}
  \caption{$\theta$= 17.9$^{\mathrm{o}}$ (projected on \\ the bottom layer)}
\end{subfigure}
\begin{subfigure}{0.5\textwidth}
  \centering
  \includegraphics[width=0.7\textwidth]{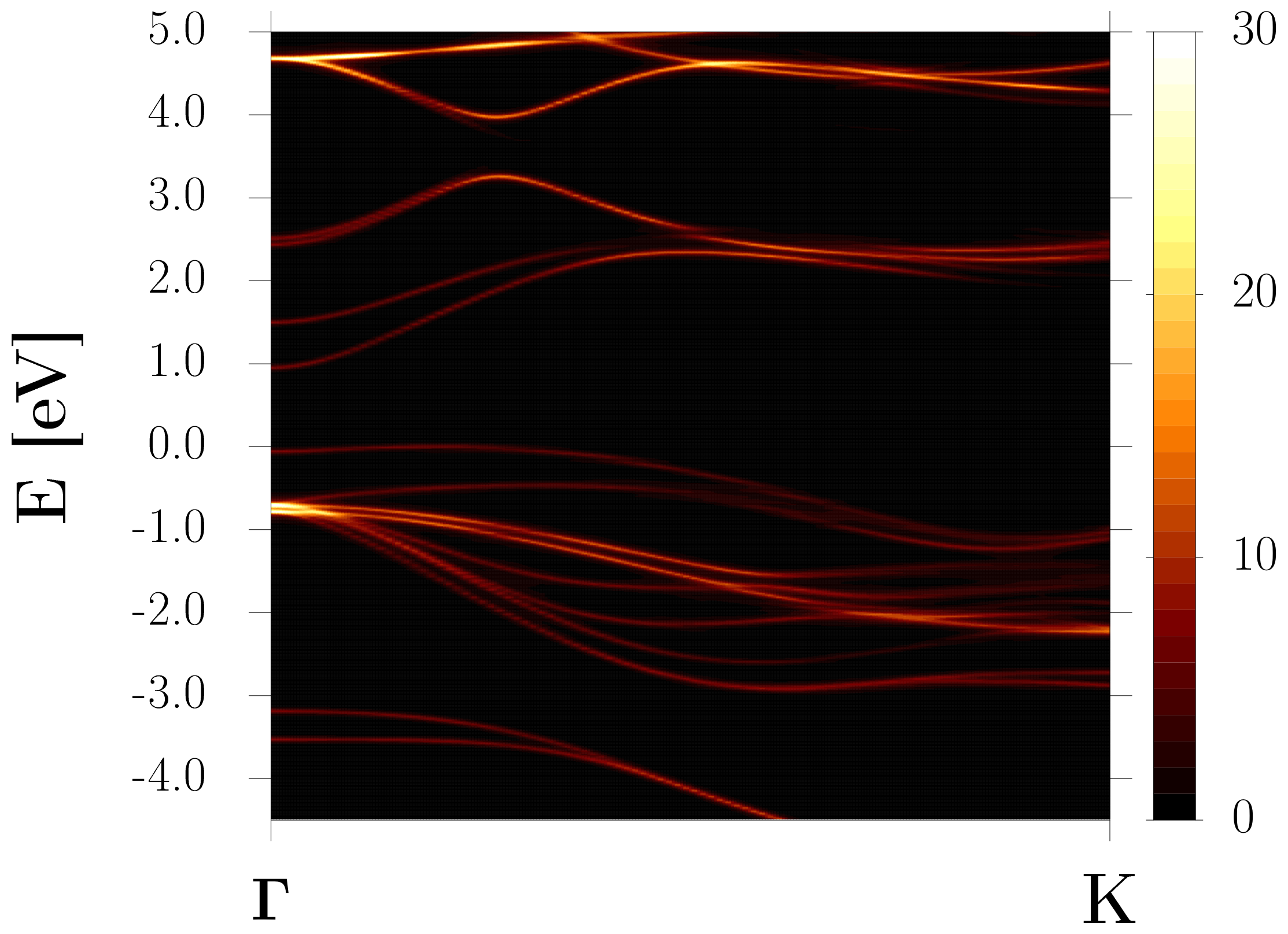}
  \caption{$\theta$= 17.9$^{\mathrm{o}}$ (projected on \\ the top layer)}
\end{subfigure}
\begin{subfigure}{0.5\textwidth}
  \centering
  \includegraphics[width=0.7\textwidth]{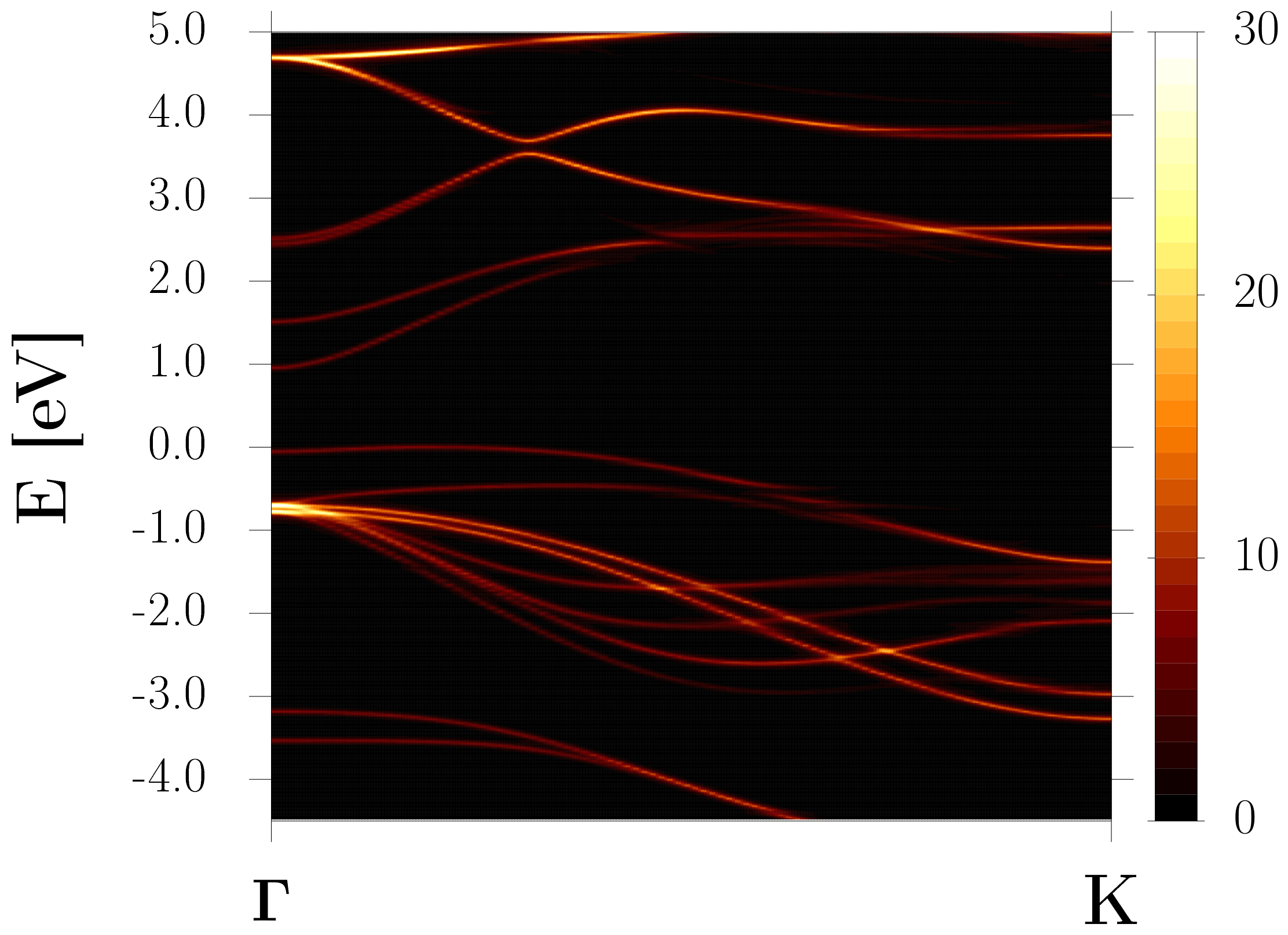}
  \caption{$\theta$= 27.8$^{\mathrm{o}}$ (projected on \\ the bottom layer)}
\end{subfigure}
\begin{subfigure}{0.5\textwidth}
  \centering
  \includegraphics[width=0.7\textwidth]{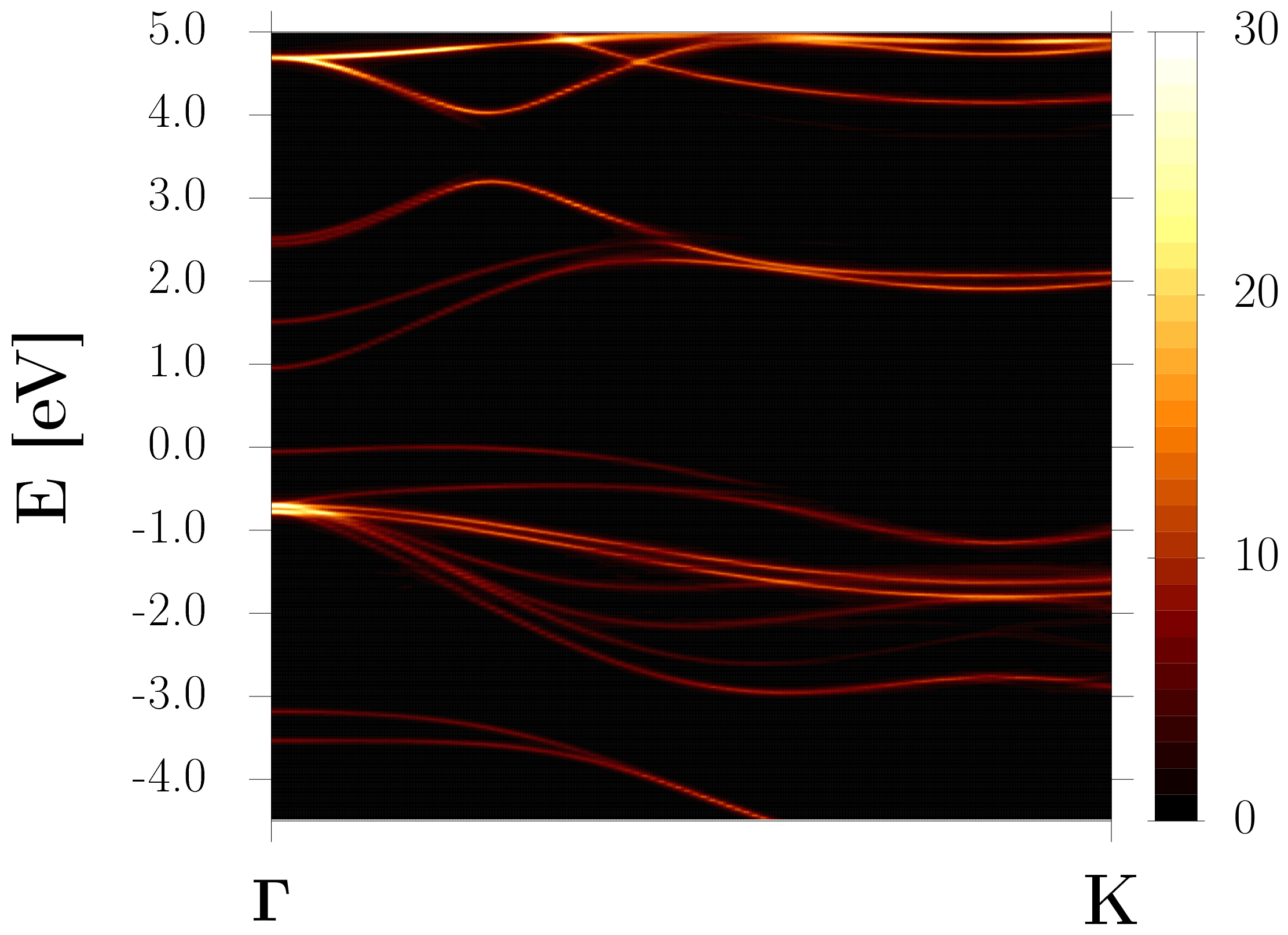}
  \caption{$\theta$= 27.8$^{\mathrm{o}}$ (projected on \\ the top layer)}
\end{subfigure}
 \caption{ Effective band structure projected on the bottom and top  InSe layers of the B-type twisted InSe bilayer. For projection on the bottom layer, the twist angle is (a) 4.4$^{\mathrm{o}}$ and (c) 17.9$^{\mathrm{o}}$ (e) 27.8$^{\mathrm{o}}$. For projection on the top layer, the twist angle is (b) 4.4$^{\mathrm{o}}$ and (d) 17.9$^{\mathrm{o}}$ (f) 27.8$^{\mathrm{o}}$. The kpoint path is according to $\Gamma$ to $\textbf{K}$ of the bottom layer. Spin-orbit coupling was not included.}
\label{fig:bandstructure_AB-stacking_top_bottom}
\end{figure}

Fig. \ref{fig:bandstructure_AB-stacking_top_bottom} shows the effective band structures of the B-type twisted InSe bilayers with twist angles of 4.4$^{\mathrm{o}}$, 17.9$^{\mathrm{o}}$ and 27.8$^{\mathrm{o}}$ projected on the bottom and top InSe layers along $\Gamma$ to $\textbf{K}$ of the bottom layer. The difference of the VBM between the bottom and top layers increases near $\textbf{K}$ as the twist angle increases (see also fig. \ref{fig:bandstructure_InSe-monolayer_rotation}).   

\newpage

\begin{figure} [H]
    \centering
    \includegraphics[width=0.6\textwidth]{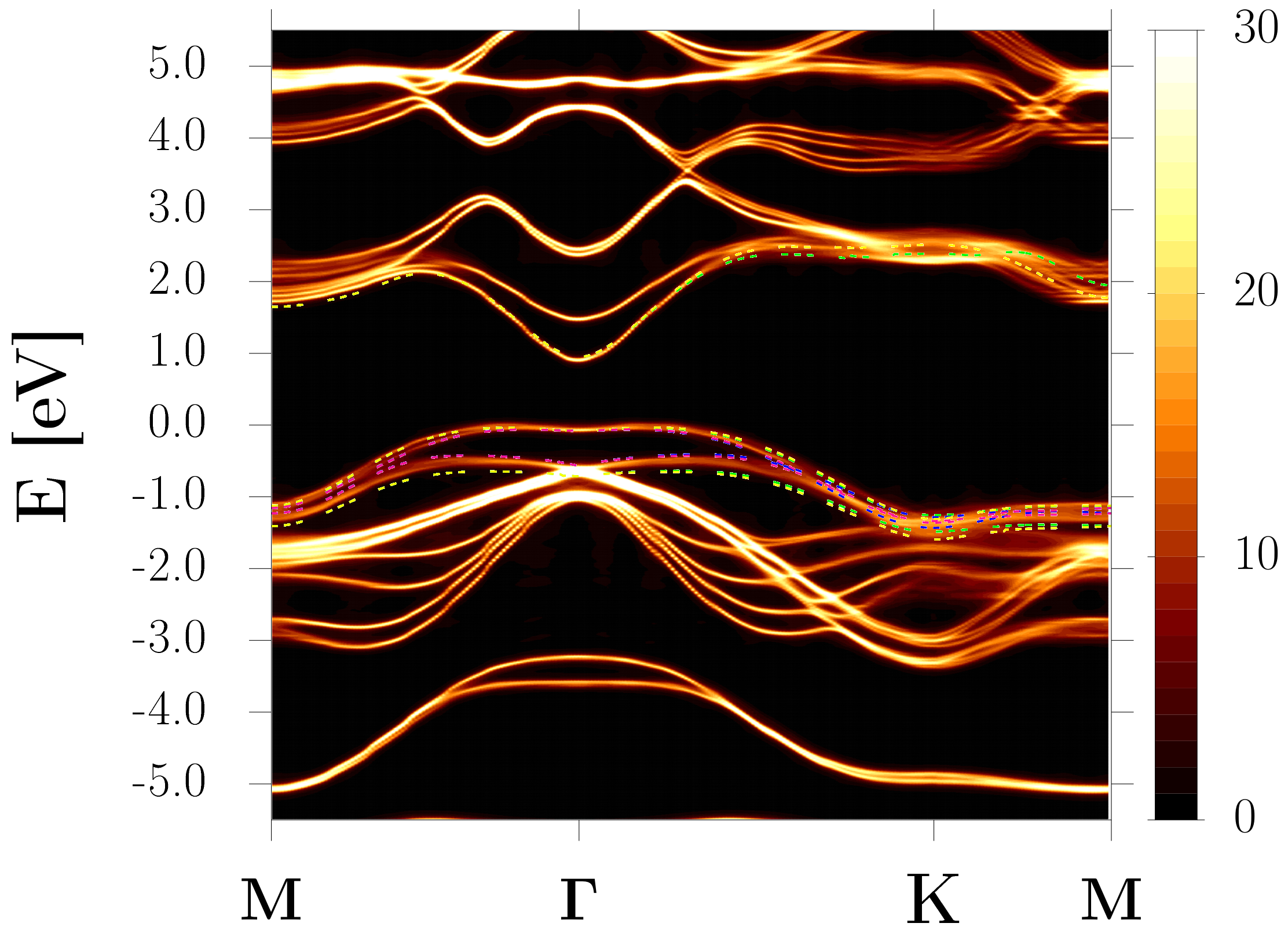}
 \caption{Effective band structure of the A-type twisted InSe bilayer with a twist angle of 4.4$^{\mathrm{o}}$. Green and yellow: same band with different spin states from the A-1 stacking, blue and pink: same band with different spin states from the A-2 stacking. Spin-orbit coupling was included.} 
\label{fig:bandstructure_twisted_InSe-bilayer-AA-stacking_ONETEP_QE_4p4deg-soc}
\end{figure}

\newpage

\begin{figure} [H]
\captionsetup[subfigure]{justification=centering}
\begin{subfigure}{0.5\textwidth}
  \centering
  \includegraphics[width=0.7\textwidth]{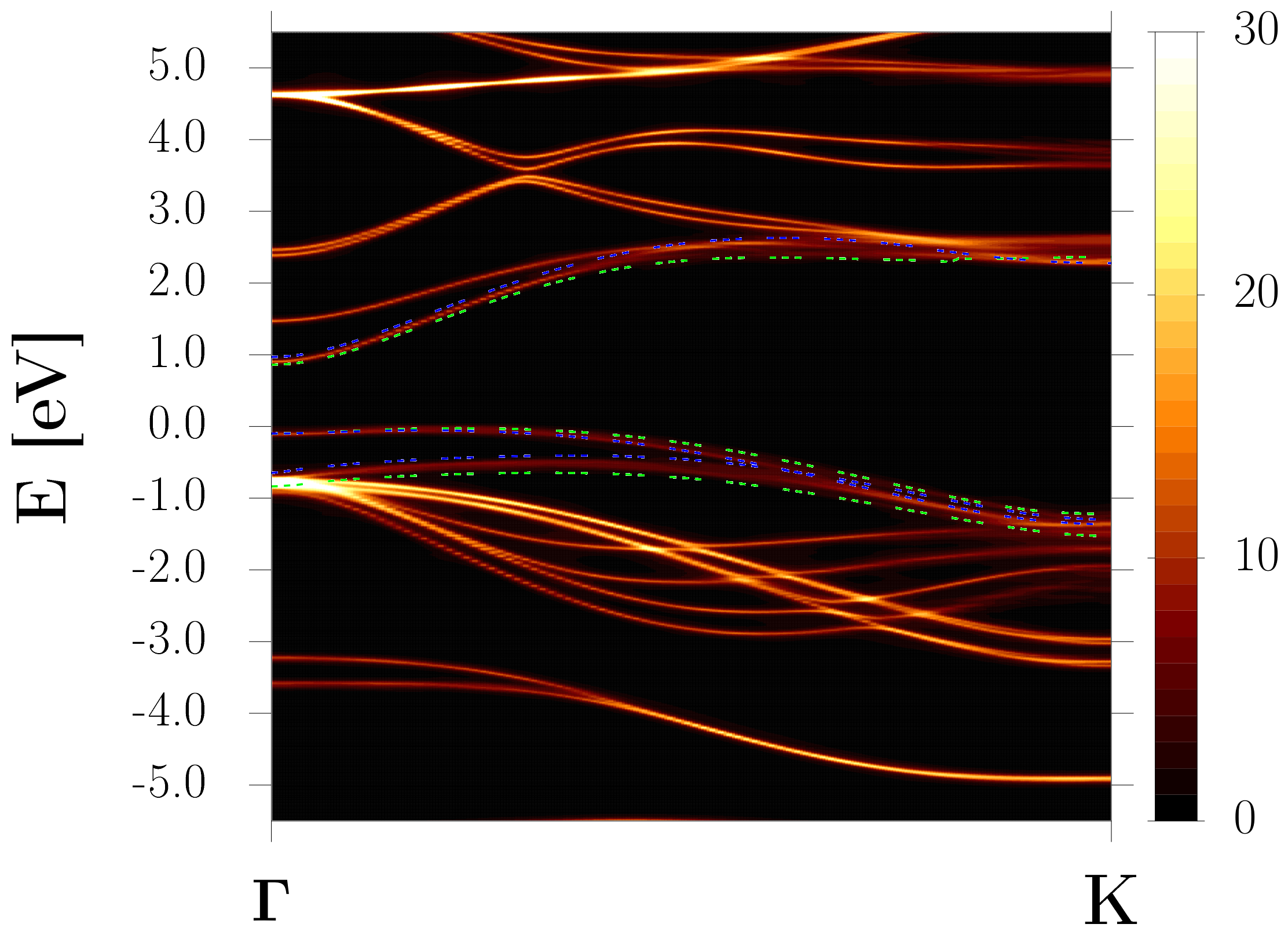}
  \caption{A type ($\theta$= 4.4$^{\mathrm{o}}$)}
\end{subfigure}
\begin{subfigure}{0.5\textwidth}
  \centering
  \includegraphics[width=0.7\textwidth]{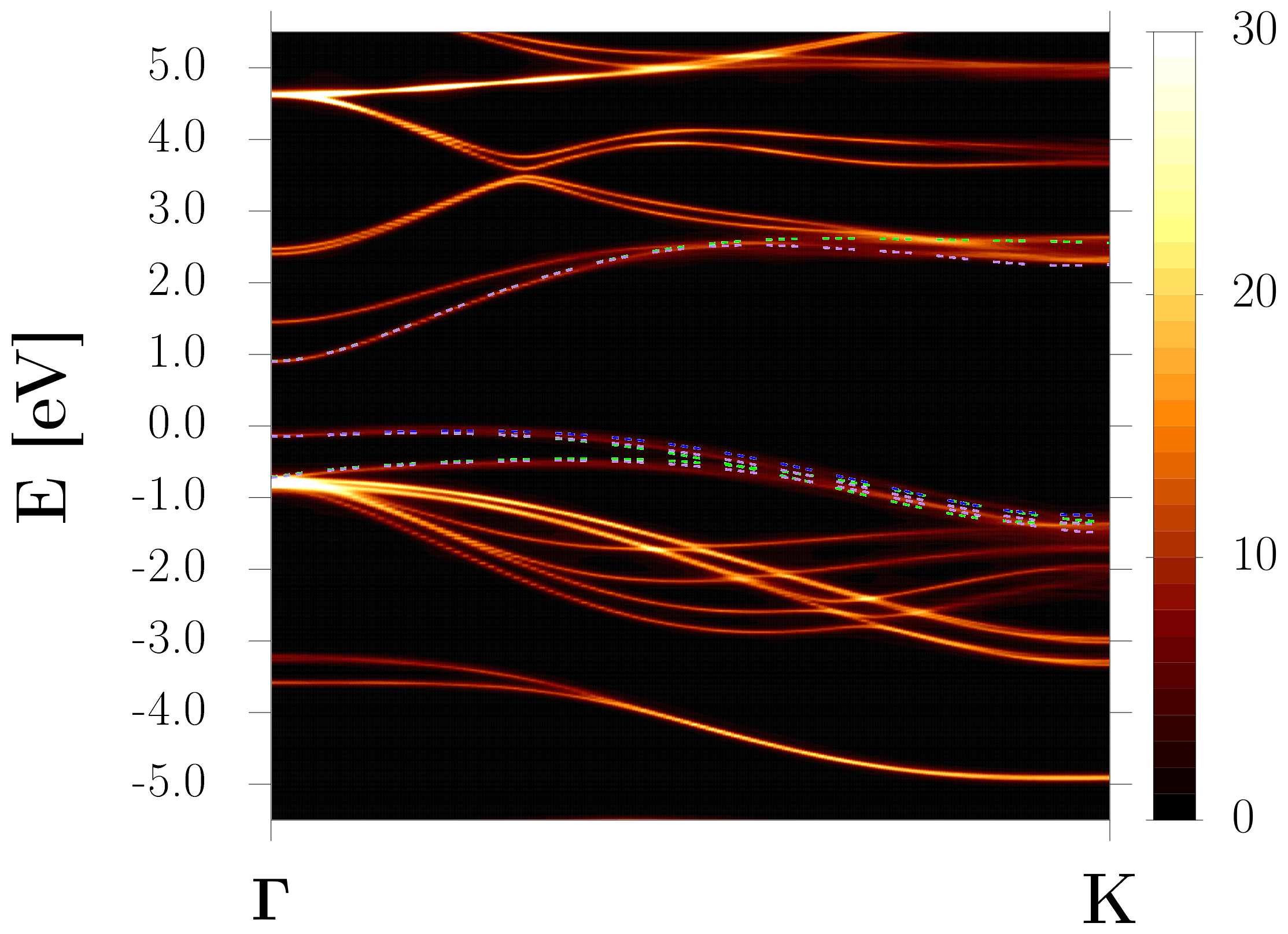}
  \caption{B type ($\theta$= 4.4$^{\mathrm{o}}$)}
\end{subfigure}
\begin{subfigure}{0.5\textwidth}
  \centering
  \includegraphics[width=0.7\textwidth]{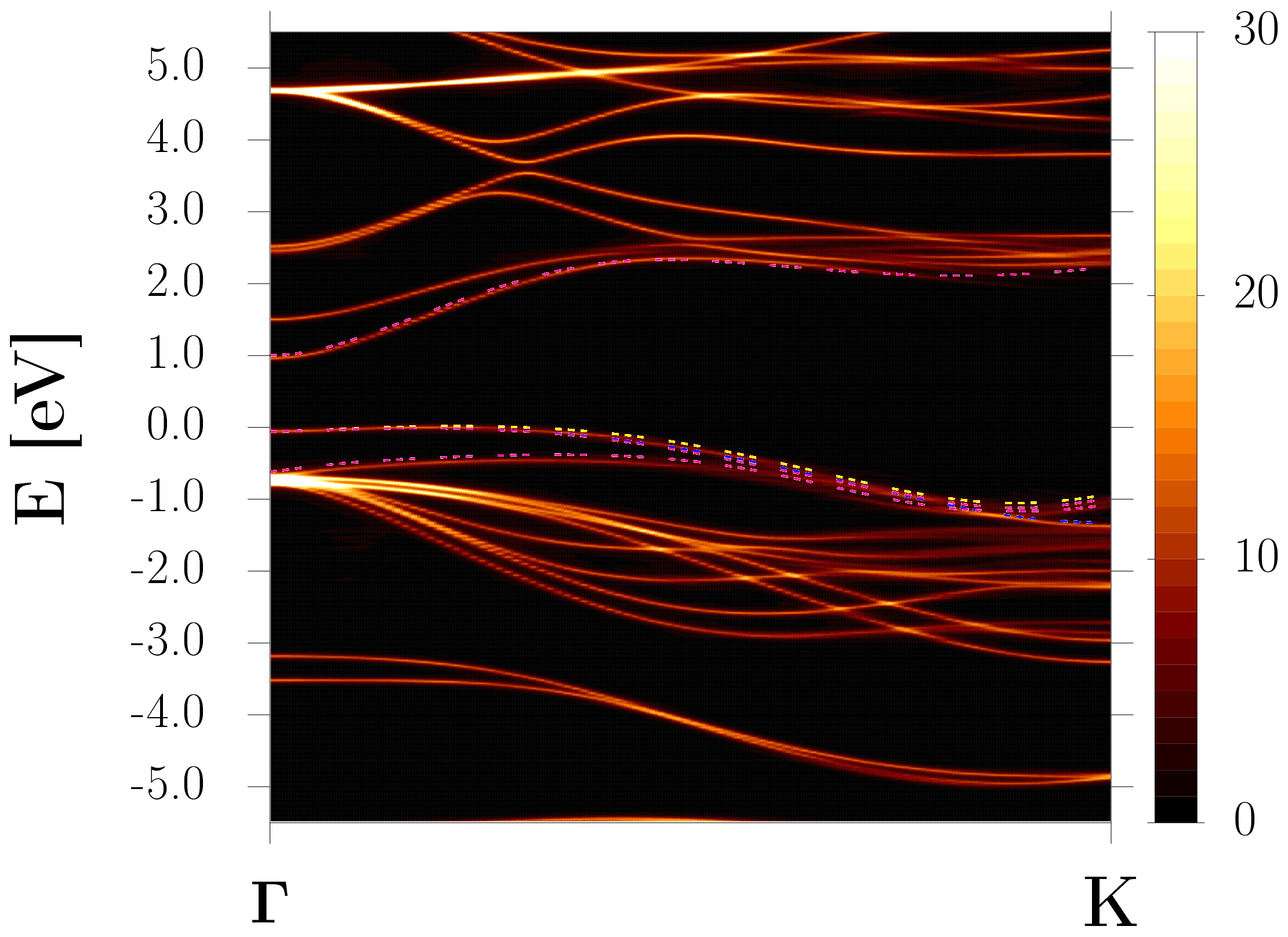}
  \caption{A type ($\theta$= 17.9$^{\mathrm{o}}$)}
\end{subfigure}
\begin{subfigure}{0.5\textwidth}
  \centering
  \includegraphics[width=0.7\textwidth]{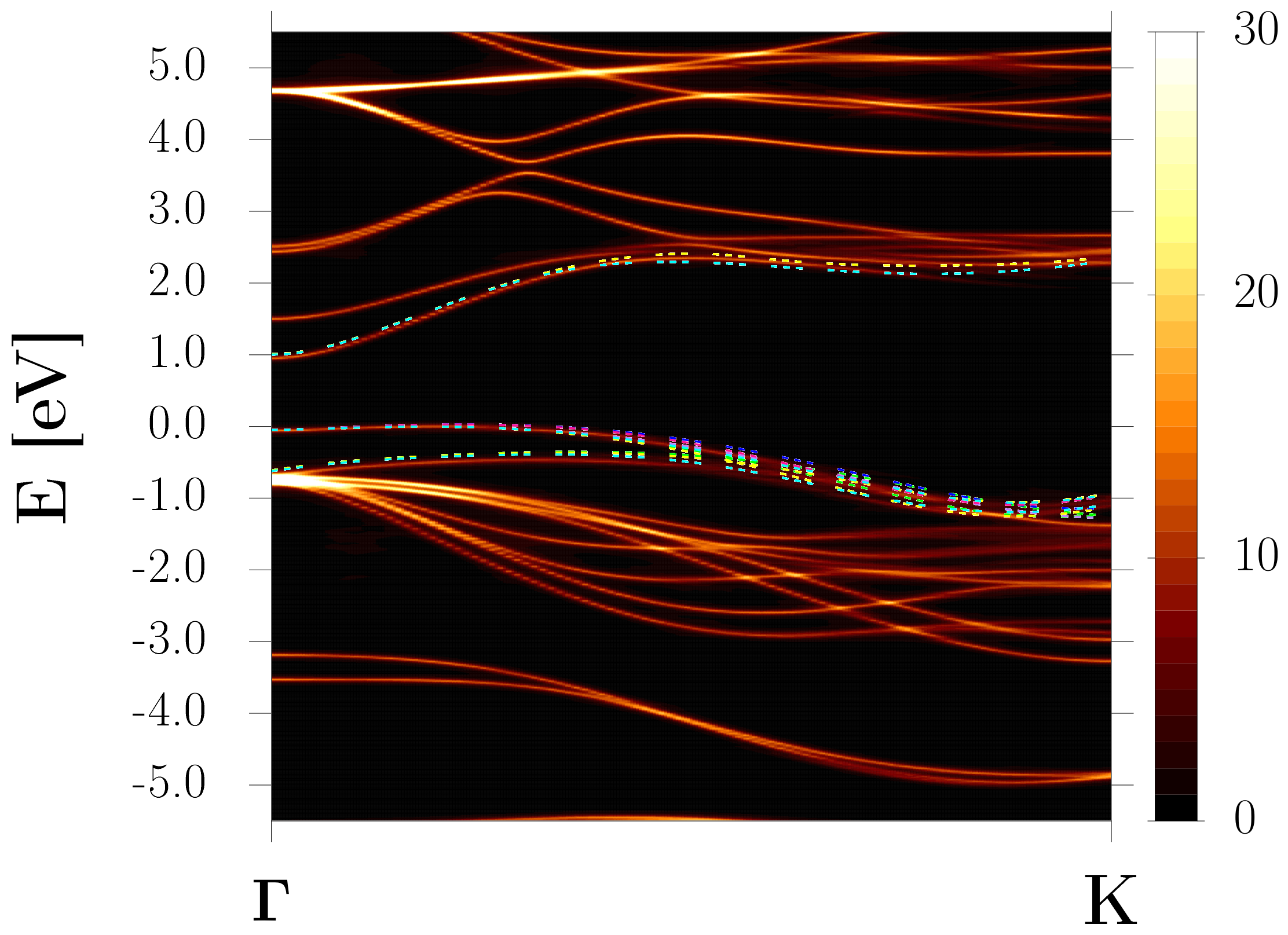}
  \caption{B type ($\theta$= 17.9$^{\mathrm{o}}$)}
\end{subfigure}
\begin{subfigure}{0.5\textwidth}
  \centering
  \includegraphics[width=0.7\textwidth]{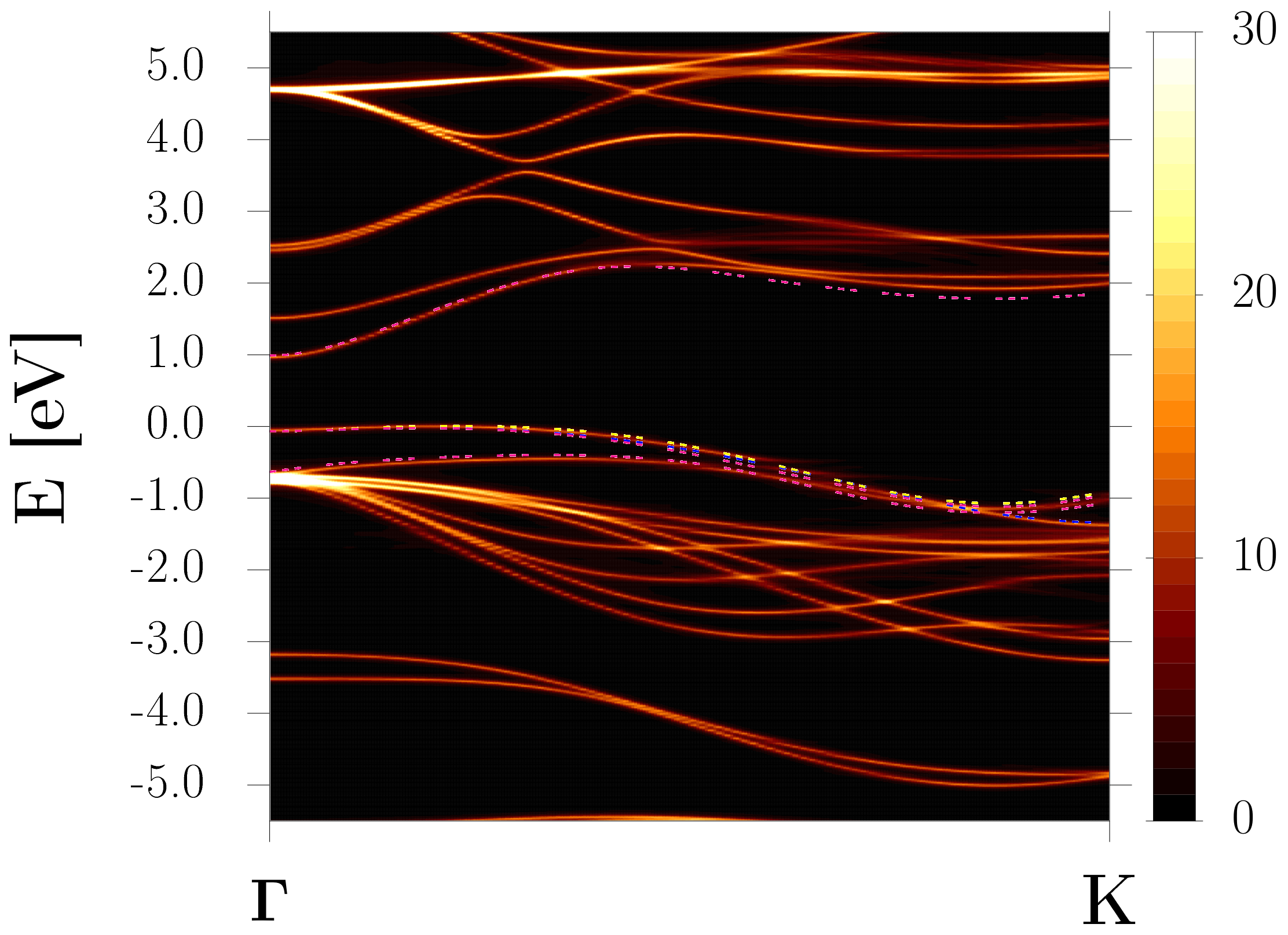}
  \caption{A type ($\theta$= 27.8$^{\mathrm{o}}$)}
\end{subfigure}
\begin{subfigure}{0.5\textwidth}
  \centering
  \includegraphics[width=0.7\textwidth]{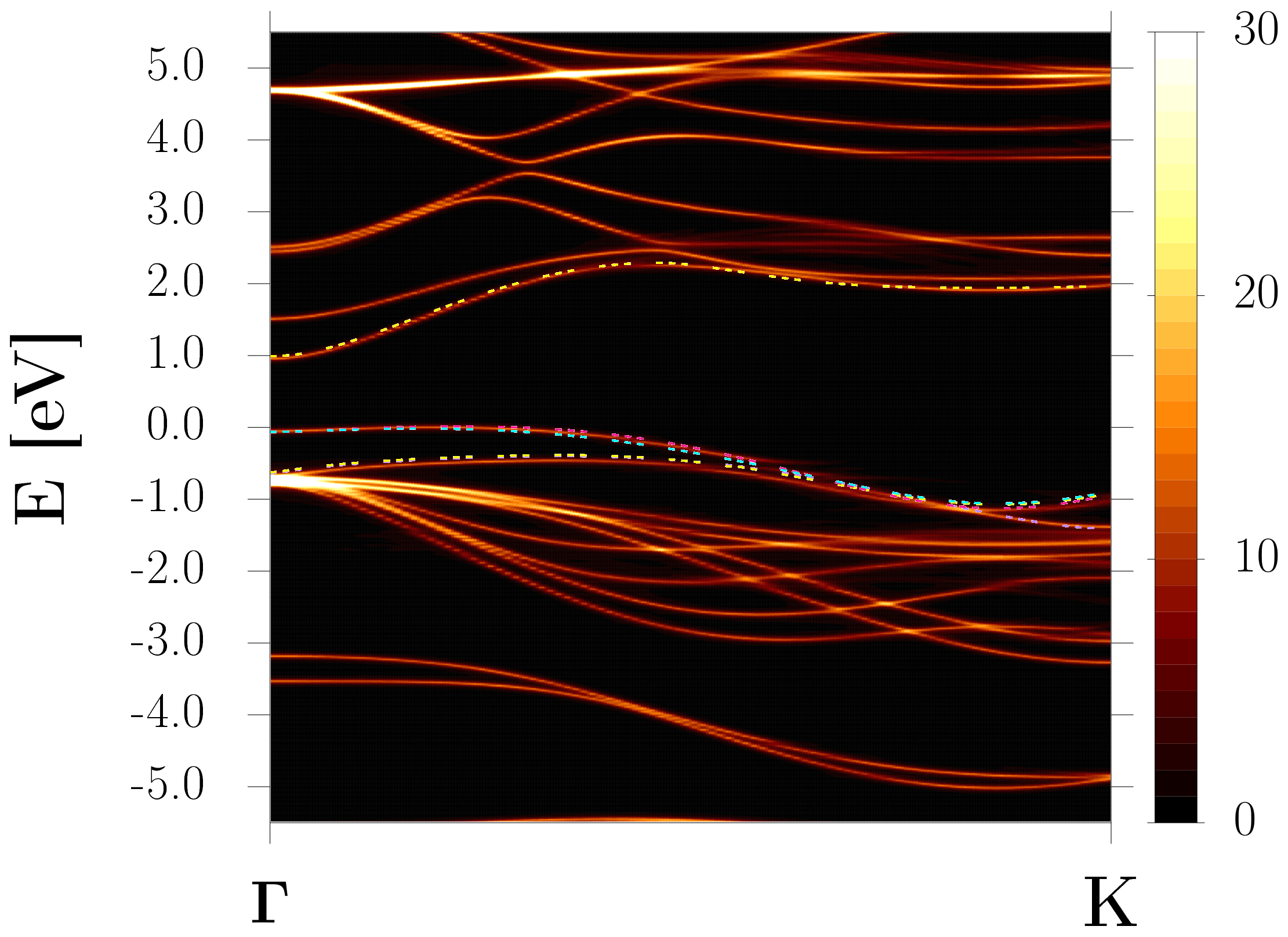}
  \caption{B type ($\theta$= 27.8$^{\mathrm{o}}$)}
\end{subfigure}
 \caption{ Effective band structures of different twisted InSe bilayers. A-type twisted InSe bilayer with a twist angle of (a) 4.4$^{\mathrm{o}}$ (c) 17.9$^{\mathrm{o}}$ (e) 27.8$^{\mathrm{o}}$. B-type twisted InSe bilayer with a twist angle of (b) 4.4$^{\mathrm{o}}$ (d)  17.9$^{\mathrm{o}}$ (f) 27.8$^{\mathrm{o}}$. Bands in the bottom layer from the primitive cell calculations are shown in green (A-1 or B-1), blue (A-2 or B-2) and purple (B-3). Bands in the top layer from the primitive cell calculations are shown in yellow (A-1 or B-1), pink (A-2 or B-2) and cyan (B-3). The interlayer distances for the primitive cell calculations are the same as the supercell calculations. Spin-orbit coupling was not included.}
\label{fig:Bandstructure_ONETEP_QE}
\end{figure}

Fig. \ref{fig:Bandstructure_ONETEP_QE} shows the effective band structure of the A-type and B-type twisted InSe bilayers with different twist angles. The contribution of bands from the top InSe layer becomes obvious as the twist angle increases.  

\newpage

\begin{figure} [H]
    \centering
    \includegraphics[width=0.52\textwidth]{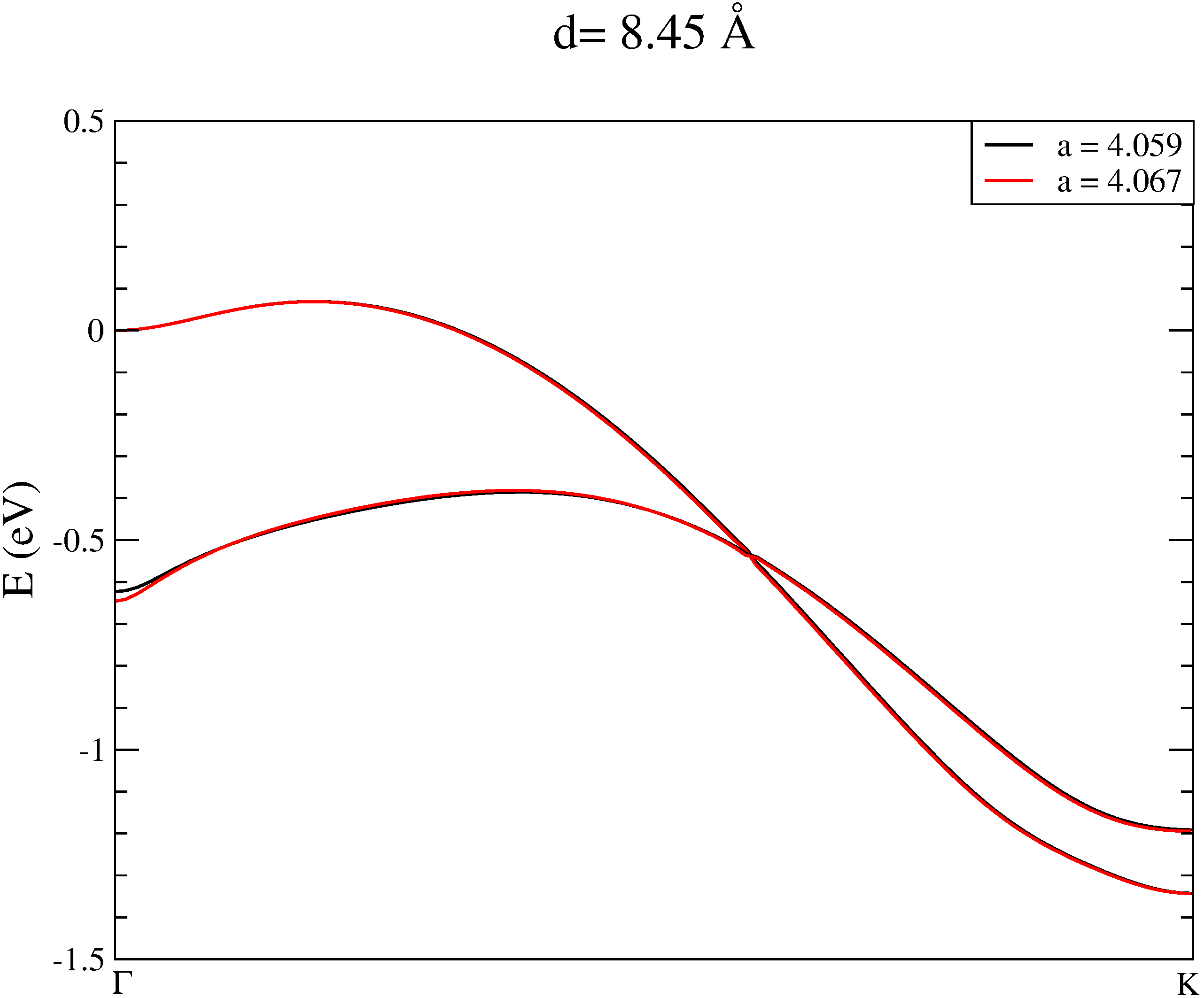}
 \caption{ Comparison of $E$ versus $k$ for the two bands below the Fermi level for the A-2 stacking with the lattice constants  of InSe monolayer (4.059 \AA) and the A-2 stacking (4.067 \AA) at the same interlayer distance. The Fermi level is shifted to the energy of the valence band maximum at $\Gamma$.} 
\label{fig:compare_lattice-constant}
\end{figure}

Fig. \ref{fig:compare_lattice-constant} shows the comparison of the VBM and the band below it between two different lattice constants for the A-2 stacking. 4.059 \AA \, is the lattice constant for twisted InSe bilayer, whereas 4.063 \AA \, is the lattice constant for the A-2 stacking. It shows that the difference of the VBM and the band below it between these two lattice constants is very small. We can ignore the effect of the lattice constant for different stacking configurations seen in some regions of twisted InSe layer when using 4.059 \AA \, as the lattice constant of twisted InSe bilayer. 

\newpage

\begin{figure} [H]
    \centering
    \includegraphics[width=0.6\textwidth]{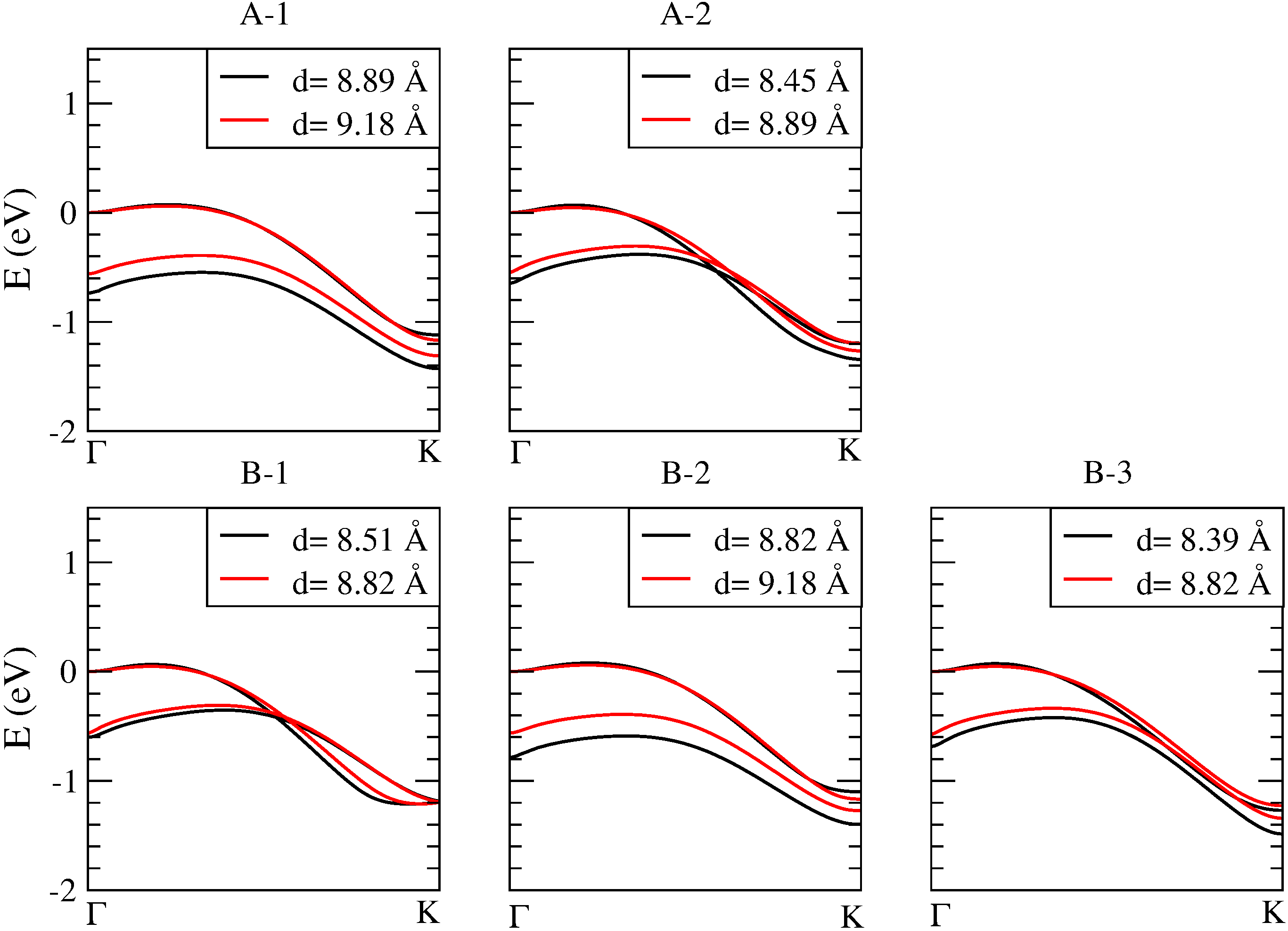}
 \caption{Comparison of $E$ versus $k$ for the two bands below the Fermi level for the A-type and B-type InSe bilayers with two different interlayer distances. Without considering corrugation, 8.89 \AA \, is the interlayer distances for the A-type twisted InSe bilayer ($\theta=4.4^{\mathrm{o}}$) and 8.82 \AA \, is the interlayer distance for the B-type twisted InSe bilayer ($\theta=4.4^{\mathrm{o}}$). The Fermi level is shifted to the energy of the valence band maximum at $\Gamma$.} 
\label{fig:compare_interlayer-distance}
\end{figure}

Fig. \ref{fig:compare_interlayer-distance} shows the comparison of the VBM and the band below it between two different interlayer distances for the A-type and B-type stacking configurations. One of the interlayer distances is the optimised interlayer distance for InSe bilayer with each type stacking configuration, the other is the optimised interlayer distance for twisted InSe bilayer with a twist angle of 4.4$^{\mathrm{o}}$. The interlayer distances for the A-type and B-type twisted InSe bilayers with a twist angle of 4.4$^{\mathrm{o}}$ are 8.89 \AA \, and 8.82 \AA, respectively (see also table \ref{InSe-bilayer_parameters}). The flatter VBM at the highest energy kpoint (red curve) associated with a larger interlayer distance, which means the effective mass for holes is larger with a larger interlayer distance (consistent with table \ref{InSe-bilayer_parameters}). The VBM and the band below is getting closer with increasing interlayer distance because of the weakening of the repulsion between these two bands.

\newpage

\begin{figure} [H]
    \centering
    \includegraphics[width=0.45\textwidth]{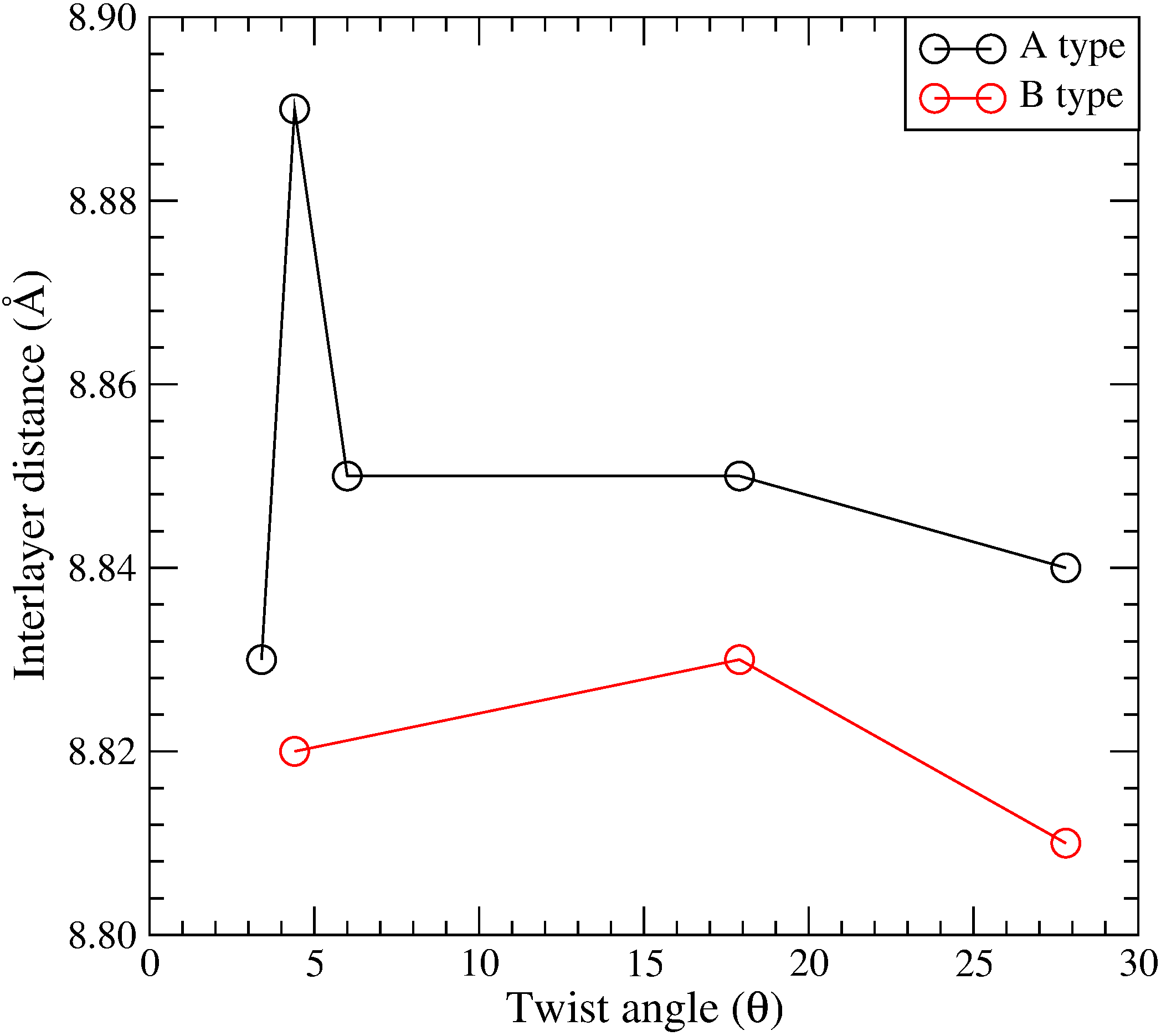}    
 \caption{ Interlayer distance versus twist angle for twisted InSe bilayers. Corrugation is not considered.} 
\label{fig:interlayer-distance_twist-angle}
\end{figure}

Figure \ref{fig:interlayer-distance_twist-angle} shows the interlayer distance versus the twist angle for twisted InSe bilayer. For the twist angles considered in this work, the interlayer distances are 8.85$\pm$0.04 \AA. 

\begin{figure} [H]
    \centering
    \includegraphics[width=0.8\textwidth]{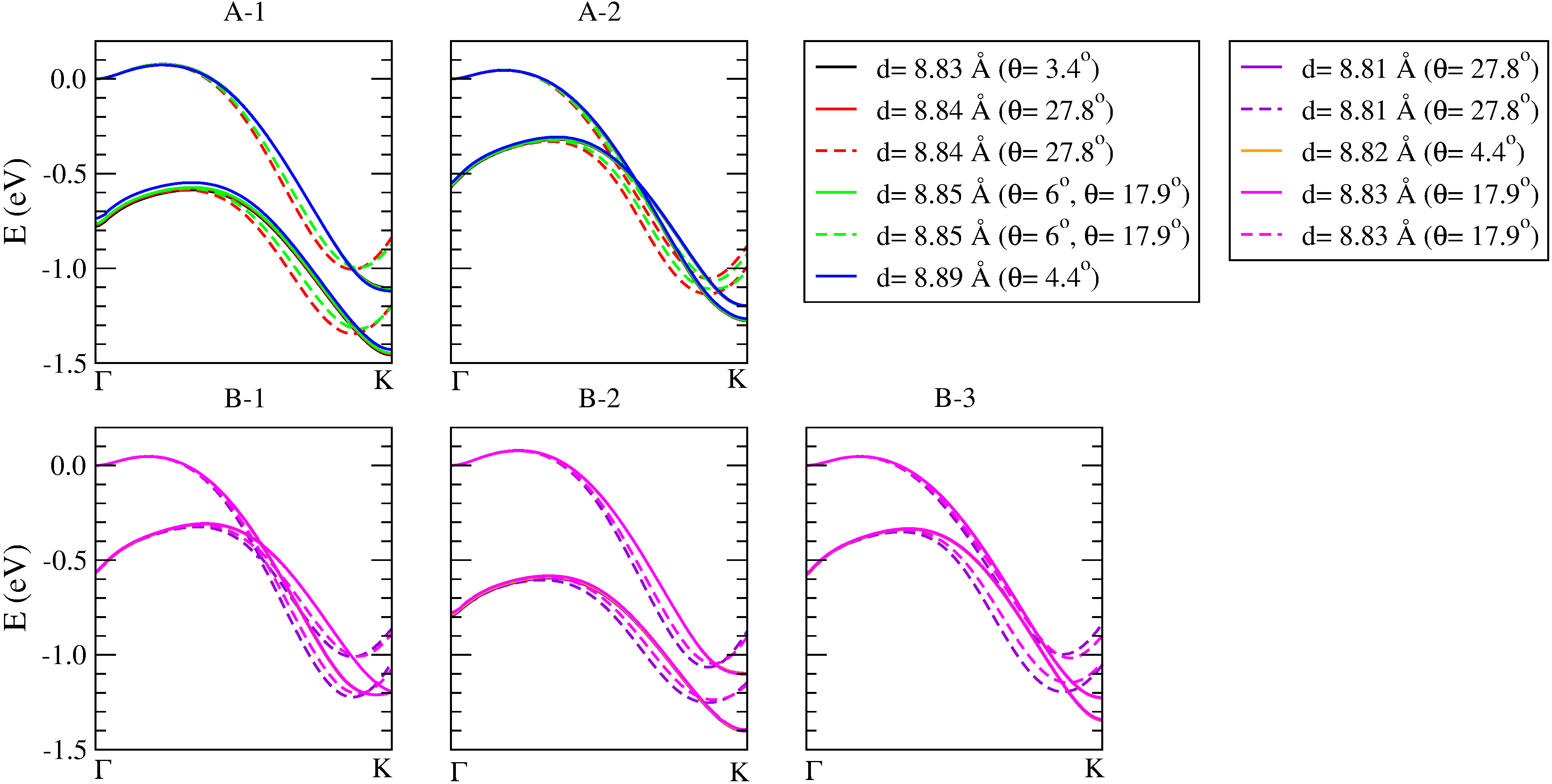}    
 \caption{ Valence band maximum and the band below it along $\Gamma$ to $\textbf{K}$ of the bottom InSe layer for the A-type and B-type InSe bilayers. Solid curves: bands from the bottom InSe layer. Dashed curves: bands from the top InSe layer. } 
\label{fig:VBM_twist-angle}
\end{figure}

Figure \ref{fig:VBM_twist-angle} shows the VBM and the band below it for InSe bilayers with different stacking configurations at different interlayer distances and twist angles. The difference of the effective mass for holes calculated around the highest energy kpoint of the VBM is very small for different twist angles considered in this work. 

\newpage

\begin{figure} [H]
\captionsetup[subfigure]{justification=centering}
\begin{subfigure}{0.5\textwidth}
  \centering
  \includegraphics[width=\textwidth]{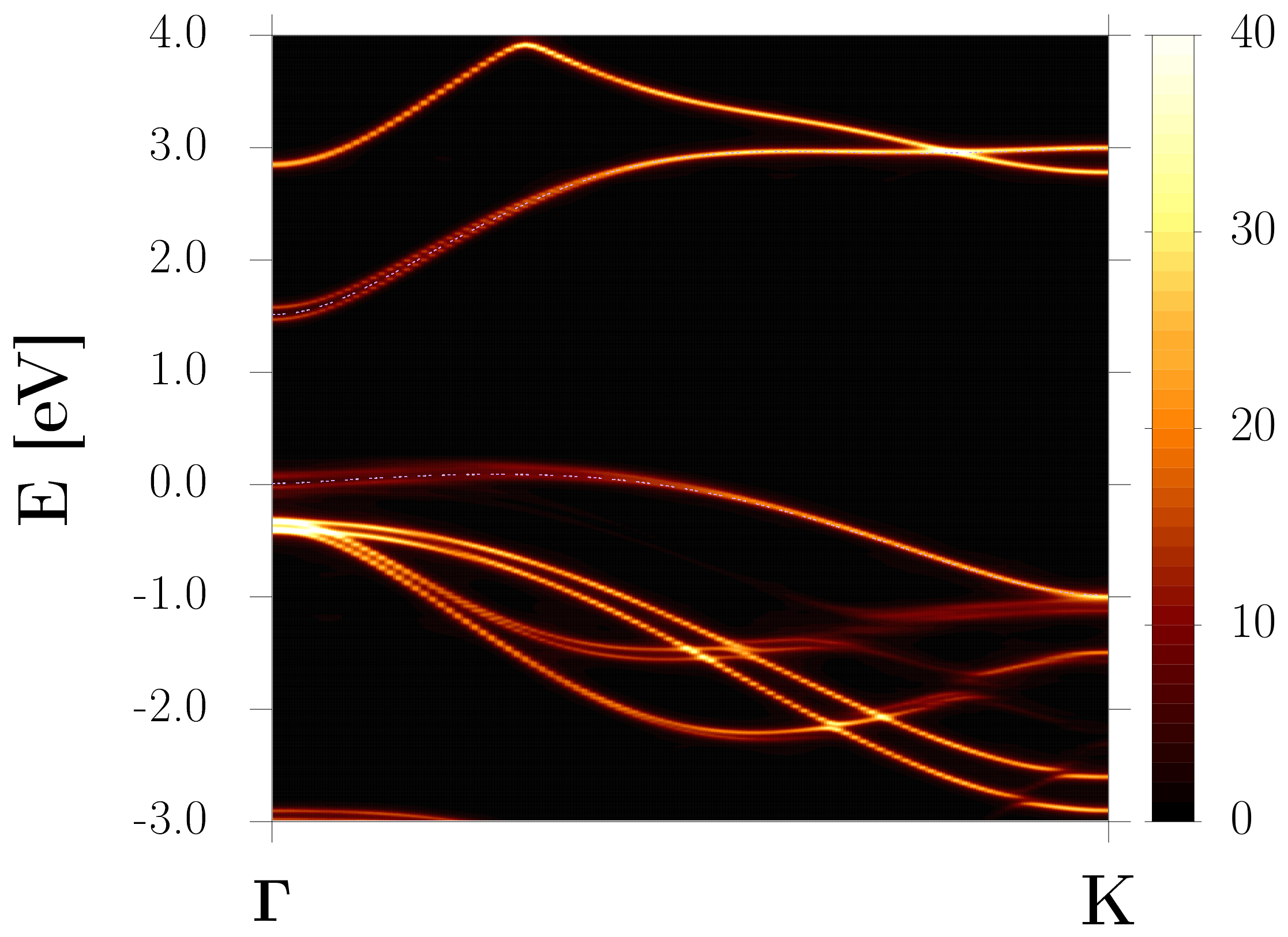}
  \caption{HBN-encapsulated A-type twisted InSe bilayer}
\end{subfigure}
\begin{subfigure}{0.5\textwidth}
  \centering
  \includegraphics[width=\textwidth]{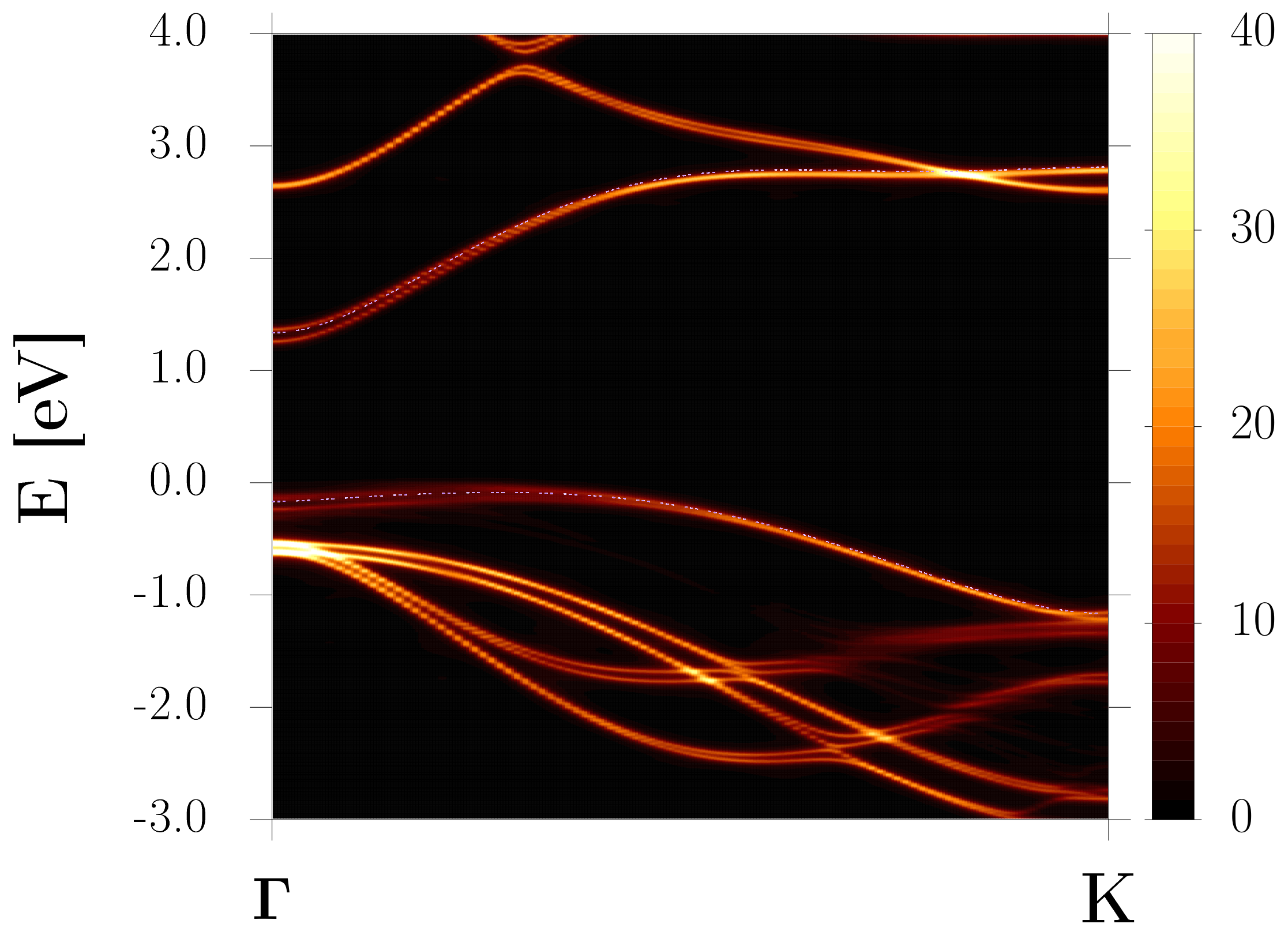}
  \caption{HBN-encapsulated B-type twisted InSe bilayer}
\end{subfigure}
 \caption{ (a) Effective band structure of the hBN-encapsulated (a) A-type (b) B-type twisted InSe bilayer with a twist angle of 4.4$^{\mathrm{o}}$ from the supercell calculation. Purple curves refer to the bands in InSe monolayer from the primitive cell calculation. Spin-orbit coupling was not included.}
\label{fig:bandstructure_InSe-hBN-InSe_ONETEP_QE}
\end{figure}

Fig. \ref{fig:bandstructure_InSe-hBN-InSe_ONETEP_QE} shows the superposition of the VBM and the CBM in InSe monolayer from the primitive cell calculations (purpe curves) on the effective band structure of the hBN-encapsulated A-type and B-type twisted InSe bilayers from the supercell calculations. These bands from the primitive cell calculation match well with the same bands in the effective band structures from the supercell calculations for both the hBN-encapsulated A-type and B-type twisted InSe bilayers. This further verifies the argument that the insertion of the hBN layer causing the separation of two InSe layers.

\newpage

\section{Calculation of exciton binding energy} 

According to the previous literature \cite{Viner2021, Danovich2018, Ruiz-Tijerina2020}, the intralayer and interlayer exciton binding energies are calculated as following procedure. The Hamiltonian of the system is
\begin{equation}\label{eq:exciton_binding-energy_Hamiltonian}
[-\frac{\hbar^{2}}{2\mu}\nabla_{\rho}^{2}+U(\rho)] \: \psi(\rho)= E_{\mathrm{b}}\psi(\rho),
\end{equation}

\noindent with $\mu= \frac{m_{\mathrm{e}}m_{\mathrm{h}}}{m_{\mathrm{e}}+m_{\mathrm{h}}}$ and 
\begin{equation}
\psi_{m}(\rho,\theta)=\sum_{j=1}^{N}A_{m,j}\frac{e^{im\theta}}{\sqrt{2\pi}}(\beta\rho)^{|m|}e^{-\beta_{j}\rho}. 
\end{equation}

\noindent $\beta^{-1}$ can be any values and $\beta_{j}= \beta_{1}e^{\xi(j-1)}$ ($\xi=(N-1)^{-1}\mathrm{log}(\beta_{N}/\beta_{1})$, $\beta_{N}^{-1} \leq r_{*} \ll \beta_{1}^{-1}$) is within the range of possible exciton Bohr radius. The Hamiltonian problem becomes
\begin{equation}
[H_{m}-E_{\mathrm{b}}S_{m}]A_{m}= 0,
\end{equation}

\noindent with $S_{m,\mu\nu}= \beta^{2|m|}\frac{\Gamma(2|m|+2)}{(\beta_{\mu}+\beta_{\nu})^{2|m|+2}}$ in analytical form and $H_{m}= K_{m}+U_{m}$, where $K_{m,\mu\nu}$ in analytical form is

\begin{equation}
K_{m,\mu\nu}=-\frac{\hbar^{2}}{2\mu}[-(2|m|+1)\frac{\beta^{2|m|}\beta_{\mu}\Gamma(2|m|+1)}{(\beta_{\mu}+\beta_{\nu})^{2|m|+1}}
+\frac{\beta^{2|m|}\beta_{\mu}^{2}\Gamma(2|m|+2)}
{(\beta_{\mu}+\beta_{\nu})^{2|m|+2}}],
\end{equation}

\noindent and $U_{m,\mu\nu}= -\frac{\beta^{2|m|}}{(2\pi)^{2}}\int_{0}^{\infty} dq \: q \: f_{\mu\nu}(-q)U_{\lambda}(q)$. With different $\lambda$, $U_{m,\mu\nu}$ can refer to intralayer or interlayer potential energy matrix elements with considering the screening effect from other layer. $U_{m,\mu\nu}$ can be obtained from the Rytova-Keldysh potential $U_{\lambda}(\rho)=-\frac{\pi e^{2}}{2r_{*}\tilde{\epsilon}} \left[ H_{0} \left( \frac{\rho}{r_{*}} \right)-Y_{0} \left( \frac{\rho}{r_{*}} \right) \right]$ ($H_{0}$: zeroth Struve functions and $Y_{0}$: zeroth Bessel functions of the second kind). 

\begin{equation}
U_{\mathrm{intra}}(q)= \frac{2\pi[1+r_{*}q(1-e^{-2q\tilde{d}})]}{\tilde{\epsilon}q[1+2qr_{*}+q^{2}r_{*}^{2}(1-e^{-2q\tilde{d}})]},
\end{equation}

\begin{equation}
U_{\mathrm{inter}}(q)= \frac{2\pi e^{-qd}}{\tilde{\epsilon}q[(1+r_{*}q)^{2}-r_{*}^{2}q^{2}e^{-2q\tilde{d}}]}.
\end{equation}

\begin{equation}\label{eq:exciton_binding-energy_U}
f_{\mu\nu}(q)=\frac{2\pi}{(\beta_{\mu}+\beta_{\nu})^{2|m|+2}}\Gamma(2|m|+2){}_{2}F_{1}(|m|+1,|m|+3/2;1;-q^{2}(\beta_{\mu}+\beta_{\nu})^{-2}).
\end{equation}

\noindent The definition of parameters are $\tilde{d}= d\sqrt{\epsilon_{\parallel}/\epsilon_{\perp}}$ ($d$ is the interlayer distance), $\tilde{\epsilon}= \sqrt{\epsilon_{\parallel}\epsilon_{\perp}}$ ( $\epsilon$ is the dielectric constant of environment) and $r_{*}=2\pi\kappa/\tilde{\epsilon}$ ($\kappa$ is the in-plane polarizability) for the screening length. 

Furthermore, the intralayer exciton binding energy of twisted InSe/hBN/InSe can be calculated according to the Coulomb potential for the monolayer in analytical form below:

\begin{equation}\label{eq:U_ML}
\begin{aligned}
U_{m, \mu \nu}^{\mathrm{ML}} =& -\frac{e^{2}\beta^{2|m|}}{r_{*}^{2}\tilde{\epsilon}} [ \frac{\Gamma(2|m|+3)}{(\beta_{\mu}+\beta_{\nu})^{2|m|+3}}F_{3,2} ( |m|+1, |m|+\frac{3}{2}, |m|+2;\frac{3}{2}; -\frac{r_{*}^{-2}}{(\beta_{\mu}+\beta_{\nu})^{2}} ) \\
&-4^{|m|}r_{*}^{2|m|+3}\mathrm{cos}(m\pi)\Gamma^{2}(|m|+1)F_{2,1} ( |m|+1, |m|+1; \frac{1}{2}; -r_{*}^{2}(\beta_{\mu}+\beta_{\nu})^{2}) ],
\end{aligned}
\end{equation}

\end{document}